\documentclass[sigconf, prologue, table]{acmart}

\AtBeginDocument{%
  \providecommand\BibTeX{{%
    \normalfont B\kern-0.5em{\scshape i\kern-0.25em b}\kern-0.8em\TeX}}}

\setcopyright{acmcopyright}
\copyrightyear{2022}
\acmYear{2022}
\acmDOI{XXXXXXX.XXXXXXX}

\acmConference[Conference acronym 'XX]{Make sure to enter the correct
  conference title from your rights confirmation emai}{June 03--05,
  2018}{Woodstock, NY}
\acmPrice{15.00}
\acmISBN{978-1-4503-XXXX-X/18/06}

\usepackage{xcolor}
\usepackage{caption}
\usepackage{cellspace}

\usepackage{relsize}

\usepackage{algpseudocode}
\usepackage{algorithm}
\usepackage{tikz}
\usepackage{tikz-qtree}

\usepackage{comment}

\usepackage{amsmath}
\usepackage{bm}
\usepackage{bm}
\usepackage{amsthm}
\usepackage{mathtools}
\usepackage{etoolbox}
\usepackage{xstring} 

\usepackage{stmaryrd}
\usepackage[normalem]{ulem}
\usepackage{subcaption}
\usepackage{booktabs}
\usepackage[disable]
{todonotes}

\usepackage{graphicx}
\usepackage{listings}
\usepackage{mdframed}
\lstdefinestyle{psql}
{
backgroundcolor=\color{black!15!white},
tabsize=2,
basicstyle=\small\upshape\ttfamily,
language=SQL,
morekeywords={PROVENANCE,BASERELATION,INFLUENCE,COPY,ON,TRANSPROV,TRANSSQL,TRANSXML,CONTRIBUTION,COMPLETE,TRANSITIVE,NONTRANSITIVE,EXPLAIN,SQLTEXT,GRAPH,IS,ANNOT,THIS,XSLT,MAPPROV,cxpath,OF,TRANSACTION,SERIALIZABLE,COMMITTED,INSERT,INTO,WITH,SCN,UPDATED,LENS,SCHEMA_MATCHING,string,WINDOW,max,OVER,PARTITION,FIRST_VALUE,WITH},
extendedchars=false,
keywordstyle=\bfseries,
mathescape=true,
escapechar=@,
sensitive=true
}
\usepackage{etoolbox}
\usepackage{wrapfig}
\usepackage{fancyvrb}
\usepackage{caption}
\usepackage{subcaption}
\usepackage{braket}
\usepackage[inline]{enumitem}
\usepackage{xspace}
\usepackage{hyperref}
\usepackage{url}
\usepackage{cleveref}
\usepackage{color}
\usepackage{adjustbox}

\graphicspath{ {figures/} }

\usepackage{outlines}
\usepackage{enumitem}

\newcommand{\xplural}{s\xspace}
\xspaceaddexceptions{\xplural}

\newcommand{\draft}{0} 
\ifnum\draft=0

\newcommand{\BG}[1]{\todo[inline]{\textbf{Boris says:$\,$} #1}}
\newcommand{\SF}[1]{\todo{\textbf{Su says:$\,$} #1}}
\newcommand{\OK}[1]{\todo[color=gray]{\textbf{Oliver says:$\,$} #1}}
\newcommand{\AH}[1]{\todo[inline, backgroundcolor=cyan, caption={}]{\textbf{Aaron says:$\,$} #1}}
\newcommand{\AR}[1]{\todo[inline,color=green]{\textbf{Atri says:$\,$} #1}}

\else
\newcommand{\BG}[1]{}
\newcommand{\SF}[1]{}
\newcommand{\OK}[1]{}
\newcommand{\AH}[1]{}
\newcommand{\AR}[1]{}
\fi

\DeclareMathAlphabet{\mathbbold}{U}{bbold}{m}{n}

\newtheorem{Theorem}{Theorem}[section]
\newtheorem{Definition}[Theorem]{Definition}
\newtheorem{Lemma}[Theorem]{Lemma}
\newtheorem{Proposition}[Theorem]{Proposition}
\newtheorem{Corollary}[Theorem]{Corollary}
\newtheorem{Example}[Theorem]{Example}
\newtheorem{hypo}[Theorem]{Conjecture}
\newtheorem{Problem}[Theorem]{Problem}

\newtoggle{shrinkspace}
\dimdef{\abovecapshrink}{-0.25cm}
\dimdef{\belowcapshrink}{-0.53cm}
\newcommand{\savecaptionspace}[3]{\iftoggle{shrinkspace}{\setlength{\abovecaptionskip}{#2}#1\vspace{#3}}{#1}}

\newcommand{\tup}{t}
\newcommand{\rel}{R}

\newcommand{\db}{D}
\newcommand{\query}{Q}
\newcommand{\qhard}{\query_{hard}}
\newcommand{\join}{\mathlarger\Join}
\newcommand{\select}{\sigma}
\newcommand{\project}{\pi}
\newcommand{\union}{\cup}

\newcommand{\attr}[1]{attr\left(#1\right)}

\newcommand{\termStepOne}{Lineage Computation\xspace}
\newcommand{\abbrStepOne}{LC\xspace}
\newcommand{\termStepTwo}{Expectation Computation\xspace}
\newcommand{\abbrStepTwo}{EC\xspace}
\newcommand{\expectProblem}{\textsc{Expected Result Multiplicity Problem}\xspace}

\newcommand{\abbrSMB}{SMB\xspace}

\newcommand{\domain}{\func{Dom}}
\newcommand{\func}[1]{\textsc{#1}\xspace}
\newcommand{\isInd}[1]{\func{isInd}\inparen{#1}}
\newcommand{\polyf}{\func{poly}}

\newcommand{\degree}{\func{deg}}
\newcommand{\size}{\func{size}}
\newcommand{\depth}{\func{depth}}
\newcommand{\topord}{\func{TopOrd}}
\newcommand{\smbOf}[1]{\func{\abbrSMB}\inparen{#1}}

\newcommand{\udom}{\mathcal{U}}
\newcommand{\domK}{K}
\newcommand{\semK}{\mathcal{K}}

\newcommand{\semN}{\mathbb{N}}
\newcommand{\semNX}{\mathbb{N}[\vct{X}]}
\newcommand{\onesymbol}{\mathbbold{1}}
\newcommand{\zerosymbol}{\mathbbold{0}}
\newcommand{\multsymb}{\otimes}
\newcommand{\addsymbol}{\oplus}
\newcommand{\addK}{\addsymbol_{\semK}}
\newcommand{\multK}{\multsymb_{\semK}}
\newcommand{\oneK}{\onesymbol_{\semK}}
\newcommand{\zeroK}{\zerosymbol_{\semK}}

\newcommand{\idb}{{\Omega}}
\newcommand{\pd}{{\mathcal{P}}}
\newcommand{\pdassign}{\mathcal{P}}
\newcommand{\pdb}{\mathcal{D}}
\newcommand{\dbbase}{\db_\idb}
\newcommand{\dbbaseName}{deterministic bounding database\xspace}
\newcommand{\pxdb}{\pdb_{\semNX}}
\newcommand{\pndb}{\pdb_{\semN}}

\newcommand{\bound}{c}

\newcommand{\tupset}{D}
\newcommand{\gentupset}{\overline{D}}
\newcommand{\world}{[0,\bound]}
\newcommand{\worldvec}{\vct{W}}
\newcommand{\worlds}{\world^\tupset}
\newcommand{\bpd}{\mathcal{P}}
\newcommand{\block}{B}

\newcommand{\onebidbworlds}[1]{\bigtimes_{\tup\in #1}\inset{0, \bound_\tup}}

\newcommand{\abbrOneBIDB}{\text{Binary-BIDB}\xspace}
\newcommand{\abbrPDB}{\textnormal{PDB}\xspace}
\newcommand{\abbrBPDB}{\textnormal{bag-PDB}\xspace}
\newcommand{\abbrTIDB}{\textnormal{TIDB}\xspace}
\newcommand{\abbrCTIDB}{\textnormal{$\bound$-TIDB}\xspace}
\newcommand{\abbrBIDB}{\textnormal{BIDB}\xspace}
\newcommand{\ti}{TIDB\xspace}
\newcommand{\tis}{TIDBs\xspace}
\newcommand{\bi}{BIDB\xspace}
\newcommand{\bis}{BIDBs\xspace}
\newcommand{\abbrNXPDB}{$\semNX$-encoded PDB\xspace}

\newcommand{\probAllTup}{\vct{\prob}}

\newcommand{\domN}{\mathbb{N}}

\newcommand{\expct}{\mathop{\mathbb{E}}}
\newcommand{\probOf}{Pr}
\newcommand{\abs}[1]{\left|#1\right|}
\newcommand{\suchthat}{\;s.t.\;} 
\newcommand{\comprehension}[2]{\left\{\;#1\;|\;#2\;\right\}}

\newcommand{\pbox}[1]{\left[#1\right]}
\newcommand{\pbrace}[1]{\left\{#1\right\}}
\newcommand{\inparen}[1]{\left({#1}\right)}
\newcommand{\inset}[1]{\left\{{#1}\right\}}
\newcommand{\intuple}[1]{\left\langle{#1}\right\rangle}

\newcommand{\prob}{p}

\newcommand{\pVar}{X}
\newcommand{\kElem}{k}
\newcommand{\randWorld}{W}
\newcommand{\rvworld}{\vct{\randWorld}}
\newcommand{\randDB}{\vct{\db}}

\newcommand{\polyinput}[2]{\left(#1,\ldots, #2\right)}
\newcommand{\numvar}{n}
\newcommand{\numblock}{m}
\newcommand{\vct}[1]{{\bf #1}}

\newcommand{\hideg}{K}
\newcommand{\poly}{\Phi}
\newcommand{\monomial}[1]{M_{{#1}}}
\newcommand{\genpoly}{\phi}

\newcommand{\polyOf}[1]{\poly[#1]}
\newcommand{\polyqdt}[3]{\polyOf{#1,#2,#3}}
\newcommand{\apolyqdt}{\polyqdt{\query}{\tupset}{\tup}}
\newcommand{\nxpolyqdt}{\polyqdt{\query}{\db_{\semNX}}{\tup}}

\newcommand{\polyX}{\poly\inparen{\vct{\pVar}}}
\newcommand{\rpoly}{\widetilde{\poly}}
\newcommand{\rmonomial}[1]{\widetilde{\monomial{#1}}}
\newcommand{\refpoly}[1]{\poly_{#1R}}

\newcommand{\vset}{V}
\newcommand{\edgeSet}{E}
\newcommand{\gtype}[1]{\inparen{#1}}
\newcommand{\esetType}[1]{\edgeSet^{\gtype{#1}}}
\newcommand{\graph}[1]{G^{(#1)}}
\newcommand{\numocc}[2]{\#\left(#1,#2\right)}
\newcommand{\eset}[1]{E^{(#1)}_S} 

\newcommand{\dtrm}[1]{Det\left(#1\right)}
\newcommand{\tuple}[1]{\left<#1\right>}
\newcommand{\indicator}[1]{\onesymbol_{#1}}

\newcommand{\circuit}{\vari{C}}
\newcommand{\circuitset}[1]{\vari{CSet}\inparen{#1}}
\newcommand{\circmult}{\times}
\newcommand{\circplus}{+}
\newcommand{\rinput}{\vari{R}}
\newcommand{\linput}{\vari{L}}

\newcommand{\subcircuit}{\vari{S}}
\newcommand{\gate}{\vari{g}}
\newcommand{\lwght}{\vari{Lweight}}
\newcommand{\rwght}{\vari{Rweight}}
\newcommand{\prt}{\vari{partial}}
\newcommand{\degval}{\vari{degree}}
\newcommand{\type}{\vari{type}}
\newcommand{\val}{\vari{val}}
\newcommand{\var}{\textsc{var}\xspace}
\newcommand{\tnum}{\textsc{num}\xspace}

\newcommand{\cost}{\func{Cost}}

\newcommand{\qClass}{\mathcal{Q}}
\newcommand{\raPlus}{\ensuremath{\mathcal{RA}^{+}}\xspace}

\newcommand{\bigO}[1]{O\inparen{#1}}
\newcommand{\littleo}[1]{o\inparen{#1}}
\newcommand{\bigOmega}[1]{\Omega\inparen{#1}}
\newcommand{\littleomega}[1]{\omega\inparen{#1}}

\newcommand{\sharpphard}{\#{\sf P}-hard\xspace}
\newcommand{\sharpwone}{\#{\sf W}[1]\xspace}
\newcommand{\sharpwzero}{\#{\sf W}[0]\xspace}
\newcommand{\sharpwonehard}{\#{\sf W}[1]-hard\xspace}
\newcommand{\ptime}{{\sf PTIME}\xspace}
\newcommand{\timeOf}[1]{T_{#1}}
\newcommand{\qruntime}[1]{T_{det}\inparen{#1}}
\newcommand{\optquery}[1]{{#1}} 
\newcommand{\qruntimenoopt}[1]{T_{det}\inparen{#1}}
\newcommand{\jointime}[1]{T_{join}(#1)}
\newcommand{\kmatchtime}{T_{match}\inparen{k, G}}

\newcommand{\randvar}{\vari{Y}}

\newcommand{\samplesize}{N}

\newcommand{\empmean}{\overline{\vct{\randvar}}}

\newcommand{\error}{\epsilon}
\newcommand{\conf}{\delta}

\newcommand{\algname}[1]{\textsc{#1}\xspace}
\newcommand{\approxq}{\algname{Approximate$\rpoly$}}
\newcommand{\onepass}{\algname{OnePass}}
\newcommand{\sampmon}{\algname{SampleMonomial}}

\newcommand{\lincirc}{\algname{LineageCircuit}}
\newcommand{\ceil}[1]{\left\lceil #1 \right\rceil}
\newcommand{\vari}[1]{\texttt{#1}\xspace}
\newcommand{\accum}{\vari{acc}}
\newcommand{\numsamp}{\vari{N}}
\newcommand{\numedge}{m}

\newcommand{\etree}{\vari{T}}
\newcommand{\stree}{\vari{S}}
\newcommand{\lchild}{\vari{L}}
\newcommand{\rchild}{\vari{R}}

\newcommand{\vpartial}{\vari{partial}}

\newcommand{\expansion}[1]{\vari{E}(#1)}
\newcommand{\elist}[1]{\vari{List}\pbox{#1}}

\newcommand{\monom}{\vari{v}}
\newcommand{\encMon}{\monom_{\vari{m}}}

\newcommand{\coef}{\vari{c}}

\newcommand{\rmod}{Mod}
\newcommand{\reprs}{\mathcal{M}}
\newcommand{\repr}{M}

\newcommand{\assign}{\psi}
\newcommand{\support}[1]{supp({#1})}

\newcommand{\eps}{\epsilon}

\newcommand{\mypar}[1]{\smallskip\noindent\textbf{{#1}.}}

\newcommand{\caseheading}[1]{\smallskip \noindent \textbf{#1}.~}

\allowdisplaybreaks

\newcommand{\multc}[2]{\overline{\mathcal{M}}\left({#1},{#2}\right)}

	\newcommand{\patternshift}[1]{\hspace*{-0.5mm}\raisebox{-0.35mm}{#1}\hspace*{-0.5mm} }
	\tikzset{
		default_node/.style={align=center, inner sep=0pt},
		pattern_node/.style={fill=gray!50, draw=black, semithick, inner sep=0pt, minimum size = 2pt, circle},
		tree_node/.style={default_node, draw=black, black, circle, text width=0.5cm, font=\bfseries, minimum size=0.65cm},
		gen_tree_node/.style={default_node, draw, circle, text width=0.5cm, font=\bfseries, minimum size=0.65cm},
		highlight_color/.style={black}, wght_color/.style={black},
		highlight_treenode/.style={tree_node, draw=black, black},
		edge from parent path={(\tikzparentnode) -- (\tikzchildnode)}
		}
	\newcommand{\ed}{\patternshift{
					\begin{tikzpicture}[every path/.style={thick, draw}]
						\node at (0, 0)[pattern_node](bottom){};
						\node [above=0.07cm of bottom, pattern_node] (top){};
						\draw (top) -- (bottom);
					\end{tikzpicture}
				}
			 }

	\newcommand{\kmatch}{\ed\cdots\ed^\kElem}
	\newcommand{\twodis}{\patternshift{
						\begin{tikzpicture}[every path/.style={thick, draw}]
							\node at (0, 0) [pattern_node] (bottom1) {};
							\node[above=0.07cm of bottom1, pattern_node] (top1) {} edge (bottom1);
							\node at (0.14, 0) [pattern_node] (bottom2) {};
							\node [above=0.07cm of bottom2, pattern_node] (top2) {} edge (bottom2);
						\end{tikzpicture}
						}
					}
	\newcommand{\twopath}{\patternshift{
						\begin{tikzpicture}[every path/.style={thick, draw}]
							\node at (0, 0.08) [pattern_node] (top){};
							\node [below left=0.095cm and 0.05cm of top, pattern_node](left){};
							\node[below right=0.095cm and 0.05cm of top, pattern_node](right){};
							\draw (top) -- (left);
							\draw (top) -- (right);
						\end{tikzpicture}
					}
				}
	\newcommand{\threedis}{\patternshift{
						\begin{tikzpicture}[every path/.style={thick, draw}]
							\node at (0, 0) [pattern_node] (bottom1) {};
							\node[above=0.07cm of bottom1, pattern_node] (top1) {} edge (bottom1);
							\node at (0.14, 0) [pattern_node] (bottom2) {};
							\node [above=0.07cm of bottom2, pattern_node] (top2) {} edge (bottom2);
							\node at (0.28, 0) [pattern_node] (bottom3) {};
							\node [above=0.07cm of bottom3, pattern_node] (top3) {} edge (bottom3);
						\end{tikzpicture}
					}
				}
	\newcommand{\tri}{\patternshift{
						\begin{tikzpicture}[every path/.style={ thick, draw}]
							\node at (0, 0.08) [pattern_node] (top){};
							\node [below left=0.08cm and 0.01cm of top, pattern_node](left){} edge (top);
							\node[below right=0.08cm and 0.01cm of top, pattern_node](right){} edge (top) edge (left);
						\end{tikzpicture}
					}
				}
	\newcommand{\twopathdis}{\ed~\twopath}
	\newcommand{\threepath}{\patternshift{
						\begin{tikzpicture}[every path/.style={thick, draw}]
							\node at (0, 0) [pattern_node] (node1a) {};
							\node [above=0.07cm of node1a, pattern_node] (node1b) {} edge (node1a);
							\node [right=0.099cm of node1a, pattern_node] (node2b) {}; 
							\node [above=0.07cm of node2b, pattern_node] (node3b) {} edge (node2b);
							\draw (node1b) -- (node3b);
						\end{tikzpicture}
					}
				}
	\newcommand{\oneint}{\patternshift{
						\begin{tikzpicture}[level/.style={sibling distance=0.14cm, level distance=0.15cm}, every path/.style={thick, draw}]
							\node at (0, 0) [pattern_node] {} [grow=down]
								child{node [pattern_node]{}}
								child {node [pattern_node] {}}
								child{node [pattern_node] {}};
						\end{tikzpicture}
					}
				}
\newcommand{\sg}[1]{S^{(#1)}}

\crefname{example}{ex.}{ex.}
\Crefname{example}{Ex.}{Ex.}
\Crefname{figure}{Fig.}{Fig.}
\Crefname{section}{Sec.}{Sec.}
\Crefname{definition}{Def.}{Def.}
\Crefname{theorem}{Thm.}{Thm.}
\Crefname{lemma}{Lem.}{Lem.}
\crefname{equation}{eq.}{eq.}
\Crefname{equation}{Eq.}{Eq.}

\begin{document}

\title{Computing Expected Multiplicities for Bag-TIDBs with Bounded Multiplicities}

\author{Su Feng}\email{sfeng14@hawk.iit.edu}
\author{Boris Glavic}\email{bglavic@iit.edu}
\affiliation{%
 \institution{Illinois Institute of Technology , USA}
 \city{Chicago}
 \state{New York}
 \country{USA}
 }
\author{Aaron Huber}\email{ahuber@buffalo.edu}
\author{Oliver Kennedy}\email{okennedy@buffalo.edu}
\author{Atri Rudra}\email{atri@buffalo.edu}
\affiliation{%
 \institution{University at Buffalo, USA}
  \city{Buffalo}
   \state{New York}
   \country{USA}
}
\renewcommand{\shortauthors}{Huber, Kennedy, Rudra, et al.}

\begin{abstract}
We study the problem of computing a query result tuple's expected multiplicity for probabilistic databases under bag semantics (where each tuple is associated with a multiplicity) exactly and approximately.
Specifically, we are interested in the fine-grained complexity of this problem for \abbrCTIDB\xplural, i.e., probabilistic databases where tuples are independent probabilistic events and the multiplicity of each tuple is bound by a constant $\bound$.
Unfortunately, our results imply that computing expected multiplicities for \abbrCTIDB\xplural 
introduces super-linear overhead over the corresponding deterministic query evaluation algorithms (under certain complexity hardness conjectures).
  Next, we develop a sampling algorithm that computes a $(1 \pm \epsilon)$-approximation of the expected multiplicity of an output tuple in time linear in the runtime of the corresponding deterministic query for any   positive relational algebra ($\raPlus$) query over \abbrCTIDB\xplural and for a non-trivial subclass of block-independent databases. 


\end{abstract}





\maketitle

\lstset{language=sql}

\section{Introduction}\label{sec:intro}

We explore the problem of computing the expectation of the multiplicity of a tuple in the result of a query over a \abbrCTIDB (tuple independent database), a type of probabilistic database with bag semantics where the multiplicity of a tuple is a random variable with range $[0,\bound]\stackrel{\text{def}}{=}\{0,1,\dots,\bound\}$ for some fixed constant $\bound$, and multiplicities assigned to any two tuples are independent of each other.
Formally, a \abbrCTIDB,
$\pdb = \inparen{\worlds, \bpd}$ is defined over a \dbbaseName (i.e. a `base' set of tuples) $\tupset$ and a probability distribution $\bpd$ over all possible worlds generated by assigning each tuple $\tup \in \tupset$ a multiplicity in the range $[0,\bound]$.
Any such world can be encoded as a vector (of length $\numvar=\abs{\tupset}$) from $\worlds$, such that the multiplicity of each $\tup \in \tupset$ is stored at a distinct index.
A given world $\worldvec \in\worlds$ can be interpreted as follows: for each $\tup \in \tupset$, $\worldvec_{\tup}$ is the multiplicity of $\tup$ in $\worldvec$.
We note that encoding a possible world as a vector, while non-standard, is equivalent to encoding it as a bag of tuples (\Cref{app:subsec:background-nxdbs}). 
Given that tuple multiplicities are independent events, the  probability distribution $\bpd$ can be expressed compactly by assigning each tuple a  probability distribution over $[0,\bound]$. Let $\prob_{\tup,j}$ denote the probability that tuple $\tup$ is assigned multiplicity $j$. The probability of a world $\worldvec$ is then $\prod_{\tup \in \tupset} \prob_{\tup,j(t)}$ for $j(t) = \worldvec_{\tup}$.
%
In this work, we consider \emph{queries with bag semantics} over such bag probabilistic databases.
We denote by $\query\inparen{\worldvec}\inparen{\tup}$ the multiplicity of a result tuple $\tup$ in query $\query$ over possible world $\worldvec\in\worlds$.
We now formally state our problem of computing the expected multiplicity: 

\begin{Problem}\label{prob:expect-mult}
Given \abbrCTIDB $\pdb = \inparen{\worlds, \bpd}$, $\raPlus$ query\footnote{
An $\raPlus$ query is a query expressed in positive relational algebra, i.e., using only the operators selection ($\select$), projection ($\project$), natural join ($\join$) and union ($\union$).
}
 $\query$, and result tuple $\tup$, compute the expectation $\expct_{\rvworld\sim\bpd}\pbox{\query\inparen{\rvworld}\inparen{\tup}}$.
\end{Problem}

\begin{figure}[t!]
  \begin{align*}
  	&\begin{aligned}[t]
	  &\polyqdt{\project_A(\query)}{\gentupset}{\tup}\inparen{\vct{X}} =\\
	  &~\sum_{\tup': \project_A(\tup') = \tup} \polyqdt{\query}{\gentupset}{\tup'}\inparen{\vct{X}}
	  \end{aligned}
	  &
	  &\begin{aligned}[t]
		  &\polyqdt{\query_1 \union \query_2}{\gentupset}{\tup}\inparen{\vct{X}} =\\
		  &~ \polyqdt{\query_1}{\gentupset}{\tup}\inparen{\vct{X}} + \polyqdt{\query_2}{\gentupset}{\tup}\inparen{\vct{X}}\\
	  \end{aligned}\\
	  &\begin{aligned}
		  &\polyqdt{\select_\theta(\query)}{\gentupset}{\tup}\inparen{\vct{X}} =\\
		  &~ \begin{cases}
		    \polyqdt{\query}{\gentupset}{\tup}\inparen{\vct{X}} & \text{if }\theta(\tup) \\
		    0                       & \text{otherwise}.
		    \end{cases}
	  \end{aligned}
	  &
       &\begin{aligned}
	          &\polyqdt{\query_1 \join \query_2}{\gentupset}{\tup}\inparen{\vct{X}} =\\
	          &~~\polyqdt{\query_1}{\gentupset}{\project_{\attr{\query_1}}{\tup}}\inparen{\vct{X}}\\
	          &~\cdot\polyqdt{\query_2}{\gentupset}{\project_{\attr{\query_2}}{\tup}}\inparen{\vct{X}}
          \end{aligned}
	\end{align*}
	\savecaptionspace{
	\caption{Lineage polynomial semantics for any $\raPlus$ query $\query$, arbitrary \dbbaseName $\gentupset$ (with variables $\vct{X}=\inparen{X_\tup}_{\tup \in\gentupset}$), where for $\rel\in\gentupset$, $\tup\in\rel$, the base case is $\polyqdt{\rel}{\gentupset}{\tup}\inparen{\vct{X}} = X_\tup$.}
	\label{fig:nxDBSemantics}
	}{\abovecapshrink}{\belowcapshrink}
\end{figure}

As also observed in \cite{https://doi.org/10.48550/arxiv.2201.11524,DBLP:journals/sigmod/GuagliardoL17}, computing the expected multiplicity of a result tuple in a bag probabilistic database is the analog of computing the marginal probability in a set \abbrPDB.

\mypar{Hardness of Set Query Semantics and Bag Query Semantics}
Set query evaluation semantics over $1$-\abbrTIDB\xplural have been studied extensively, and their data complexity has, in general been shown 
to be \sharpphard\cite{10.1145/1265530.1265571}.
In an independent work, Grohe et. al.~\cite{https://doi.org/10.48550/arxiv.2201.11524} studied bag-\abbrTIDB\xplural with unbounded multiplicities, which requires 
a succinct representation of probability distributions over infinitely many multiplicities and proved a dichotomy for
computing the probability of an output tuple's multiplicity being bounded by given  $s$.

\mypar{Our Setup} In contrast to~\cite{https://doi.org/10.48550/arxiv.2201.11524}, we consider \abbrCTIDB\xplural, i.e., the multiplicity of input tuples is bound by a constant $\bound$.
Then, 
there exists a trivial \ptime algorithm for computing the expectation of a result tuple's multiplicity~(\Cref{prob:expect-mult}) for any fixed $\raPlus$ query  due to linearity of expectation (see~\Cref{sec:intro-poly-equiv}).
Since the {\em data complexity} of~\Cref{prob:expect-mult} is in \ptime, 
we then explore the question of  
 computing expectation using fine-grained and parameterized complexity, where we are interested in the exponent of polynomial runtime.\footnote{While 
  \cite{https://doi.org/10.48550/arxiv.2201.11524} also observes that computing the expectation of an output tuple multiplicity is in \ptime, it does not investigate the fine-grained complexity of this problem.}

Specifically, in this work we ask if~\Cref{prob:expect-mult} can be solved in time linear in the runtime of an analogous deterministic query, which we make more precise shortly. 
  If true, this opens up the way for deployment of \abbrCTIDB\xplural in practice. We expand on the practical implications of this problem later in the section but for now we stress that in practice, $\bound$ is indeed constant and most often $\bound=1$.
  That is, although production database systems use bag semantics for query evaluation, allowing duplicate intermediate or output tuples, input tuples in real world datasets are still frequently unique.
  To analyze this question we denote by $\timeOf{}^*(\query,\pdb, \bound)$ the optimal runtime complexity of computing~\Cref{prob:expect-mult} over \abbrCTIDB $\pdb$ and query $\query$.

\begin{table*}[t!]
\centering
\begin{tabular}{|p{0.43\textwidth}|p{0.12\textwidth}|p{0.35\textwidth}|}
\hline
\textbf{Lower bound on $\timeOf{}^*(\qhard^k,\pdb,1)$} & \textbf{Num.} $\bpd$s
  & \textbf{Hardness Assumption}\\
\hline
$\Omega\inparen{\inparen{\qruntime{\optquery{\qhard^k}, \tupset, \bound}}^{1+\eps_0}}$ for {\em some} $\eps_0>0$ & Single & Triangle Detection hypothesis\\
$\omega\inparen{\inparen{\qruntime{\optquery{\qhard^k}, \tupset, \bound}}^{C_0}}$ for {\em all} $C_0>0$ & Multiple &$\sharpwzero\ne\sharpwone$\\
$\Omega\inparen{\inparen{\qruntime{\optquery{\qhard^k}, \tupset, \bound}}^{c_0\cdot k/\log{k}}}$ for {\em some} $c_0>0$ & Multiple & Exponential Time Hypothesis (ETH)\\
\hline
\end{tabular}
\savecaptionspace{
\caption{Our lower bounds for $\qhard^k$ parameterized by $k$  over $1$-TIDB  $\pdb$.  
Those with `Multiple' in the second column need the algorithm to be able to handle multiple $\bpd$.  See~\Cref{sec:hard} for further details.}
\label{tab:lbs}
}{0cm}{-0.73cm}
\end{table*}

\mypar{Our lower bound results}
Let $\qruntime{\query,\gentupset,\bound}$ (see~\Cref{sec:gen} for a formal definition) denote the runtime for query $\query$ over any deterministic database 
that maps each tuple in $\gentupset$ to a multiplicity in $[0,\bound]$.
Our question is whether or not it is always true that for every $\query$,  $\timeOf{}^*\inparen{\query, \pdb, \bound}\leq \bigO{\qruntime{\optquery{\query}, \tupset, \bound}}$.  We remark that the issue of query optimization is orthogonal to this question (recall that an $\raPlus$ query explicitly encodes order of operations) since we want to answer the above question for {\em all} $\query$. \emph{Specifically, if there is an equivalent and more efficient query $\query'$, we allow both deterministic and probabilistic query processing access to $\query'$}.

Unfortunately the answer to the above question  is no--
 \Cref{tab:lbs} shows our results.
Specifically, depending on what hardness result/conjecture we assume, we get various weaker or stronger versions of {\em no} as an answer to our question.  To make some sense of the  lower bounds in \Cref{tab:lbs}, we note that it is not too hard to show that $\timeOf{}^*(\query,\pdb, \bound) \le  \bigO{\inparen{\qruntime{\optquery{\query}, \tupset, \bound}}^k}$, where $k$ is the join width of $\query$ (our notion of join width 
is essentially the degree of the corresponding lineage polynomial we introduce in \Cref{sec:intro-poly-equiv}).
%
What our lower bound in the third row says, is that for a specific family of hard queries, one cannot get more than a polynomial improvement (for fixed $k$) over essentially the trivial algorithm for~\Cref{prob:expect-mult}, assuming the Exponential Time Hypothesis (ETH)~\cite{eth}.
We also note that existing results\footnote{This claim follows when we set $\query$ to the query that counts the number of $k$-cliques over database $\tupset$ that encodes a graph. Precisely the same bounds as in the three rows of~ \Cref{tab:lbs} (with $n$ replacing $\qruntime{\optquery{\query}, \tupset, \bound}$) follow from the same complexity assumptions we make: triangle detection hypothesis (by definition), $\sharpwzero\ne\sharpwone$~\cite{10.5555/645413.652181} and Strong ETH~\cite{CHEN20061346}. For the last result we can replace $k/\log{k}$ by just $k$.
 } 
 imply the claimed lower bounds if we replace the $\qruntime{\optquery{\query}, \tupset, \bound}$ by just $\numvar = |\tupset|$.
 Our contribution is to identify a family of hard queries where deterministic query processing is `easy' but computing the expected multiplicities is hard.

\mypar{Our upper bound results} We introduce a $(1\pm \epsilon)$-approximation algorithm that computes ~\Cref{prob:expect-mult} in time $O_\epsilon\inparen{\qruntime{\optquery{\query}, \tupset, \bound}}$.  This means, when we are okay with approximation, that we solve~\Cref{prob:expect-mult} in time linear in the size of the deterministic query\BG{What is the size of the deterministic query?}.  
In contrast, known approximation techniques (\cite{DBLP:conf/icde/OlteanuHK10,DBLP:journals/jal/KarpLM89}) in set-\abbrPDB\xplural need time $\Omega(\qruntime{\optquery{\query}, \tupset, \bound}^{2k})$
(see \Cref{sec:karp-luby}).
Further, our approximation algorithm works for a more general notion of bag \abbrPDB\xplural beyond \abbrCTIDB\xplural
(see \Cref{subsec:tidbs-and-bidbs}).

\subsection{Polynomial Equivalence}\label{sec:intro-poly-equiv}
A common encoding of probabilistic databases (e.g., in \cite{IL84a,4497507,DBLP:conf/vldb/AgrawalBSHNSW06} and many others) annotates tuples with lineages, propositional formulas that describe the set of possible worlds that the tuple appears in.  The bag semantics analog is a provenance/lineage polynomial (see~\Cref{fig:nxDBSemantics}) $\apolyqdt$~\cite{DBLP:conf/pods/GreenKT07}, a polynomial with non-zero integer coefficients and exponents, over variables $\vct{X}$ encoding input tuple multiplicities. The lineage polynomial for result tuple $t_{out}$ evaluates to $t_{out}$'s multiplicity in a given possible world when each $X_{t_{in}}$ is replaced by the multiplicity of $t_{in}$ in the possible world.

 We now state the problem of computing the expectation of tuple multiplicity in terms of lineage polynomials (which is equivalent to \Cref{prob:bag-pdb-poly-expected}-- see \Cref{prop:expection-of-polynom}):
\begin{Problem}[Expected Multiplicity of Lineage Polynomials]\label{prob:bag-pdb-poly-expected}
Given an $\raPlus$ query $\query$, \abbrCTIDB $\pdb$ and result tuple $\tup$, 
compute $\expct_{\vct{W}\sim \pdassign}\pbox{\apolyqdt\inparen{\worldvec}}$).
\end{Problem}

We drop $\query$, $\tupset$, and $\tup$ from $\apolyqdt$ when they are clear from the context or not relevant to the discussion.
All of our results rely on working with a {\em reduced} form $\rpoly$ of the lineage polynomial $\poly$. As we show, for the $1$-\abbrTIDB case, computing the expected multiplicity (over bag query semantics) is {\em exactly} the same as evaluating $\rpoly$ over the probabilities that define the $1$-\abbrTIDB.  
Further, only light extensions (see \Cref{def:reduced-poly-one-bidb}) are required to support block independent disjoint probabilistic databases~\cite{DBLP:conf/icde/OlteanuHK10}. 

Next, we motivate the reduced polynomial $\rpoly$.
Consider the query $\query_1$ defined as follows over the bag relations of \Cref{fig:two-step}:

\begin{lstlisting}[frame=single,framerule=0pt]
SELECT DISTINCT 1 FROM T $t_1$, R r, T $t_2$
WHERE $t_1$.Point = r.Point$_1$ AND $t_2$.Point = r.Point$_2$
\end{lstlisting}

It can be verified that $\poly\inparen{A, B, C, E, U, Y, Z}$ for the sole result tuple of $\query_1$ is $AUB + BYE + BZC$. Now consider the product query $\query_1^2 = \query_1 \times \query_1$.
The lineage polynomial for $\query_1^2$ is $\poly_1^2\inparen{A, B, C, E, U, Y, Z}$
$$
=A^2U^2B^2 + B^2Y^2E^2 + B^2Z^2C^2 + 2AUB^2YE + 2AUB^2ZC + 2B^2YEZC.
$$
To compute $\expct\pbox{\poly_1^2}$ we can use linearity of expectation and push the expectation through each summand.  To keep things simple, let us focus on the monomial $\monomial{1}(A,U,B) \stackrel{\text{def}}{=} A^2U^2B^2$ as the procedure is the same for all other monomials of $\poly_1^2$.  Let $\randWorld_X$ be the random variable corresponding to a variable $X$. Because the distinct variables in the product are independent, we can push expectation through them yielding $\expct\pbox{\randWorld_A^2\randWorld_U^2\randWorld_B^2}=\expct\pbox{\randWorld_A^2}\expct\pbox{\randWorld_U^2}\expct\pbox{\randWorld_B^2}$.  Since $\randWorld_A, \randWorld_B\in \inset{0, 1}$ we can simplify to $\expct\pbox{\randWorld_A}\expct\pbox{\randWorld_U^2}\expct\pbox{\randWorld_B}$ by the fact that for any $W\in \inset{0, 1}$, $W^2 = W$.  Observe that if $W_U\in\inset{0, 1}$, then we further would have $\expct\pbox{\randWorld_A}\expct\pbox{\randWorld_U}\expct\pbox{\randWorld_B} = \prob_A\cdot\prob_U\cdot\prob_B$ 
(denoting $\probOf\pbox{\randWorld_X = 1} \stackrel{\text{def}}{=} \prob_X$).  However, in this example, we get stuck with $\expct\pbox{\randWorld_U^2}$, since $\randWorld_U\in\inset{0, 1, 2}$ and for $\randWorld_U \gets 2$, $\randWorld_U^2 \neq \randWorld_U$.

The simple insight to get around this issue is to note that the random variables $\randWorld_U$ and $\randWorld_{U_1}+2\randWorld_{U_2}$ have exactly the same distribution, where $\randWorld_{U_1},\randWorld_{U_2}\in\inset{0,1}$ and $\probOf\pbox{\randWorld_{U_j} = 1} = \probOf\pbox{\randWorld_{U} = j}$. Thus, the idea is to replace the variable $U$ by $U_1+2U_2$ (where $U_j$ corresponds to the event that $U$ has multiplicity $j$) yielding
%
%
\[\monomial{1,R}\inparen{A, U_1, U_2, B} \stackrel{\text{def}}{=}  \monomial{1}\inparen{A,(U_1+2U_2),B}.\]

Given that $U$ can only have multiplicity of $1$ or $2$ but not both, 
given world vectors $(\randWorld_A,\randWorld_{U_1},\randWorld_{U_2},\randWorld_B)\in\inset{0,1}^4$, we have $\expct\pbox{\randWorld_{U_1}\randWorld_{U_2}}=0$. Further, since the world vectors are Binary vectors, we have $\expct\pbox{\monomial{1,R}}=\expct\pbox{\randWorld_{A}}\expct\pbox{\randWorld_{U_1}}\expct\pbox{\randWorld_{B}}+$  $4\expct\pbox{\randWorld_{A}}\expct\pbox{\randWorld_{U_2}}\expct\pbox{\randWorld_{B}}\stackrel{\text{def}}{=}\rmonomial{1}\inparen{p_A,\probOf\inparen{U=1},\probOf\inparen{U=2},p_B}$.  We only did the argument for a single monomial but by linearity of expectation we can apply the same argument to all monomials in $\poly_1^2$. Generalizing this argument to arbitrary $\poly$ leads us to consider its following `reduced' version:

\begin{Definition}\label{def:reduced-poly}
For any polynomial $\poly\inparen{\inparen{X_\tup}_{\tup\in\tupset}}$ define the reformulated polynomial $\refpoly{}\inparen{\inparen{X_{\tup, j}}_{\tup\in\tupset, j\in\pbox{\bound}}}
$ to be the polynomial $\refpoly{}$ = $\poly\inparen{\inparen{\sum_{j\in\pbox{\bound}}j\cdot X_{\tup, j}}_{\tup\in\tupset}}
$ and ii) define the \emph{reduced polynomial} $\rpoly\inparen{\inparen{X_{\tup, j}}_{\tup\in\tupset, j\in\pbox{\bound}}}
$ to be the polynomial resulting from converting $\refpoly{}$ into the standard monomial basis\footnote{
  This is the representation, typically used in set-\abbrPDB\xplural, where the polynomial is reresented as sum of `pure' products. See \Cref{def:smb} for a formal definition.
} (\abbrSMB),
removing all monomials containing the term $X_{\tup, j}X_{\tup, j'}$ for any $\tup\in\tupset, j\neq j'\in\pbox{c}$, and setting each \emph{variable}'s exponents $e > 1$ to $1$.
\end{Definition}
As we have essentially argued earlier, the expecation for our specific example is $\rpoly_1^2(\probOf\inparen{A=1}, \ldots, 
\probOf\inparen{Z=1})$.
 This equivalence generalizes  to {\em all} $\raPlus$ queries on \abbrCTIDB\xplural (proof in \Cref{subsec:proof-exp-poly-rpoly}):
\begin{Lemma}\label{lem:tidb-reduce-poly}
For any \abbrCTIDB $\pdb$, $\raPlus$ query $\query$, and lineage polynomial
 $\poly\inparen{\vct{X}}$
 , it holds that $
	\expct_{\vct{W} \sim \pdassign}\pbox{\poly\inparen{\vct{W}}} = \rpoly\inparen{\probAllTup}
$, where $\probAllTup = \inparen{\prob_{\tup,j}}_{\tup\in\tupset,j\in\pbox{\bound}}.$  
\end{Lemma}


\subsection{Our Techniques}
\mypar{Lower Bound Proof Techniques}
To prove the lower bounds in \Cref{tab:lbs} we show that for the same $\query_1$ from the example above, for an arbitrary `product width' $k$, the query $\qhard^k = \query_1^k$ is able to encode various hard graph-counting problems (assuming $\bigO{\numvar}$ tuples rather than the $\bigO{1}$ tuples in \Cref{fig:two-step}).
We do so by considering an arbitrary graph $G$ (analogous to relation $\boldsymbol{R}$ of $\query_1$) and analyzing how the coefficients in the (univariate) polynomial $\widetilde{\poly}\left(p,\dots,p\right)$ relate to counts of subgraphs in $G$ that are isomorphic to various subgraphs with $k$ edges. E.g., for the last two rows in \cref{tab:lbs}, we exploit the fact that the coefficient corresponding to $\prob^{2k}$ in $\rpoly\inparen{\prob,\ldots,\prob}$ of $\qhard^k$ is proportional to the number of $k$-matchings in $G$,
 a known hard problem in parameterized/fine-grained complexity literature.

\mypar{Upper Bound Techniques}
Our negative results (\Cref{tab:lbs}) indicate that \abbrCTIDB{}s (even for $\bound=1$) cannot achieve comparable performance to deterministic databases for exact results (under complexity assumptions). In fact, under well established hardness conjectures, one cannot (drastically) improve upon the trivial algorithm to exactly compute the expected multiplicities for $1$-\abbrTIDB\xplural. A natural followup is whether we can do better if we are willing to settle for an approximation to the expected multiplities.

\usetikzlibrary{shapes.geometric}
\usetikzlibrary{shapes.arrows}
\usetikzlibrary{shapes.misc}
\renewcommand{\aboverulesep}{0pt}
\renewcommand{\belowrulesep}{0pt}

\begin{figure*}[t!]
	\centering
	\resizebox{\textwidth}{!}{%
	\begin{tikzpicture}
		\node[cylinder, text width=0.28\textwidth, align=center, draw=black, text=black, cylinder uses custom fill, cylinder body fill=blue!10, aspect=0.12, minimum height=2.5cm, minimum width=2.5cm, cylinder end fill=blue!50, shape border rotate=90] (cylinder) at (0, 0) {
		\tabcolsep=0.1cm
		\begin{tabular}[t]{>{\small}c | >{\small}c | >{\small}c}
				\multicolumn{2}{c}{$\boldsymbol{T}$}\\
				Point & $\Phi$  \\
				\midrule
	                     $e_1$    & $A$ \\
	                     $e_2$     & $B$ \\
	                     $e_3$      & $C$ \\
	                     $e_4$      & $E$ \\
			\end{tabular}\hspace{0.15cm}
			\tabcolsep=0.05cm
			\begin{tabular}[t]{>{\footnotesize}c | >{\footnotesize}c | >{\footnotesize}c | >{\footnotesize}c}
				\multicolumn{3}{c}{$\boldsymbol{R$}}\\
				$\text{Point}_1$ & $\text{Point}_2$ & $\Phi$\\
				\midrule
	                    $e_1$         & $e_2$         & $U$\\
	                    $e_2$         & $e_4$          & $Y$\\
	                    $e_2$         & $e_3$          & $Z$\\
			\end{tabular}};
			\node[below=0.2 cm of cylinder]{{\LARGE$ \pdb$}};
		\node[single arrow, right=0.25 of cylinder, draw=black, fill=black!65, text=white, minimum height=0.75cm, minimum width=0.25cm](arrow1) {\textbf{\abbrStepOne}};
		\node (arrow1Label) at (3, 1.4) {$\query_2$};
		\usetikzlibrary{arrows.meta}
			\draw[line width=0.5mm, dashed, arrows = -{Latex[length=3mm, open]}] ([yshift=0cm, xshift=-0.75cm]arrow1Label)->([yshift=0cm, xshift=0cm]arrow1);
		\node[rectangle, right=0.175 of arrow1, draw=black, text=black, fill=purple!10, minimum height=2.5cm, minimum width=2cm](rect) {
		\tabcolsep=0.075cm
		 \begin{tabular}{>{\footnotesize}c | >{\centering\arraybackslash\footnotesize}m{1.95cm} | >{\centering\arraybackslash\footnotesize}m{3.95cm}}
	            Point    & $\Phi$ & Circuit\\
			\midrule
	                 $e_1$ & $AU$ &\adjustbox{valign=b}{\resizebox{!}{9mm}{
	                       \begin{tikzpicture}[thick]
	                       		\node[gen_tree_node](sink) at (0.5, 0.8){$\boldsymbol{\circmult}$};
	                       		\node[gen_tree_node](source1) at (0, 0){$A$};
	                       		\node[gen_tree_node](source2) at (1, 0){$U$};
	                       		\draw[->](source1)--(sink);
	                       		\draw[->] (source2)--(sink);
					\end{tikzpicture}$\inparen{1}$
					}}\\
	                       $e_2$ & $B(Y + Z)$ Or $BY+ BZ$&
	            			\adjustbox{valign=m}{
	                       \resizebox{!}{14mm} {
						\begin{tikzpicture}[thick]
							\node[gen_tree_node] (a1) at (1, 0){$Y$};
							\node[gen_tree_node] (b1) at (2, 0){$Z$};
							\node[gen_tree_node] (a2) at (0.75, 0.8){$B$};
							\node[gen_tree_node] (b2) at (1.5, 0.8){$\boldsymbol{\circplus}$};
							\node[gen_tree_node] (a3) at (1.1, 1.6){$\boldsymbol{\circmult}$};
							\draw[->] (a1) -- (b2);
							\draw[->] (b1) -- (b2);
							\draw[->] (a2) -- (a3);
							\draw[->] (b2) -- (a3);
						\end{tikzpicture}$\inparen{2}$
					}} \adjustbox{valign=m}{Or}
					\adjustbox{valign=m}{
	                       \resizebox{!}{14mm} {
					\begin{tikzpicture}[thick]
						\node[gen_tree_node] (a2) at (0, 0){$Y$};
						\node[gen_tree_node] (b2) at (1, 0){$B$};
						\node[gen_tree_node] (c2) at (2, 0){$Z$};
						\node[gen_tree_node] (a1) at (0.5, 0.8){$\boldsymbol{\circmult}$};
						\node[gen_tree_node] (b1) at (1.5, 0.8){$\boldsymbol{\circmult}$};
						\node[gen_tree_node] (a0) at (1.0, 1.6){$\boldsymbol{\circplus}$};
						\draw[->] (a2) -- (a1);
						\draw[->] (b2) -- (a1);
						\draw[->] (b2) -- (b1);
						\draw[->] (c2) -- (b1);
						\draw[->] (a1) -- (a0);
						\draw[->] (b1) -- (a0);
					\end{tikzpicture}
				}}\\
	          \end{tabular}
		};
		\node[below=0.2cm of rect]{{\LARGE $\query_2(\pdb)\inparen{\tup}\equiv \poly\inparen{\vct{X}}$}};
		\node[single arrow, right=0.25 of rect, draw=black, fill=black!65, text=white, minimum height=0.75cm, minimum width=0.25cm](arrow2) {\textbf{\abbrStepTwo}};
		\node[rectangle, right=0.25 of arrow2, rounded corners, draw=black, fill=red!10, text=black, minimum height=2.5cm, minimum width=2cm](rrect) {
		\tabcolsep=0.09cm
		\begin{tabular}{>{\small}c | >{\arraybackslash\normalsize}c}
			Point & $\mathbb{E}[\poly(\vct{X})]$\\
			\midrule
			$e_1$ & $\inparen{\prob_{A_1} +\prob_{A_2}}\cdot\left(\prob_{U_1} + 2\prob_{U_2}\right)$\\
			$e_2$ & $\inparen{\prob_{B_1} + \prob_{B_2}}\inparen{\prob_{Y_1}+2\prob_{Y_2} + \prob_{Z_1} + 2\prob_{Z_2}}$\\
		\end{tabular}
			};
		\node[below=0.2cm of rrect]{{\LARGE $\expct\pbox{\poly(\vct{X})}$}};
	\end{tikzpicture}
	}
	\savecaptionspace{
	\caption{Intensional Query Evaluation Model $(\query_2 = \project_{\text{Point}}$ $\inparen{T\join_{\text{Point} = \text{Point}_1}R}$ where, for table $R,~\bound = 2$, while for $T,~\bound = 1.)$}
	\label{fig:two-step}
	}{-0.35cm}{-0.43cm}
\end{figure*}

We adopt the two-step intensional model of query evaluation used in set-\abbrPDB\xplural, as illustrated in \Cref{fig:two-step}:
(i) \termStepOne (\abbrStepOne): Given input $\tupset$ and $\query$, output every tuple $\tup$ that possibly satisfies $\query$, annotated with its lineage polynomial $\poly(\vct{X})
$;
(ii) \termStepTwo (\abbrStepTwo): Given $\poly(\vct{X})$ for each tuple, compute a $(1\pm \eps)$-approximation $\expct_{\randWorld\sim\bpd}\pbox{\poly(\vct{\randWorld})}$.
Let $\timeOf{\abbrStepOne}(\query,\tupset,\circuit)$ denote the runtime of \abbrStepOne when it outputs $\circuit$ (a representation of $\poly$ as an arithmetic circuit --- more on this representation in~\Cref{sec:expression-trees}).
Denote by $\timeOf{\abbrStepTwo}(\circuit, \epsilon)$ (recall $\circuit$ is the output of \abbrStepOne) the runtime of \abbrStepTwo when $\poly$ is input as $\circuit$. Then to answer if we can compute a $(1\pm \eps)$-approximation to the expected multiplicity in time linear to the deterministic query, it is enough to answer the following:

\begin{Problem}[\abbrCTIDB linear time approximation]\label{prob:big-o-joint-steps}
Given \abbrCTIDB $\pdb$, $\raPlus$ query $\query$,
is there a $(1\pm\epsilon)$-approximation of $\expct_{\rvworld\sim\bpd}$\allowbreak$\pbox{\query\inparen{\rvworld}\inparen{\tup}}$ for all result tuples $\tup$ where there exists
$\circuit : \timeOf{\abbrStepOne}(\query,\tupset, \circuit) + \timeOf{\abbrStepTwo}(\circuit, \epsilon) \le$\allowbreak$ O_\epsilon(\qruntime{\optquery{\query}, \tupset, \bound})$?
\end{Problem}

A key insight of this paper is that the representation of $\circuit$ matters.
For example, if we insist that $\circuit$ represent the lineage polynomial in \abbrSMB, the answer to the above question in general is no, since then we will need $\abs{\circuit}\ge \Omega\inparen{\inparen{\qruntime{\optquery{\query}, \tupset, \bound}}^k}$, where $|\circuit|$ is the size of circuit $\circuit$.
Hence, just $\timeOf{\abbrStepOne}(\query,\tupset,\circuit)$ is too large.
However, systems can directly emit compact, factorized representations of $\poly(\vct{X})$ (e.g., as a consequence of the standard projection push-down optimization~\cite{DBLP:books/daglib/0020812}).
Accordingly, this work uses (arithmetic) circuits\footnote{
  An arithmetic circuit is a DAG with variable/numeric source gates and multiplication/addition internal gates.
}
as the representation system of $\poly(\vct{X})$, and we show in \Cref{sec:circuit-depth} an $\bigO{\qruntime{\optquery{\query}, \tupset, \bound}}$ algorithm for constructing the lineage polynomial for all result tuples of an $\raPlus$ query $\query$ (or more precisely, a circuit $\circuit$ with $\numvar$ sinks, one per output tuple).
 Since a representation $\circuit^*$ exists where $\timeOf{\abbrStepOne}(\query,\tupset,\circuit^*)\le \bigO{\qruntime{\optquery{\query}, \tupset, \bound}}$ and
the size of $\circuit^*$ is bounded by $\qruntime{\optquery{\query}, \tupset, \bound}$ (i.e., $|\circuit^*| \le \bigO{\qruntime{\optquery{\query}, \tupset, \bound}}$) (see~\Cref{sec:circuit-runtime}), we can focus on the complexity of \abbrStepTwo.
Thus, to solve \Cref{prob:big-o-joint-steps}, it is \emph{sufficient} to solve: 
\begin{Problem}\label{prob:intro-stmt}
Given any circuit $\circuit$ that encodes $\Phi\inparen{\vct{X}}$ for all result tuples $\tup$ (one sink per $\tup$) for \abbrCTIDB $\pdb$ and $\raPlus$ query $\query$, does there exist an algorithm that computes a $(1\pm\epsilon)$-approximation of $\expct_{\rvworld\sim\bpd}\pbox{\query\inparen{\rvworld}\inparen{\tup}}$ (for all result tuples $\tup$) in $\bigO{|\circuit|}$ time?
\end{Problem}

We will formalize the notions of circuits and hence, \Cref{prob:intro-stmt} in \Cref{sec:expression-trees}. For an upper bound on approximating the expected count, it is easy to check that if all the probabilties are constant then (with an additive adjustment) $\poly\left(\prob_1,\dots, \prob_n\right)$ is a constant factor approximation of $\rpoly$ (where we assume $\tupset=[n]$).
This is illustrated in the following example using $\query_1^2$ from earlier.  To aid in presentation we again limit our focus to $\monomial{1,R}(A,U,B)$, assume $\bound = 2$ for variable $U$ and $\bound = 1$ for all other variables.  Recall $\prob_X$ denotes $\probOf\pbox{X = 1}$.
Then we have:
%
$\monomial{1,R}\inparen{\vct{X}} = A^2\inparen{U_1^2 + 4U_1U_2 + 4U_2^2}B^2 =A^2U_1^2B^2 + 4A^2U_1U_2B^2+4A^2U_2^2B^2$, which in turn implies:

\[	 \monomial{1,R}\inparen{\probAllTup} -4\prob_A^2\prob_{U_1}\prob_{U_2}\prob_B^2=\prob_A^2\prob_{U_1}^2\prob_B^2 +   4\prob_A^2\prob_{U_2}^2\prob_B^2.\]
Noting that $\rmonomial{1}\inparen{\vct{X}} = AU_1B+4AU_2B$,
if we assume that all probability values are in $[p_0,1]$ for some $p_0>0$, 
we get that $\monomial{1,R}\inparen{\vct{\prob}} - 4\prob_A^2\prob_{U_1}\prob_{U_2}\prob_B^2$ is in the range $\pbox{p_0^3\cdot\rmonomial{1}\inparen{\vct{\prob}}, \rmonomial{1}\inparen{\vct{\prob}}}$.
Note however, that this is \emph{not a tight approximation}.
In~\Cref{sec:algo} we demonstrate that a $(1\pm\epsilon)$ (multiplicative) approximation with competitive performance is achievable.
To get an $(1\pm \epsilon)$-multiplicative approximation and solve~\Cref{prob:intro-stmt}, using \circuit we uniformly sample monomials from the equivalent \abbrSMB representation of $\poly$ (without materializing the \abbrSMB representation) and `adjust' their contribution to $\widetilde{\poly}\left(\vct{p}\right)$.


\mypar{Applications}
Recent work in heuristic data cleaning~\cite{yang:2015:pvldb:lenses,DBLP:journals/vldb/SaRR0W0Z17,DBLP:journals/pvldb/RekatsinasCIR17,DBLP:journals/pvldb/BeskalesIG10} emits a \abbrPDB when insufficient data exists to select the `correct' data repair.
Probabilistic data cleaning is a crucial innovation, as the alternative is to arbitrarily select one repair and `hope' that queries receive meaningful results.
Although \abbrPDB queries instead convey the trustworthiness of results~\cite{kumari:2016:qdb:communicating}, they are impractically slow~\cite{feng:2019:sigmod:uncertainty,feng:2021:sigmod:efficient}, even in approximation (see \Cref{sec:karp-luby}).
Bags, as we consider, are sufficient for production use, where bag-relational algebra is already the default for performance reasons.
Our results show that bag-\abbrPDB\xplural can be competitive, laying the groundwork for probabilistic functionality in production database engines.


\mypar{Paper Organization} We present background and notation in \Cref{sec:background}. We prove our main hardness results in \Cref{sec:hard} and present our approximation algorithm in \Cref{sec:algo}.
Finally, we discuss related work in \Cref{sec:related-work} and conclude in \Cref{sec:concl-future-work}.  All proofs are in the appendix.



\section{Background and Notation}\label{sec:background}

\subsection{Polynomial Definition and Terminology}
Given an index set $S$ and variables $X_\tup$ for $\tup\in S$, a (general) polynomial $\genpoly$ over $\inparen{X_\tup}_{\tup \in S}$ with individual degree $\hideg <\infty$ 
is formally defined as: 
\begin{align}
  \label{eq:sop-form}
\genpoly\inparen{\inparen{X_\tup}_{\tup\in S}}=\sum_{\vct{d}=\inparen{d_\tup}_{\tup\in S}\in[0,\hideg]^{S}} c_{\vct{d}}\cdot \prod_{\tup\in S}X_\tup^{d_\tup}&&\text{ where } c_{\vct{d}}\in \semN.
\end{align}

\begin{Definition}[Standard Monomial Basis]\label{def:smb}
The term $\prod_{\tup\in S} X_\tup^{d_\tup}$ in \Cref{eq:sop-form} is a {\em monomial}. A polynomial $\genpoly\inparen{\vct{X}}$ is in standard monomial basis (\abbrSMB) when we keep only the terms with $c_{\vct{d}}\ne 0$ from \Cref{eq:sop-form}.
\end{Definition}
Unless othewise noted, we consider all polynomials to be in \abbrSMB representation. 
When it is unclear, we use $\smbOf{\genpoly}$ 
to denote the \abbrSMB form of a polynomial $\genpoly$.
%
%
We call a polynomial $\poly\inparen{\vct{X}}$ a \emph{\abbrCTIDB-lineage polynomial} (or simply lineage polynomial), if it is clear from context that there exists an $\raPlus$ query $\query$, \abbrCTIDB $\pdb$, and result tuple $\tup$ such that $\poly\inparen{\vct{X}} = \apolyqdt\inparen{\vct{X}}.$
\subsection{\abbrOneBIDB}\label{subsec:one-bidb}
\label{subsec:tidbs-and-bidbs}

\noindent A block independent database \abbrBIDB $\pdb'$ models a set of worlds each of which consists of a subset of the \dbbaseName $\tupset'$, where $\tupset'$ is partitioned into $\numblock$ blocks $\block_i$ and the random variables $\worldvec_\tup$ for $\tup\in\block_i$ and $\tup\in\block_j$ are independent  for $i\ne j$.  $\pdb'$ further constrains that $\worldvec_\tup$ all $\tup\in\block_i$ for the same $i\in\pbox{\numblock}$ of $\tupset'$ be disjoint events.  
We define next a specific construction of \abbrBIDB that is useful for our work.

\begin{Definition}[\abbrOneBIDB]\label{def:one-bidb}
Define a \emph{\abbrOneBIDB} to be the pair $\pdb' = \inparen{\bigtimes_{\tup\in\tupset'}\inset{0, \bound_\tup}, \bpd'},$  where $\tupset'$ is the \dbbaseName such that each $\tup \in \tupset'$ has a multiplicity in $\inset{0, \bound_\tup}$, with $\bound_\tup\in\mathbb{N}$.  $\tupset'$ is partitioned into $\numblock$ independent blocks $\block_i,$ for $i\in\pbox{\numblock}$, of disjoint tuples.  $\bpd'$ is characterized by the vector $\inparen{\prob_\tup}_{\tup\in\tupset'}$ where for every block $\block_i$, $\sum_{\tup \in \block_i}\prob_\tup \leq 1$.  For $W\in\onebidbworlds{\tupset'}$ and  $i\in\pbox{\numblock}$, let $\prob_i(W) = \begin{cases}
	1 - \sum_{\tup\in\block_i}\prob_\tup	&	\text{if }W_\tup = 0\text{ for all }\tup\in\block_i\\
	0									&	\text{if there exists } \tup \neq \tup'\in\block_i; W_\tup, W_{\tup'}\neq 0\\
	\prob_\tup							&	W_\tup \ne 0 \text{ for one unique } t\in B_i.\\
	\end{cases}$
	
\noindent$\bpd'$ is the probability distribution across all worlds such that, given $W\in\bigtimes_{\tup \in \tupset'}\inset{0,\bound_\tup}$, $\probOf\pbox{\worldvec = W} = \prod_{i\in [m]}\prob_{i}(W)$.
\end{Definition}

Lineage polynomials for arbitrary \dbbaseName $\gentupset'$ are constructed in a manner analogous to $1$-\abbrTIDB\xplural (see \Cref{fig:nxDBSemantics}), differing only in the base case.  
In a $1$-\abbrTIDB, each tuple contributes a multiplicity of 0 or 1, and $\polyqdt{\rel}{\gentupset}{\tup} = X_\tup$. 
In a \abbrOneBIDB, each tuple $\tup\in\tupset'$ contributes its corresponding multiplicity: 
$\polyqdt{\rel}{\gentupset}{\tup} = c_\tup\cdot X_\tup$.  See \Cref{fig:lin-poly-bidb} for full details.

\abbrOneBIDB are powerful enough to encode \abbrCTIDB:
\begin{Proposition}[\abbrCTIDB reduction]\label{prop:ctidb-reduct}
Given \abbrCTIDB $\pdb =$\newline $\inparen{\worlds, \bpd}$, let $\pdb' = \inparen{\onebidbworlds{\tupset'}, \bpd'}$ be the \emph{\abbrOneBIDB} obtained in the following manner: for each $\tup\in\tupset$, create block $\block_\tup = \inset{\intuple{\tup, j}_{j\in\pbox{\bound}}}$ of disjoint tuples, 
where $\bound_{\intuple{\tup, j}} = j$ for each $\intuple{\tup, j}$ in $\tupset'$.
  The probability distribution $\bpd'$ is the characterized by the vector $\vct{p} = \inparen{\inparen{\prob_{\tup, j}}_{\tup\in\tupset, j\in\pbox{\bound}}}$. 
  Then, $\mathcal{P}$ and $\mathcal{P}'$ are equivalent.
\end{Proposition} 

We now define the reduced polynomial $\rpoly'$ of a \abbrOneBIDB.

\begin{Definition}[$\rpoly'$]\label{def:reduced-poly-one-bidb}
Given a polynomial $\poly'\inparen{\vct{X}}$ generated from a \abbrOneBIDB, let $\rpoly'\inparen{\vct{X}}$ denote the reduced $\poly'\inparen{\vct{X}}$ derived as follows:  i) compute $\smbOf{\poly'\inparen{\vct{X}}}$ eliminating monomials with cross terms $X_{\tup}X_{\tup'}$ for $\tup\neq \tup' \in \block_i$ and  ii) reduce all \emph{variable} exponents $e > 1$ to $1$.
\end{Definition}

We have the following generalization of \Cref{lem:tidb-reduce-poly}:
\begin{Lemma}\label{lem:bin-bidb-phi-eq-redphi}
Given any \emph{\abbrOneBIDB} $\pdb'$, $\raPlus$ query $\query$, and lineage polynomial
 $\poly'\inparen{\vct{X}}=\poly'\pbox{\query,\tupset',\tup}\inparen{\vct{X}}$, it holds that \newline$
	\expct_{\vct{W} \sim \pdassign'}\pbox{\poly'\inparen{\vct{W}}} = \rpoly'\inparen{\probAllTup}.
$ 
\end{Lemma}

Let $\abs{\poly'}$ be the number of operators in $\poly'$. Then:

\begin{Corollary}\label{cor:expct-sop}
If $\poly'$ is a \abbrOneBIDB lineage polynomial already in \abbrSMB, then the expectation of $\poly$, i.e., $	\expct_{\vct{W} \sim \pdassign'}\pbox{\poly'\inparen{\vct{W}}}$ 
can be computed in $\bigO{\abs{\poly'}}$ time.
\end{Corollary}

\subsection{Formalizing \Cref{prob:intro-stmt}}\label{sec:expression-trees}
We focus on the problem of computing $\expct_{\worldvec\sim\pdassign}\pbox{\poly\inparen{\vct{\randWorld}}}$ from now on. 
\Cref{prob:intro-stmt} asks if there exists a linear time approximation algorithm in the size of a given circuit \circuit that encodes $\poly\inparen{\vct{X}}$.  Recall that in this work we
 represent lineage polynomials via {\em arithmetic circuits}~\cite{arith-complexity}, a standard way to represent polynomials over fields (particularly in the field of algebraic complexity), which we use for polynomials over $\mathbb N$ in the obvious way.  Since we are specifically using circuits to model lineage polynomials, we can refer to these circuits as lineage circuits.  However, when the meaning is clear, we will drop the term lineage and only refer to them as circuits.

\begin{Definition}[Circuit]\label{def:circuit}
A circuit $\circuit$ is a Directed Acyclic Graph (DAG) with source gates (in degree of $0$) drawn from either $\domN$ or $\vct{X} = \inparen{X_\tup}_{\tup\in\tupset}$ and one sink gate for each result tuple.  Internal gates have binary input and are either sum ($\circplus$) or product ($\circmult$) gates.
Each gate has the following members: \type, \vari{input}, 
\vpartial, \degval, \vari{Lweight}, and \vari{Rweight}, where \type is the value type $\{\circplus, \circmult, \var, \tnum\}$ and \vari{input} the list of inputs. Source gates have an extra member \val for the value.  $\circuit_\linput$ ($\circuit_\rinput$) denotes the left (right) input of \circuit.
\end{Definition}

\colorlet{figray}{black!65}
\colorlet{fillred}{red!45}
\colorlet{fillblue}{blue!45}
\colorlet{fillbrown}{green!45}
\begin{wrapfigure}{r}{0.2\textwidth}
		\centering
		\begin{tikzpicture}[thick, scale=0.67, every node/.style={transform shape}]
			\node[tree_node] (a1) at (0, 0) {$\boldsymbol{X}$};
			\node[tree_node] (b1) at (1.5, 0) {$\boldsymbol{2}$};
			\node[tree_node] (c1) at (3, 0) {$\boldsymbol{Y}$};
			\node[tree_node] (d1) at (4.5, 0) {$\boldsymbol{-1}$};

			\node[tree_node] (a2) at (0.75, 0.75) {$\boldsymbol{\circmult}$};
			\node[tree_node] (b2) at (2.25, 0.75) {$\boldsymbol{\circmult}$};
			\node[tree_node, fill = fillblue] (c2) at (3.75, 0.75) {$\boldsymbol{\circmult}$};

			\node[tree_node, fill = fillred] (a3) at (0.55, 1.5) {$\boldsymbol{\circplus}$};
			\node[tree_node, fill = fillbrown] (b3) at (3.75, 1.5) {$\boldsymbol{\circplus}$};

			\node[tree_node] (a4) at (2.25, 2.25) {$\boldsymbol{\circmult}$};
			\node[draw=none, figray
			](a1l) at (0, -0.75) {$1$};
			\node[draw=none, figray] (b1l) at (1.5, -0.75) {$2$};
			\node[draw=none, figray] (c1l) at (3, -0.75) {$1$};
			\node[draw=none, figray] (d1l) at (4.5, -0.75) {$1$};
			
			\node[draw=none, figray] (a2l) at (0.75, 0) {$2$};
			\node[draw=none, figray] (b2l) at (2.25, 0) {$2$};
			\node[draw=none, figray] (c2l) at (3.75, 0) {$1$};
			\node[draw=none, figray] (a3l) at (-0.2, 1.5) {$3$};
			\node[draw=none, figray] (b3l) at (4.5, 1.5) {$3$};
			\node[draw=none, figray] (a4l) at (2.8, 2.7) {$9$};
			\draw[figray] (a1l) -- (a1);
			\draw[figray] (b1) -- (b1l);
			\draw[figray] (c1) -- (c1l);
			\draw[figray] (d1) -- (d1l);
			\draw[figray] (a2) -- (a2l);
			\draw[figray] (b2) -- (b2l);
			\draw[figray] (c2) -- (c2l);
			\draw[figray] (a3) -- (a3l);
			\draw[figray] (b3) -- (b3l);
			\draw[figray] (a4) -- (a4l);
			\draw[->] (a1) -- (a2);
			\draw[->] (a1) -- (a3);
			\draw[->] (b1) -- (a2);
			\draw[->] (b1) -- (b2);
			\draw[->] (c1) -- (c2);
			\draw[->] (c1) -- (b2);
			\draw[->] (d1) -- (c2);
			\draw[->] (a2) -- (b3);
			\draw[->] (b2) -- (a3);
			\draw[->] (c2) -- (b3);
			\draw[->] (a3) -- (a4);
			\draw[->] (b3) -- (a4);
			\draw[->] (a4) -- (2.25, 2.75);
		\end{tikzpicture}
		\savecaptionspace{
		\caption{Circuit encoding of $(X + 2Y)(2X - Y)$.}
		\label{fig:circuit}
		}{-0.025cm}{-0.58cm}
	\end{wrapfigure}

We refer to the structure when the underlying DAG is a tree (with edges pointing towards the root) as an expression tree \etree.  The circuits $\inparen{1}$ and $\inparen{2}$ in column $\poly$ of \Cref{fig:two-step} are both expression trees.  
Members not described in~\Cref{def:circuit} are defined and used in the appendix.  


The function $\polyf\inparen{\cdot}$ (\Cref{def:poly-func}) maps a circuit to its corresponding polynomial.  We next define its inverse:

\begin{Definition}[Circuit Set]\label{def:circuit-set}
$\circuitset{\polyX}$ is the set of all possible circuits $\circuit$ such that $\polyf(\circuit) = \polyX$.
\end{Definition}


\Cref{fig:circuit} depicts a circuit \circuit in $\circuitset{2X^2+3XY-2Y^2}$.   Light-text annotations and the colors
can be ignored until~\Cref{sec:algo}. 
%
 We are now ready to formally state the final version of \Cref{prob:intro-stmt}.


\begin{Definition}[The Expected Result Multiplicity Problem]\label{def:the-expected-multipl}
Let $\pdb$ be an arbitrary \abbrCTIDB and $\vct{X}$ be the set of variables annotating tuples in $\tupset$.  Fix an $\raPlus$ query $\query$ and a result tuple $\tup$.
  The \expectProblem is defined as follows:
\begin{flalign*}
&\textbf{Input}: \circuit \in \circuitset{\poly\pbox{\query, \tupset, \tup}\inparen{\vct{X}}}\\
&\textbf{Output}: \expct_{\vct{W} \sim \bpd}\pbox{\poly\pbox{\query, \tupset, \tup}\inparen{\vct{W}}}.&
\end{flalign*}
\end{Definition}


\subsection{Deterministic Query Runtimes}\label{sec:gen}
In~\Cref{sec:intro}, we introduced the function $T_{det}\inparen{\cdot}$ to analyze the runtime complexity of~\Cref{prob:expect-mult}.
To decouple our results from any specific join algorithm, we first lower bound the cost of a join.

\begin{Definition}[Join Cost]
\label{def:join-cost}
Denote by $\jointime{R_1, \ldots, R_m}$ the runtime of an algorithm for computing the $m$-ary join $R_1 \bowtie \ldots \bowtie R_m$.
We require only that the algorithm must enumerate its output, i.e., that $\jointime{R_1, \ldots, R_m} \geq |R_1 \bowtie \ldots \bowtie R_m|$.  
\end{Definition}

Worst-case optimal join algorithms~\cite{skew,ngo-survey} and query evaluation via factorized databases~\cite{factorized-db} (as well as work on FAQs~\cite{DBLP:conf/pods/KhamisNR16} and AJAR~\cite{ajar}) can be modeled as $\raPlus$ queries (though the query size is data dependent).
For these algorithms, $\jointime{R_1, \ldots, R_n}$ is linear in the {\em AGM bound}~\cite{AGM}.
 Our cost model for general query evaluation follows:
{\small
\begin{flalign*}
&\begin{aligned}
    &\qruntimenoopt{R,\gentupset,\bound} = |\gentupset.R| 
    & 
    &\qquad\qquad\qruntimenoopt{\sigma \query, \gentupset,\bound} = \qruntimenoopt{\query,\gentupset,\bound}\\
\end{aligned}&\\
 &\qruntimenoopt{\pi \query, \gentupset,\bound} = \qruntimenoopt{\query,\gentupset,\bound} +\abs{\query(\gentupset)}&\\
 &\qruntimenoopt{\query \cup \query', \gentupset,\bound} = \qruntimenoopt{\query, \gentupset,\bound} + \qruntimenoopt{\query', \gentupset,\bound} + \abs{\query\inparen{\gentupset}}+\abs{\query'\inparen{\gentupset}}& \\
&\qruntimenoopt{\query_1 \bowtie \ldots \bowtie \query_m, \gentupset,\bound} = &\\
&\qquad\qruntimenoopt{\query_1, \gentupset,\bound} + \ldots + \qruntimenoopt{\query_m,\gentupset,\bound} + 
       \jointime{\query_1(\gentupset), \ldots, \query_m(\gentupset)}&\\
\end{flalign*}
}
\vspace{-0.93cm}

\noindent
Under this model, an $\raPlus$ query $\query$ evaluated over over any deterministic database 
that maps each tuple in $\gentupset$ to a multiplicity in $[0,\bound]$ 
has runtime $O\inparen{\qruntimenoopt{Q,\gentupset, \bound}}$.
We assume that full table scans are used for every base relation access. We can model index scans by treating an index scan query $\sigma_\theta(R)$ as a base relation.

\Cref{lem:circ-model-runtime} and \Cref{lem:tlc-is-the-same-as-det} show that for any $\raPlus$ query $\query$ and \dbbaseName $\tupset$, there exists a circuit $\circuit^*$ such that $\timeOf{\abbrStepOne}(Q,\tupset,\circuit^*)$ and $|\circuit^*|$ are both $O(\qruntimenoopt{\optquery{\query}, \tupset,\bound})$, as we assumed when moving from \Cref{prob:big-o-joint-steps} to \Cref{prob:intro-stmt}.  Lastly, we can handle FAQs/AJAR queries and factorized databases by allowing for optimization. 
%
%
%
%






\section{Hardness of Exact Computation}
\label{sec:hard}
In this section, we will prove the hardness results claimed in Table~\ref{tab:lbs} for a specific (family) of hard instances $(\qhard^k,\pdb)$ for \Cref{prob:bag-pdb-poly-expected} where $\pdb$ is a $1$-\abbrTIDB.
Note that this implies hardness for \abbrCTIDB\xplural $\inparen{\bound\geq1}$
as well as \abbrOneBIDB. 
Our hardness results are based on (exactly) counting the number of (not necessarily induced) subgraphs in $G$ isomorphic to $H$. Let $\numocc{G}{H}$ denote this quantity.  We think of $H$ as being of constant size and $G$ as growing.  
In particular, we will consider computing the following counts (given $G$ in its adjacency list representation): $\numocc{G}{\tri}$ (the number of triangles), $\numocc{G}{\threedis}$ (the number of $3$-matchings), and the latter's generalization $\numocc{G}{\kmatch}$ (the number of $k$-matchings).  We use $\kmatchtime$ to denote the optimal runtime of computing $\numocc{G}{\kmatch}$ exactly.  Our results in \Cref{sec:multiple-p} are based on the following known (conditional) hardness results:

\begin{Theorem}[\cite{k-match}]
\label{thm:k-match-hard}
Given positive integer $k$ and undirected graph $G=(\vset,\edgeSet)$ with no self-loops or parallel edges,  $\kmatchtime\ge \littleomega{f(k)\cdot |\edgeSet|^c}$ for any function $f$ and any constant $c$ independent of $\abs{E}$ and $k$ (assuming $\sharpwzero\ne\sharpwone$). 
\end{Theorem}

\begin{Theorem}[~\cite{10.1109/FOCS.2014.22}]\label{conj:known-algo-kmatch}
Given positive integer $k$ and undirected graph $G=(\vset,\edgeSet)$,  $\kmatchtime\ge |\vset|^{\Omega\inparen{k/\log{k}}}$ (assuming ETH).
\end{Theorem}

The above result is saying is that, assuming Exponential Time Hypothesis (ETH), one can only hope for a slightly super-polynomial improvement over the trivial algorithm to compute $\numocc{G}{\kmatch}$.

Our hardness result in Section~\ref{sec:single-p} is based on the following conjectured hardness result:
%
\begin{hypo}
\label{conj:graph}
There exists a constant $\eps_0>0$ such that given an undirected graph $G=(\vset,\edgeSet)$, computing $\numocc{G}{\tri}$ exactly cannot be done in time $o\inparen{|\edgeSet|^{1+\eps_0}}$.
\end{hypo}
%
The so called {\em Triangle detection hypothesis} (cf.~\cite{triang-hard}), which states that detecting the presence of triangles in $G$ takes time $\Omega\inparen{|\edgeSet|^{4/3}}$, implies that in Conjecture~\ref{conj:graph} we can take $\eps_0\ge \frac 13$.

All of our hardness results rely on a simple lineage polynomial encoding of the edges of a graph.
To prove our hardness result, consider a graph $G=(\vset, \edgeSet)$, where $|\edgeSet| = m$, $\vset = [\numvar]$. Our lineage polynomial has a variable $X_i$ for every $i$ in $[\numvar]$.
Consider the polynomial
$\poly_{G}(\vct{X}) = \sum\limits_{(i, j) \in \edgeSet} X_i \cdot X_j.$
The hard polynomial for our problem will be a suitable power $k\ge 3$ of the polynomial above:
\begin{Definition}\label{def:qk}
For any graph $G=(V,\edgeSet)$ and $\kElem\ge 1$, define
\[\poly_{G}^\kElem(X_1,\dots,X_n) = \left(\sum\limits_{(i, j) \in \edgeSet} X_i \cdot X_j\right)^\kElem.\]
\end{Definition}

\noindent Returning to \Cref{fig:two-step}, it can be seen that $\poly_{G}^\kElem(\vct{X})$ is the lineage polynomial from query $\qhard^k$, which we define next. 
\mdfdefinestyle{underbrace}{topline=false, rightline=false, bottomline=false, leftline=false, backgroundcolor=black!15!white, innerbottommargin=0pt}
\begin{mdframed}[style=underbrace]
\begin{lstlisting}
SELECT COUNT(*) FROM $\underbrace{Q_1\text{ JOIN }Q_1\text{ JOIN}\cdots\text{JOIN }Q_1}_{k\rm\ times}$
\end{lstlisting}          
\end{mdframed}
In the above, $\query_1$ is as defined in  \Cref{sec:intro}, which is the same as $\qhard^1$. 
%
We next define the instances for $T$ and $R$ that lead to the lineage polynomial in~\Cref{def:qk} as follows. Relation $T$ has $n$ tuples corresponding to each vertex for $i$ in $[n]$, each with probability $\prob$ and $R$ has tuples corresponding to the edges $\edgeSet$ (each with a probability of $1$).\footnote{Technically, $\poly_{G}^\kElem(\vct{X})$ should have variables corresponding to tuples in $R$ as well, but since they always are present with probability $1$, we drop those. Our argument also works when all the tuples in $R$ also are present with probability $\prob$ but to simplify notation we assign probability $1$ to edges.}
In other words, the \dbbaseName $\tupset$ contains the set of $\numvar$ unary tuples in $T$ (which corresponds to $\vset$) and $\numedge$ binary tuples in $R$ (which corresponds to $\edgeSet$).
Note that this implies that $\poly_{G}^\kElem$ is indeed a $1$-\abbrTIDB lineage polynomial. 

Next, we note that the runtime for answering $\qhard^k$ on deterministic database $\tupset$, as defined above, is $O_k\inparen{\numedge}$ (i.e. deterministic query processing is `easy' for this query):
\begin{Lemma}\label{lem:tdet-om}
For $\qhard^k,\tupset$  as above, 
$\qruntimenoopt{\qhard^k, \tupset, \bound}$ is $O_k\inparen{\numedge}$.
\end{Lemma}

\subsection{Multiple Distinct $\prob$ Values}
\label{sec:multiple-p}
We are now ready to present one of our main hardness result.
%

\begin{Theorem}\label{thm:mult-p-hard-result}
Let $\prob_0,\ldots,\prob_{2k}$ be $2k + 1$ distinct values in $(0, 1]$.  Then computing $\rpoly_G^\kElem(\prob_i,\dots,\prob_i)$ (for all $i\in [2k+1]$) for arbitrary $G=(\vset,\edgeSet)$
needs time $\bigOmega{\kmatchtime}$, if $\kmatchtime\ge \omega\inparen{\abs{\edgeSet}}$.
\end{Theorem}
%
Note that the second (and third) row(s) of \Cref{tab:lbs} follow from 
 \Cref{thm:mult-p-hard-result}, \Cref{lem:tdet-om}, and \Cref{thm:k-match-hard} (\Cref{conj:known-algo-kmatch} resp.).
 \Cref{thm:mult-p-hard-result} follows by observing that $\rpoly_G^\kElem(\prob,\dots,\prob)=\prob^{2k}\cdot \numocc{G}{\kmatch} +r(p)$, where $r(p)$ is a polynomial of degree at most $2k-1$ (with coefficients that just depend on $G$). By polynomial interpolation, knowing the values $\rpoly_G^\kElem(\prob_i,\dots,\prob_i)$ (over all $i\in [2k+1]$) allows us to compute all the coefficients, including $\numocc{G}{\kmatch}$.

 Note that \Cref{thm:mult-p-hard-result} needs one to be able to compute the expected multiplicities over $(2k+1)$ distinct values of $p_i$, each of which corresponds to distinct $\bpd$ (for the same $\tupset$), which explain the `Multiple' entries in the second column of the second and third rows in \Cref{tab:lbs}. Next, we argue how to get rid of this latter requirement.



\subsection{Single $\prob$ value}
\label{sec:single-p}

While \Cref{thm:mult-p-hard-result} shows that computing $\rpoly(\prob,\dots,\prob)$ for multiple values of $\prob$ in general is hard it does not rule out the possibility that one can compute this value exactly for a {\em fixed} value of $\prob$. 
One can compute $\rpoly(\prob,\dots,\prob)$ exactly in linear time for $\prob\in \inset{0,1}$. Next, we show that these are the only two possibilities:

\begin{Theorem}\label{th:single-p-hard}
Fix $\prob\in (0,1)$. Then assuming \Cref{conj:graph}, any algorithm that computes $\rpoly_{G}^3(\prob,\dots,\prob)$ for arbitrary $G = (\vset, \edgeSet)$ exactly has to run in time $\Omega\inparen{\abs{\edgeSet}^{1+\eps_0}}$, where $\eps_0$ is as in \Cref{conj:graph}.
\end{Theorem}

Note that \Cref{lem:tdet-om} and \Cref{th:single-p-hard} above imply the hardness result in the first row of \Cref{tab:lbs}.
We note that \Cref{thm:k-match-hard} and \Cref{conj:known-algo-kmatch} (and the lower bounds in the second and third rows) need $k$ to be large enough (in particular, we need a family of hard queries). But the above \Cref{th:single-p-hard} (and the lower bound in first row of Table~\ref{tab:lbs}) holds for $k=3$ (and hence for a fixed query). 

Unlike the proof of \Cref{thm:mult-p-hard-result}, in this case we have to pay close attention to all the coefficients of $\rpoly_{G}^3(\prob,\ldots, \prob)$:
\begin{Lemma}\label{lem:qE3-exp}
For any $\prob$, we have:
	{\small 
	\begin{align}
		&\rpoly_{G}^3(\prob,\ldots, \prob) = \numocc{G}{\ed}\prob^2 + 6\numocc{G}{\twopath}\prob^3 + 6\numocc{G}{\twodis}\prob^4 + 6\numocc{G}{\tri}\prob^3\nonumber\\
		&+ 6\numocc{G}{\oneint}\prob^4 + 6\numocc{G}{\threepath}\prob^4 + 6\numocc{G}{\twopathdis}\prob^5 + 6\numocc{G}{\threedis}\prob^6.\label{claim:four-one}
	\end{align}}
\end{Lemma}
Since $p$ is fixed, the earlier polynomial interpolation based argument does not work anymore. Next, we use the fact that the algorithm still has to compute $\rpoly_{G}^3(\prob,\ldots, \prob)$ for {\em all} graphs $G$. We focus on the graphs $\graph{\ell}$  obtained from  $G$ by replacing each edge by path of length $\ell$ (\Cref{def:Gk}). We then show


\begin{Lemma}\label{lem:lin-sys}
Fix $\prob\in (0,1)$. Given $\rpoly_{\graph{\ell}}^3(\prob,\dots,\prob)$ for $\ell\in [2]$, we can compute in $O(m)$ time a vector $\vct{b}\in\mathbb{R}^2$ such that
\begin{equation}
\label{eq:lin-eqs-single-p}
\begin{pmatrix}
1 - 3p                                     &       -(3\prob^2 - \prob^3)\\
10(3\prob^2 - \prob^3)		&       10(3\prob^2 - \prob^3)
\end{pmatrix}
\cdot
\begin{pmatrix}
\numocc{G}{\tri}]\\
\numocc{G}{\threedis}
\end{pmatrix}
=\vct{b},
\end{equation}
allowing us to compute $\numocc{G}{\tri}$ and $\numocc{G}{\threedis}$ in $O(1)$ time.
\end{Lemma}
Note that \cref{eq:lin-eqs-single-p} only depends on sub-graph counts on $G=\graph{1}$.
It can be verified that the coefficient matrix in \cref{eq:lin-eqs-single-p} is full rank for all $p\in (0,1)$. Then by solving the linear equations in \cref{eq:lin-eqs-single-p} we can compute $\numocc{G}{\tri}$, from where \Cref{conj:graph} implies \Cref{th:single-p-hard}.



\section{$1 \pm \epsilon$ Approximation Algorithm}\label{sec:algo}
We showed in~\Cref{sec:hard} that a runtime of $\bigO{\qruntime{\optquery{\query},\tupset,\bound}}$ cannot be acheived for~\Cref{prob:bag-pdb-poly-expected}.  In light of this, we desire to produce an approximation algorithm that runs in time $\bigO{\qruntime{\optquery{\query},\tupset,\bound}}$.  We do this by showing the result via circuits,
such that our $1\pm\epsilon$ approximation algorithm for this problem runs in $\bigO{\abs{\circuit}}$ for a very broad class of circuits, (thus solving~\Cref{prob:intro-stmt});  see the discussion after \Cref{lem:val-ub} for more.
The following approximation algorithm applies to bag query semantics over both
\abbrCTIDB lineage polynomials and general \abbrBIDB lineage polynomials in practice. 
Our experimental results (see~\Cref{app:subsec:experiment}), which use queries from the PDBench benchmark~\cite{pdbench} support the notion that our bounds hold for general \abbrBIDB in practice.
Proofs and pseudocode for all formal statements and algorithms
  are in \Cref{sec:proofs-approx-alg}.

\subsection{Preliminaries and some more notation}

For notational convenience, in this section we will assume that \dbbaseName $\tupset'=[n]$.
We now introduce definitions  related to circuits and polynomials that we will need to state our upper bound results. First, we introduce the expansion $\expansion{\circuit}$ of circuit $\circuit$ which 
is used in our auxiliary algorithm \sampmon for sampling monomials when computing the approximation.  

\begin{Definition}[$\expansion{\circuit}$]\label{def:expand-circuit}
For a circuit $\circuit$, we define $\expansion{\circuit}$ as a list of tuples $(\monom, \coef)$, where $\monom$ is a set of variables and $\coef \in \domN$.
$\expansion{\circuit}$ has the following recursive definition ($\circ$ is list concatenation).\\
$\expansion{\circuit} =
\begin{cases}
					\expansion{\circuit_\linput} \circ \expansion{\circuit_\rinput}		&\textbf{ if }\circuit.\type = \circplus\\
					\left\{(\monom_\linput \cup \monom_\rinput, \coef_\linput \cdot \coef_\rinput) \right.\\\left.~|~(\monom_\linput, \coef_\linput) \in \expansion{\circuit_\linput}, (\monom_\rinput, \coef_\rinput) \in \expansion{\circuit_\rinput}\right\} 		&\textbf{ if }\circuit.\type = \circmult\\
					\elist{(\emptyset, \circuit.\val)}								&\textbf{ if }\circuit.\type = \tnum\\
					\elist{(\circuit.\val, 1)}									&\textbf{ if }\circuit.\type = \var.\\
\end{cases}
$
\end{Definition}
Later on, we will denote the monomial composed of the variables in $\monom$ as $\encMon$.  As an example of $\expansion{\circuit}$, consider $\circuit$ illustrated in \Cref{fig:circuit}.  $\expansion{\circuit}$ is then $[(X, 2), (XY, -1), (XY, 4), (Y, -2)]$. This helps us redefine $\rpoly$ (see \Cref{eq:tilde-Q-bi}) in a way that makes our algorithm more transparent.

Next, we present a sequence of definitions that will be useful for our algorithm and its analysis.
\begin{Definition}[$\abs{\circuit}$]\label{def:positive-circuit}
For any circuit $\circuit$, the corresponding
{\em positive circuit}, denoted $\abs{\circuit}$, is obtained from $\circuit$ as follows. For each leaf node $\ell$ of $\circuit$ where $\ell.\type$ is $\tnum$, update $\ell.\vari{value}$ to $|\ell.\vari{value}|$.
\end{Definition}
We will overload notation and use $\abs{\circuit}\inparen{\vct{X}}$ to mean $\polyf\inparen{\abs{\circuit}}$.
Conveniently, $\abs{\circuit}\inparen{1,\ldots,1}$ gives us $\sum\limits_{\inparen{\monom, \coef} \in \expansion{\circuit}}\abs{\coef}$.

\begin{Definition}[\size($\cdot$), \depth$\inparen{\cdot}$]\label{def:size-depth}
The functions \size and \depth output the number of gates and levels respectively for input \circuit.
\end{Definition}

\begin{Definition}[$\degree(\cdot)$]\label{def:degree}\footnote{Note that the degree of $\polyf(\abs{\circuit})$ is always upper bounded by $\degree(\circuit)$ and the latter can be strictly larger (e.g. consider the case when $\circuit$ multiplies two copies of the constant $1$-- here we have $\deg(\circuit)=1$ but degree of $\polyf(\abs{\circuit})$ is $0$).}
$\degree(\circuit)$ is defined recursively:
\[\degree(\circuit)=
\begin{cases}
\max(\degree(\circuit_\linput),\degree(\circuit_\rinput)) & \text{ if }\circuit.\type=+\\
\degree(\circuit_\linput) + \degree(\circuit_\rinput)+1 &\text{ if }\circuit.\type=\times\\
1 & \text{ if }\circuit.\type = \var\\
0 & \text{otherwise}.
\end{cases}
\]
\end{Definition}

\noindent
We use the following notation for integer multiplication complexity:
\begin{Definition}[$\multc{\cdot}{\cdot}$]\footnote{We note that when doing arithmetic operations on the RAM model for input of size $N$, we have that $\multc{O(\log{N})}{O(\log{N})}=O(1)$.}
In a RAM model of word size of $W$-bits, $\multc{M}{W}$ denotes the complexity of multiplying two integers represented with $M$-bits. (For input of size $N$, we make the standard assumption $W=O(\log{N})$.)
\end{Definition}

Finally, to get linear runtime results, we will need to define another parameter modeling the (weighted) number of monomials in $\expansion{\circuit}$
that need to be `canceled' when monomials with dependent variables are removed (\Cref{subsec:one-bidb}).  
Let $\isInd{\cdot}$ be a boolean function returning true if monomial $\encMon$ is composed of independent variables and false otherwise; further, let $\indicator{\theta}$ also be a boolean function returning true if $\theta$ evaluates to true.
\begin{Definition}[Parameter $\gamma$]\label{def:param-gamma}
Given \abbrOneBIDB circuit $\circuit$, let
\[\gamma(\circuit)=\frac{\sum_{(\monom, \coef)\in \expansion{\circuit}} \abs{\coef}\cdot \indicator{\neg\isInd{\encMon}} }
{\abs{\circuit}(1,\ldots, 1)}.\]
\end{Definition}
\subsection{Our main result}\label{sec:algo:sub:main-result}
 In what follows, we solve~\Cref{prob:intro-stmt} for any fixed $\epsilon > 0$.

\mypar{Algorithm Idea}
Our approximation algorithm (\approxq) 
is based on the following observation.
Given a lineage polynomial $\poly(\vct{X})=\polyf(\circuit)$ for circuit \circuit over 
\abbrOneBIDB (recall that all \abbrCTIDB can be reduced to \abbrOneBIDB by~\Cref{prop:ctidb-reduct}), we have: 

\begin{equation}
\label{eq:tilde-Q-bi}
\rpoly\inparen{p_1,\dots,p_\numvar}=
\sum_{(\monom,\coef)\in \expansion{\circuit}} 
\indicator{\isInd{\encMon}
}\cdot \coef\cdot\prod_{X_i\in \monom} p_i.
\end{equation}

Given the above, the algorithm is a sampling based algorithm for the above sum: we sample (via \sampmon) $(\monom,\coef)\in \expansion{\circuit}$ with probability proportional
 to $\abs{\coef}$ and compute $\vari{Y}=\indicator{\isInd{\encMon}}
 \cdot \prod_{X_i\in \monom} p_i$. 
Repeating the sampling an appropriate number of times
and computing the average of $\vari{Y}$ gives us our final estimate. 

We illustrate \sampmon  (\Cref{alg:sample}), using the circuit $\circuit$ in \Cref{fig:circuit}. As a pre-processing step, \onepass (\Cref{alg:one-pass-iter}) recursively (for each sub-circuit) computes $\abs{\circuit}\inparen{1,\ldots, 1}$ in the obvious manner.
The \textcolor{gray}{gray} values in~\Cref{fig:circuit} represent the value $\abs{\circuit'}\inparen{1,\ldots, 1}$ for each sub-circuit $\circuit'$ rooted at the corresponding node. E.g. in the bottom right $\circmult$ (\textcolor{blue}{blue}) gate, the value is $1\times 1=1$ (where the left child has $Y\gets 1$ and the right child has $\abs{-1}=1$).  

  We now consider a partial run of \sampmon that samples  $\inparen{XY, -1}$ in~\Cref{fig:circuit}.  It recursively traverses \emph{both} children of the sink $\circmult$ gate.  For the  \textcolor{red}{red} and \textcolor{green}{green} children, which are both  $\circplus$ gates, we randomly choose one of their children. Specifically \sampmon then randomly picks the right child of the \textcolor{green}{green} $\circplus$ gate (i.e the \textcolor{blue}{blue} $\circmult$ gate that represents $-Y$ and is computed by recursing on both children of the \textcolor{blue}{blue} $\circmult$ gate) with probability of $\frac{1}{3}$ (where the numerator and denominator are the values computed by \onepass for the \textcolor{blue}{blue} $\circmult$ and \textcolor{green}{green} $\circplus$ gates respectively). Similarly at the left \textcolor{red}{red} $\circplus$ gate we sample the left child (representing $X$ and is computed by recursing to the leaf node $X$) with  probability $\frac{1}{3}$.  Note that the probability for choosing $\inparen{XY, -1}$ overall  is $\frac{1}{3}\cdot \frac{1}{3}=\frac{1}{9}$, which is indeed the ratio of the coefficient of $\inparen{XY, -1}$ to the sum of all coefficients in $\abs{\circuit}$, as needed.  
 All algorithm details, including those for \approxq (\Cref{alg:mon-sam}), are in \Cref{sec:proofs-approx-alg}.

\mypar{Runtime analysis} We can argue the following runtime for the \approxq (which solves \Cref{prob:intro-stmt}):
\begin{Theorem}
\label{cor:approx-algo-const-p}
Let \circuit be an arbitrary \emph{\abbrOneBIDB} circuit, define $\poly(\vct{X})=\polyf(\circuit)$, let $k=\degree(\circuit)$, and let $\gamma=\gamma(\circuit)$. Further let it be the case that $\prob_i\ge \prob_0$ for all $i\in[\numvar]$. Then for all $\epsilon'>0$, an estimate $\mathcal{E}$  of $\rpoly(\prob_1,\ldots, \prob_\numvar)$
satisfying
\begin{equation}
\label{eq:approx-algo-bound-main}
\probOf\left(\left|\mathcal{E} - \rpoly(\prob_1,\dots,\prob_\numvar)\right|> \error' \cdot \rpoly(\prob_1,\dots,\prob_\numvar)\right) \leq \conf
\end{equation}
 can be computed in time
\begin{footnotesize}
\begin{equation}
\label{eq:approx-algo-runtime}
O\left(\left(\size(\circuit) + \frac{\log{\frac{1}{\conf}}\cdot k\cdot \log{k} \cdot \depth(\circuit))}{\inparen{\error'}^2\cdot(1-\gamma)^2\cdot \prob_0^{2k}}\right)\cdot\multc{\log\left(\abs{\circuit}(1,\ldots, 1)\right)}{\log\left(\size(\circuit)\right)}\right).
\end{equation}
\end{footnotesize}
In particular, if $\prob_0>0$ and $\gamma<1$ are absolute constants then the above runtime simplifies to $$O_k\left(\left(\frac 1{\inparen{\error'}^2}\cdot\size(\circuit)\cdot \log{\frac{1}{\conf}}\right)\cdot\multc{\log\left(\abs{\circuit}(1,\ldots, 1)\right)}{\log\left(\size(\circuit)\right)}\right).$$
\end{Theorem}

The restriction on $\gamma$ is satisfied by any
$1$-\abbrTIDB (where $\gamma=0$ in the equivalent $1$-\abbrBIDB of~\Cref{prop:ctidb-reduct})
as well as for all three queries of the PDBench \abbrBIDB benchmark (\Cref{app:subsec:experiment}). 

Next,  by \Cref{prop:circuit-depth} and \Cref{lem:circ-model-runtime} for any $\raPlus$ query $\query$, there exists a circuit $\circuit^*$ for $\apolyqdt$ such that $\depth(\circuit^*)\le O_{|Q|}(\log{n})$ and $\size(\circuit^*)\le O_k\inparen{\qruntime{\query, \tupset, \bound}}$. Then, we note that \Cref{prop:ctidb-reduct} gives us an equivalent $\circuit$ from $\circuit^*$ with essentially the same size/depth and has $\gamma(\circuit)\le 1-c^{-\Omega(k)}$ (\Cref{lem:ctidb-gamma}). Finally, we argue (using the fact $\circuit$ has low depth) that $\abs{\circuit}(1,\dots,1)\le \size(\circuit)^{O_k(1)}$ (\Cref{lem:val-ub}).
The above sequence of arguments results in the following result (which answers \Cref{prob:big-o-joint-steps} in the affirmative):
\begin{Corollary}
\label{cor:approx-algo-punchline-ctidb}
Let $\query$ be an $\raPlus$ query and $\pdb$ be a \abbrCTIDB with $p_0>0$, where $p_0$ as in \Cref{cor:approx-algo-const-p}, is an absolute constant. Let $\poly(\vct{X})=\apolyqdt$ for any result tuple $\tup$ with $\deg(\poly)=k$. Then one can compute an approximation satisfying \Cref{eq:approx-algo-bound-main} in time $O_{k,|Q|,\error',\conf,\bound}\inparen{\qruntime{\optquery{\query}, \tupset, \bound}}$ (given $\query,\tupset$ and $\prob_{\tup, j}$ for each $\tup\in\tupset,~j\in\pbox{\bound}$ that defines $\bpd$).
\end{Corollary}

If we want to approximate the expected multiplicities of all $Z=O(n^k)$ result tuples $\tup$ simultaneously, we just need to run the above result with $\conf$ replaced by $\frac \conf Z$, which increases the runtime by a  factor of $O_k(\log{n})$.

\section{Related Work}\label{sec:related-work}
\textbf{Probabilistic Databases} (PDBs) have been studied predominantly for set semantics.
Approaches for probabilistic query processing (i.e., computing marginal probabilities of tuples), fall into two broad categories.
\emph{Intensional} (or \emph{grounded}) query evaluation computes the \emph{lineage} of a tuple
and then the probability of the lineage formula.
It has been shown that computing the marginal probability of a tuple is \sharpphard~\cite{valiant-79-cenrp} (by reduction from weighted model counting).
The second category, \emph{extensional} query evaluation, 
is in \ptime, but is limited to certain classes of queries.
Dalvi et al.~\cite{DS12} and Olteanu et al.~\cite{FO16} proved dichotomies for UCQs and two classes of queries with negation, respectively.
Amarilli et al. investigated tractable classes of databases for more complex queries~\cite{AB15}. 
Another line of work studies which structural properties of lineage formulas lead to tractable cases~\cite{kenig-13-nclexpdc,roy-11-f,sen-10-ronfqevpd}.
In this paper we focus on intensional query evaluation with polynomials.

Many data models have been proposed for encoding PDBs more compactly than as sets of possible worlds.
 These include tuple-independent databases~\cite{VS17} (\tis), block-independent databases (\bis)~\cite{RS07}, and \emph{PC-tables}~\cite{GT06}.
Fink et al.~\cite{FH12} study aggregate queries over a probabilistic version of the extension of K-relations for aggregate queries proposed in~\cite{AD11d} (\emph{pvc-tables}) that supports bags, and has runtime complexity linear in the size of the lineage.
However, this lineage is encoded as a tree; the size (and thus the runtime) are still superlinear in $\qruntime{\query, \tupset, \bound}$.
The runtime bound is also limited to a specific class of (hierarchical) queries, suggesting the possibility of a generalization of \cite{DS12}'s dichotomy result to \abbrBPDB\xplural for our problem (\cite{https://doi.org/10.48550/arxiv.2201.11524} presents a dichotomy result for a related problem).

Several techniques for approximating tuple probabilities have been proposed in related work~\cite{FH13,heuvel-19-anappdsd,DBLP:conf/icde/OlteanuHK10,DS07}, relying on Monte Carlo sampling, e.g.,~\cite{DS07}, or a branch-and-bound paradigm~\cite{DBLP:conf/icde/OlteanuHK10}.
Our approximation algorithm is also based on sampling.

\noindent \textbf{Compressed Encodings} are used for Boolean formulas (e.g, various types of circuits including OBDDs~\cite{jha-12-pdwm}) and polynomials (e.g., factorizations~\cite{factorized-db}) some of which have been utilized for  probabilistic query processing, e.g.,~\cite{jha-12-pdwm}.
Compact representations for which probabilities can be computed in linear time include OBDDs, SDDs, d-DNNF, and FBDD.
\cite{DM14c} studies circuits for absorptive semirings while~\cite{S18a} studies circuits that include negation (expressed as the monus operation). Algebraic Decision Diagrams~\cite{bahar-93-al} (ADDs) generalize BDDs to variables with more than two values. Chen et al.~\cite{chen-10-cswssr} introduced the generalized disjunctive normal form.
\Cref{sec:param-compl} covers more related work on fine-grained complexity.


\section{Conclusions and Future Work}\label{sec:concl-future-work}

We have studied the problem of calculating the expected multiplicity of a bag-query result tuple, 
 a problem that has a practical application in probabilistic databases over multisets. 
We show that under various parameterized complexity hardness results/conjectures computing the expected multiplicities exactly is not possible in time linear in the corresponding deterministic query processing time.
We prove that it is possible to approximate the expectation of a lineage polynomial in linear time
 in the deterministic query processing over TIDBs and BIDBs (assuming that there are few cancellations).
Interesting directions for future work include development of a dichotomy for bag \abbrPDB\xplural.  While we can handle higher moments (this follows fairly easily from our existing results-- see \Cref{sec:momemts}), more general approximations are an interesting area for exploration, including those for more general data models. 


\begin{acks}
We thank Virginia Williams for showing us \Cref{eq:3p-3tri}, which greatly simplified our earlier proof of Lemma 3.8, and for graciously allowing us to use it.  This work is supported by \grantsponsor{nsfid}{NSF}{} grants \grantnum{nsfid}{IIS-1956149}, \grantnum{nsfid}{IIS-1750460}, and \grantnum{nsfid}{IIS-2107107}.
\end{acks}

\bibliographystyle{ACM-Reference-Format}
\bibliography{main}


\begin{thebibliography}{53}


\ifx \showCODEN    \undefined \def \showCODEN     #1{\unskip}     \fi
\ifx \showDOI      \undefined \def \showDOI       #1{#1}\fi
\ifx \showISBNx    \undefined \def \showISBNx     #1{\unskip}     \fi
\ifx \showISBNxiii \undefined \def \showISBNxiii  #1{\unskip}     \fi
\ifx \showISSN     \undefined \def \showISSN      #1{\unskip}     \fi
\ifx \showLCCN     \undefined \def \showLCCN      #1{\unskip}     \fi
\ifx \shownote     \undefined \def \shownote      #1{#1}          \fi
\ifx \showarticletitle \undefined \def \showarticletitle #1{#1}   \fi
\ifx \showURL      \undefined \def \showURL       {\relax}        \fi
\providecommand\bibfield[2]{#2}
\providecommand\bibinfo[2]{#2}
\providecommand\natexlab[1]{#1}
\providecommand\showeprint[2][]{arXiv:#2}

\bibitem[Agrawal et~al\mbox{.}(2006)]%
        {DBLP:conf/vldb/AgrawalBSHNSW06}
\bibfield{author}{\bibinfo{person}{Parag Agrawal}, \bibinfo{person}{Omar
  Benjelloun}, \bibinfo{person}{Anish~Das Sarma}, \bibinfo{person}{Chris
  Hayworth}, \bibinfo{person}{Shubha~U. Nabar}, \bibinfo{person}{Tomoe
  Sugihara}, {and} \bibinfo{person}{Jennifer Widom}.}
  \bibinfo{year}{2006}\natexlab{}.
\newblock \showarticletitle{Trio: A System for Data, Uncertainty, and Lineage}.
  In \bibinfo{booktitle}{\emph{VLDB}}. \bibinfo{pages}{1151--1154}.
\newblock


\bibitem[Amarilli et~al\mbox{.}(2015)]%
        {AB15}
\bibfield{author}{\bibinfo{person}{Antoine Amarilli}, \bibinfo{person}{Pierre
  Bourhis}, {and} \bibinfo{person}{Pierre Senellart}.}
  \bibinfo{year}{2015}\natexlab{}.
\newblock \showarticletitle{Probabilities and provenance via tree
  decompositions}.
\newblock \bibinfo{journal}{\emph{PODS}} (\bibinfo{year}{2015}).
\newblock


\bibitem[Amsterdamer et~al\mbox{.}(2011)]%
        {AD11d}
\bibfield{author}{\bibinfo{person}{Yael Amsterdamer}, \bibinfo{person}{Daniel
  Deutch}, {and} \bibinfo{person}{Val Tannen}.}
  \bibinfo{year}{2011}\natexlab{}.
\newblock \showarticletitle{Provenance for Aggregate Queries}. In
  \bibinfo{booktitle}{\emph{PODS}}. \bibinfo{pages}{153--164}.
\newblock


\bibitem[Antova et~al\mbox{.}(2008)]%
        {4497507}
\bibfield{author}{\bibinfo{person}{Lyublena Antova}, \bibinfo{person}{Thomas
  Jansen}, \bibinfo{person}{Christoph Koch}, {and} \bibinfo{person}{Dan
  Olteanu}.} \bibinfo{year}{2008}\natexlab{}.
\newblock \showarticletitle{Fast and Simple Relational Processing of Uncertain
  Data}. In \bibinfo{booktitle}{\emph{2008 IEEE 24th International Conference
  on Data Engineering}}. \bibinfo{pages}{983--992}.
\newblock
\urldef\tempurl%
\url{https://doi.org/10.1109/ICDE.2008.4497507}
\showDOI{\tempurl}


\bibitem[Atserias et~al\mbox{.}(2013)]%
        {AGM}
\bibfield{author}{\bibinfo{person}{Albert Atserias}, \bibinfo{person}{Martin
  Grohe}, {and} \bibinfo{person}{D{\'{a}}niel Marx}.}
  \bibinfo{year}{2013}\natexlab{}.
\newblock \showarticletitle{Size Bounds and Query Plans for Relational Joins}.
\newblock \bibinfo{journal}{\emph{{SIAM} J. Comput.}} \bibinfo{volume}{42},
  \bibinfo{number}{4} (\bibinfo{year}{2013}), \bibinfo{pages}{1737--1767}.
\newblock
\urldef\tempurl%
\url{https://doi.org/10.1137/110859440}
\showDOI{\tempurl}


\bibitem[Bahar et~al\mbox{.}(1993)]%
        {bahar-93-al}
\bibfield{author}{\bibinfo{person}{R.~Iris Bahar}, \bibinfo{person}{Erica~A.
  Frohm}, \bibinfo{person}{Charles~M. Gaona}, \bibinfo{person}{Gary~D.
  Hachtel}, \bibinfo{person}{Enrico Macii}, \bibinfo{person}{Abelardo Pardo},
  {and} \bibinfo{person}{Fabio Somenzi}.} \bibinfo{year}{1993}\natexlab{}.
\newblock \showarticletitle{Algebraic Decision Diagrams and Their
  Applications}. In \bibinfo{booktitle}{\emph{IEEE CAD}}.
\newblock


\bibitem[Beskales et~al\mbox{.}(2010)]%
        {DBLP:journals/pvldb/BeskalesIG10}
\bibfield{author}{\bibinfo{person}{George Beskales}, \bibinfo{person}{Ihab~F.
  Ilyas}, {and} \bibinfo{person}{Lukasz Golab}.}
  \bibinfo{year}{2010}\natexlab{}.
\newblock \showarticletitle{Sampling the Repairs of Functional Dependency
  Violations under Hard Constraints}.
\newblock \bibinfo{journal}{\emph{Proc. {VLDB} Endow.}} \bibinfo{volume}{3},
  \bibinfo{number}{1} (\bibinfo{year}{2010}), \bibinfo{pages}{197--207}.
\newblock


\bibitem[Bürgisser et~al\mbox{.}(1997)]%
        {arith-complexity}
\bibfield{author}{\bibinfo{person}{Peter Bürgisser}, \bibinfo{person}{Michael
  Clausen}, {and} \bibinfo{person}{Mohammad~Amin Shokrollahi}.}
  \bibinfo{year}{1997}\natexlab{}.
\newblock \bibinfo{booktitle}{\emph{Algebraic complexity theory}}.
  Vol.~\bibinfo{volume}{315}.
\newblock \bibinfo{publisher}{Springer}.
\newblock


\bibitem[Chen and Grohe(2010)]%
        {chen-10-cswssr}
\bibfield{author}{\bibinfo{person}{Hubie Chen} {and} \bibinfo{person}{Martin
  Grohe}.} \bibinfo{year}{2010}\natexlab{}.
\newblock \showarticletitle{Constraint Satisfaction With Succinctly Specified
  Relations}.
\newblock \bibinfo{journal}{\emph{J. Comput. Syst. Sci.}} \bibinfo{volume}{76},
  \bibinfo{number}{8} (\bibinfo{year}{2010}), \bibinfo{pages}{847--860}.
\newblock


\bibitem[Chen et~al\mbox{.}(2006)]%
        {CHEN20061346}
\bibfield{author}{\bibinfo{person}{Jianer Chen}, \bibinfo{person}{Xiuzhen
  Huang}, \bibinfo{person}{Iyad~A. Kanj}, {and} \bibinfo{person}{Ge Xia}.}
  \bibinfo{year}{2006}\natexlab{}.
\newblock \showarticletitle{Strong computational lower bounds via parameterized
  complexity}.
\newblock \bibinfo{journal}{\emph{J. Comput. System Sci.}}
  \bibinfo{volume}{72}, \bibinfo{number}{8} (\bibinfo{year}{2006}),
  \bibinfo{pages}{1346--1367}.
\newblock
\showISSN{0022-0000}
\urldef\tempurl%
\url{https://doi.org/10.1016/j.jcss.2006.04.007}
\showDOI{\tempurl}


\bibitem[Curticapean(2013)]%
        {k-match}
\bibfield{author}{\bibinfo{person}{Radu Curticapean}.}
  \bibinfo{year}{2013}\natexlab{}.
\newblock \showarticletitle{Counting Matchings of Size k Is W[1]-Hard}. In
  \bibinfo{booktitle}{\emph{ICALP}}, Vol.~\bibinfo{volume}{7965}.
  \bibinfo{pages}{352--363}.
\newblock


\bibitem[Curticapean and Marx(2014)]%
        {10.1109/FOCS.2014.22}
\bibfield{author}{\bibinfo{person}{Radu Curticapean} {and}
  \bibinfo{person}{D\'{a}niel Marx}.} \bibinfo{year}{2014}\natexlab{}.
\newblock \showarticletitle{Complexity of Counting Subgraphs: Only the
  Boundedness of the Vertex-Cover Number Counts}. In
  \bibinfo{booktitle}{\emph{Proceedings of the 2014 IEEE 55th Annual Symposium
  on Foundations of Computer Science}} \emph{(\bibinfo{series}{FOCS '14})}.
  \bibinfo{publisher}{IEEE Computer Society}, \bibinfo{address}{USA},
  \bibinfo{pages}{130–139}.
\newblock
\showISBNx{9781479965175}
\urldef\tempurl%
\url{https://doi.org/10.1109/FOCS.2014.22}
\showDOI{\tempurl}


\bibitem[Dalvi and Suciu(2007a)]%
        {10.1145/1265530.1265571}
\bibfield{author}{\bibinfo{person}{Nilesh Dalvi} {and} \bibinfo{person}{Dan
  Suciu}.} \bibinfo{year}{2007}\natexlab{a}.
\newblock \showarticletitle{The Dichotomy of Conjunctive Queries on
  Probabilistic Structures}. In \bibinfo{booktitle}{\emph{PODS}}.
  \bibinfo{pages}{293--302}.
\newblock


\bibitem[Dalvi and Suciu(2007b)]%
        {DS07}
\bibfield{author}{\bibinfo{person}{N. Dalvi} {and} \bibinfo{person}{D. Suciu}.}
  \bibinfo{year}{2007}\natexlab{b}.
\newblock \showarticletitle{Efficient query evaluation on probabilistic
  databases}.
\newblock \bibinfo{journal}{\emph{VLDB}} \bibinfo{volume}{16},
  \bibinfo{number}{4} (\bibinfo{year}{2007}), \bibinfo{pages}{544}.
\newblock


\bibitem[Dalvi and Suciu(2012)]%
        {DS12}
\bibfield{author}{\bibinfo{person}{Nilesh Dalvi} {and} \bibinfo{person}{Dan
  Suciu}.} \bibinfo{year}{2012}\natexlab{}.
\newblock \showarticletitle{The dichotomy of probabilistic inference for unions
  of conjunctive queries}.
\newblock \bibinfo{journal}{\emph{JACM}} \bibinfo{volume}{59},
  \bibinfo{number}{6} (\bibinfo{year}{2012}), \bibinfo{pages}{30}.
\newblock


\bibitem[den Heuvel et~al\mbox{.}(2019)]%
        {heuvel-19-anappdsd}
\bibfield{author}{\bibinfo{person}{Maarten~Van den Heuvel},
  \bibinfo{person}{Peter Ivanov}, \bibinfo{person}{Wolfgang Gatterbauer},
  \bibinfo{person}{Floris Geerts}, {and} \bibinfo{person}{Martin Theobald}.}
  \bibinfo{year}{2019}\natexlab{}.
\newblock \showarticletitle{Anytime Approximation in Probabilistic Databases
  via Scaled Dissociations}. In \bibinfo{booktitle}{\emph{SIGMOD}}.
  \bibinfo{pages}{1295--1312}.
\newblock


\bibitem[Deutch et~al\mbox{.}(2014)]%
        {DM14c}
\bibfield{author}{\bibinfo{person}{Daniel Deutch}, \bibinfo{person}{Tova Milo},
  \bibinfo{person}{Sudeepa Roy}, {and} \bibinfo{person}{Val Tannen}.}
  \bibinfo{year}{2014}\natexlab{}.
\newblock \showarticletitle{Circuits for Datalog Provenance}. In
  \bibinfo{booktitle}{\emph{ICDT}}. \bibinfo{pages}{201--212}.
\newblock


\bibitem[Feng et~al\mbox{.}(2021)]%
        {feng:2021:sigmod:efficient}
\bibfield{author}{\bibinfo{person}{Su Feng}, \bibinfo{person}{Boris Glavic},
  \bibinfo{person}{Aaron Huber}, {and} \bibinfo{person}{Oliver Kennedy}.}
  \bibinfo{year}{2021}\natexlab{}.
\newblock \showarticletitle{Efficient Uncertainty Tracking for Complex Queries
  with Attribute-level Bounds}. In \bibinfo{booktitle}{\emph{SIGMOD}}.
\newblock


\bibitem[Feng et~al\mbox{.}(2019)]%
        {feng:2019:sigmod:uncertainty}
\bibfield{author}{\bibinfo{person}{Su Feng}, \bibinfo{person}{Aaron Huber},
  \bibinfo{person}{Boris Glavic}, {and} \bibinfo{person}{Oliver Kennedy}.}
  \bibinfo{year}{2019}\natexlab{}.
\newblock \showarticletitle{Uncertainty Annotated Databases - A Lightweight
  Approach for Approximating Certain Answers}. In
  \bibinfo{booktitle}{\emph{SIGMOD}}.
\newblock


\bibitem[Fink et~al\mbox{.}(2012)]%
        {FH12}
\bibfield{author}{\bibinfo{person}{Robert Fink}, \bibinfo{person}{Larisa Han},
  {and} \bibinfo{person}{Dan Olteanu}.} \bibinfo{year}{2012}\natexlab{}.
\newblock \showarticletitle{Aggregation in probabilistic databases via
  knowledge compilation}.
\newblock \bibinfo{journal}{\emph{PVLDB}} \bibinfo{volume}{5},
  \bibinfo{number}{5} (\bibinfo{year}{2012}), \bibinfo{pages}{490--501}.
\newblock


\bibitem[Fink et~al\mbox{.}(2013)]%
        {FH13}
\bibfield{author}{\bibinfo{person}{Robert Fink}, \bibinfo{person}{Jiewen
  Huang}, {and} \bibinfo{person}{Dan Olteanu}.}
  \bibinfo{year}{2013}\natexlab{}.
\newblock \showarticletitle{Anytime approximation in probabilistic databases}.
\newblock \bibinfo{journal}{\emph{VLDBJ}} \bibinfo{volume}{22},
  \bibinfo{number}{6} (\bibinfo{year}{2013}), \bibinfo{pages}{823--848}.
\newblock


\bibitem[Fink and Olteanu(2016)]%
        {FO16}
\bibfield{author}{\bibinfo{person}{Robert Fink} {and} \bibinfo{person}{Dan
  Olteanu}.} \bibinfo{year}{2016}\natexlab{}.
\newblock \showarticletitle{Dichotomies for Queries with Negation in
  Probabilistic Databases}.
\newblock \bibinfo{journal}{\emph{TODS}} \bibinfo{volume}{41},
  \bibinfo{number}{1} (\bibinfo{year}{2016}), \bibinfo{pages}{4:1--4:47}.
\newblock


\bibitem[Flum and Grohe(2002)]%
        {10.5555/645413.652181}
\bibfield{author}{\bibinfo{person}{J\"{o}rg Flum} {and} \bibinfo{person}{Martin
  Grohe}.} \bibinfo{year}{2002}\natexlab{}.
\newblock \showarticletitle{The Parameterized Complexity of Counting Problems}.
  In \bibinfo{booktitle}{\emph{Proceedings of the 43rd Symposium on Foundations
  of Computer Science}} \emph{(\bibinfo{series}{FOCS '02})}.
  \bibinfo{publisher}{IEEE Computer Society}, \bibinfo{address}{USA},
  \bibinfo{pages}{538}.
\newblock
\showISBNx{0769518222}


\bibitem[Flum and Grohe(2006)]%
        {param-comp}
\bibfield{author}{\bibinfo{person}{J{\"o}rg Flum} {and} \bibinfo{person}{Martin
  Grohe}.} \bibinfo{year}{2006}\natexlab{}.
\newblock \showarticletitle{Parameterized Complexity Theory}. In
  \bibinfo{booktitle}{\emph{Texts in Theoretical Computer Science. An EATCS
  Series}}.
\newblock


\bibitem[Garcia{-}Molina et~al\mbox{.}(2009)]%
        {DBLP:books/daglib/0020812}
\bibfield{author}{\bibinfo{person}{Hector Garcia{-}Molina},
  \bibinfo{person}{Jeffrey~D. Ullman}, {and} \bibinfo{person}{Jennifer Widom}.}
  \bibinfo{year}{2009}\natexlab{}.
\newblock \bibinfo{booktitle}{\emph{Database Systems - The Complete Book {(2.}
  ed.)}}.
\newblock \bibinfo{publisher}{Pearson Education}.
\newblock


\bibitem[Green et~al\mbox{.}(2007)]%
        {DBLP:conf/pods/GreenKT07}
\bibfield{author}{\bibinfo{person}{Todd~J. Green}, \bibinfo{person}{Gregory
  Karvounarakis}, {and} \bibinfo{person}{Val Tannen}.}
  \bibinfo{year}{2007}\natexlab{}.
\newblock \showarticletitle{Provenance semirings}. In
  \bibinfo{booktitle}{\emph{PODS}}. \bibinfo{pages}{31--40}.
\newblock


\bibitem[Green and Tannen(2006)]%
        {GT06}
\bibfield{author}{\bibinfo{person}{Todd~J Green} {and} \bibinfo{person}{Val
  Tannen}.} \bibinfo{year}{2006}\natexlab{}.
\newblock \showarticletitle{Models for incomplete and probabilistic
  information}.
\newblock In \bibinfo{booktitle}{\emph{EDBT}}. \bibinfo{pages}{278--296}.
\newblock


\bibitem[Grohe et~al\mbox{.}(2022)]%
        {https://doi.org/10.48550/arxiv.2201.11524}
\bibfield{author}{\bibinfo{person}{Martin Grohe}, \bibinfo{person}{Peter
  Lindner}, {and} \bibinfo{person}{Christoph Standke}.}
  \bibinfo{year}{2022}\natexlab{}.
\newblock \bibinfo{title}{Probabilistic Query Evaluation with Bag Semantics}.
\newblock
\newblock
\urldef\tempurl%
\url{https://doi.org/10.48550/ARXIV.2201.11524}
\showDOI{\tempurl}


\bibitem[Guagliardo and Libkin(2017)]%
        {DBLP:journals/sigmod/GuagliardoL17}
\bibfield{author}{\bibinfo{person}{Paolo Guagliardo} {and}
  \bibinfo{person}{Leonid Libkin}.} \bibinfo{year}{2017}\natexlab{}.
\newblock \showarticletitle{Correctness of SQL Queries on Databases with
  Nulls}.
\newblock \bibinfo{journal}{\emph{SIGMOD Rec.}} \bibinfo{volume}{46},
  \bibinfo{number}{3} (\bibinfo{year}{2017}), \bibinfo{pages}{5--16}.
\newblock


\bibitem[Imieli\'nski and Lipski~Jr(1984)]%
        {IL84a}
\bibfield{author}{\bibinfo{person}{Tomasz Imieli\'nski} {and}
  \bibinfo{person}{Witold Lipski~Jr}.} \bibinfo{year}{1984}\natexlab{}.
\newblock \showarticletitle{Incomplete Information in Relational Databases}.
\newblock \bibinfo{journal}{\emph{JACM}} \bibinfo{volume}{31},
  \bibinfo{number}{4} (\bibinfo{year}{1984}), \bibinfo{pages}{761--791}.
\newblock


\bibitem[Impagliazzo et~al\mbox{.}(2001)]%
        {eth}
\bibfield{author}{\bibinfo{person}{Russell Impagliazzo},
  \bibinfo{person}{Ramamohan Paturi}, {and} \bibinfo{person}{Francis Zane}.}
  \bibinfo{year}{2001}\natexlab{}.
\newblock \showarticletitle{Which Problems Have Strongly Exponential
  Complexity?}
\newblock \bibinfo{journal}{\emph{J. Comput. System Sci.}}
  \bibinfo{volume}{63}, \bibinfo{number}{4} (\bibinfo{year}{2001}),
  \bibinfo{pages}{512--530}.
\newblock
\showISSN{0022-0000}
\urldef\tempurl%
\url{https://doi.org/10.1006/jcss.2001.1774}
\showDOI{\tempurl}


\bibitem[Jha and Suciu(2012)]%
        {jha-12-pdwm}
\bibfield{author}{\bibinfo{person}{Abhay~Kumar Jha} {and} \bibinfo{person}{Dan
  Suciu}.} \bibinfo{year}{2012}\natexlab{}.
\newblock \showarticletitle{Probabilistic Databases With Markoviews}.
\newblock \bibinfo{journal}{\emph{PVLDB}} \bibinfo{volume}{5},
  \bibinfo{number}{11} (\bibinfo{year}{2012}), \bibinfo{pages}{1160--1171}.
\newblock


\bibitem[Joglekar et~al\mbox{.}(2016)]%
        {ajar}
\bibfield{author}{\bibinfo{person}{Manas~R. Joglekar}, \bibinfo{person}{Rohan
  Puttagunta}, {and} \bibinfo{person}{Christopher R{\'{e}}}.}
  \bibinfo{year}{2016}\natexlab{}.
\newblock \showarticletitle{{AJAR:} Aggregations and Joins over Annotated
  Relations}. In \bibinfo{booktitle}{\emph{Proceedings of the 35th {ACM}
  {SIGMOD-SIGACT-SIGAI} Symposium on Principles of Database Systems, {PODS}
  2016, San Francisco, CA, USA, June 26 - July 01, 2016}},
  \bibfield{editor}{\bibinfo{person}{Tova Milo} {and}
  \bibinfo{person}{Wang{-}Chiew Tan}} (Eds.). \bibinfo{publisher}{{ACM}},
  \bibinfo{pages}{91--106}.
\newblock
\urldef\tempurl%
\url{https://doi.org/10.1145/2902251.2902293}
\showDOI{\tempurl}


\bibitem[Karp et~al\mbox{.}(1989)]%
        {DBLP:journals/jal/KarpLM89}
\bibfield{author}{\bibinfo{person}{Richard~M. Karp}, \bibinfo{person}{Michael
  Luby}, {and} \bibinfo{person}{Neal Madras}.} \bibinfo{year}{1989}\natexlab{}.
\newblock \showarticletitle{Monte-Carlo Approximation Algorithms for
  Enumeration Problems}.
\newblock \bibinfo{journal}{\emph{J. Algorithms}} \bibinfo{volume}{10},
  \bibinfo{number}{3} (\bibinfo{year}{1989}), \bibinfo{pages}{429--448}.
\newblock


\bibitem[Kenig et~al\mbox{.}(2013)]%
        {kenig-13-nclexpdc}
\bibfield{author}{\bibinfo{person}{Batya Kenig}, \bibinfo{person}{Avigdor Gal},
  {and} \bibinfo{person}{Ofer Strichman}.} \bibinfo{year}{2013}\natexlab{}.
\newblock \showarticletitle{A New Class of Lineage Expressions over
  Probabilistic Databases Computable in P-Time}. In
  \bibinfo{booktitle}{\emph{SUM}}, Vol.~\bibinfo{volume}{8078}.
  \bibinfo{pages}{219--232}.
\newblock


\bibitem[Khamis et~al\mbox{.}(2016)]%
        {DBLP:conf/pods/KhamisNR16}
\bibfield{author}{\bibinfo{person}{Mahmoud~Abo Khamis},
  \bibinfo{person}{Hung~Q. Ngo}, {and} \bibinfo{person}{Atri Rudra}.}
  \bibinfo{year}{2016}\natexlab{}.
\newblock \showarticletitle{FAQ: Questions Asked Frequently}. In
  \bibinfo{booktitle}{\emph{PODS}}. \bibinfo{pages}{13--28}.
\newblock


\bibitem[Kopelowitz and Williams(2020)]%
        {triang-hard}
\bibfield{author}{\bibinfo{person}{Tsvi Kopelowitz} {and}
  \bibinfo{person}{Virginia~Vassilevska Williams}.}
  \bibinfo{year}{2020}\natexlab{}.
\newblock \showarticletitle{Towards Optimal Set-Disjointness and
  Set-Intersection Data Structures}. In \bibinfo{booktitle}{\emph{ICALP}},
  Vol.~\bibinfo{volume}{168}. \bibinfo{pages}{74:1--74:16}.
\newblock


\bibitem[Kumari et~al\mbox{.}(2016)]%
        {kumari:2016:qdb:communicating}
\bibfield{author}{\bibinfo{person}{Poonam Kumari}, \bibinfo{person}{Said
  Achmiz}, {and} \bibinfo{person}{Oliver Kennedy}.}
  \bibinfo{year}{2016}\natexlab{}.
\newblock \showarticletitle{Communicating Data Quality in On-Demand Curation}.
  In \bibinfo{booktitle}{\emph{QDB}}.
\newblock


\bibitem[Ngo(2018)]%
        {ngo-survey}
\bibfield{author}{\bibinfo{person}{Hung~Q. Ngo}.}
  \bibinfo{year}{2018}\natexlab{}.
\newblock \showarticletitle{Worst-Case Optimal Join Algorithms: Techniques,
  Results, and Open Problems}. In \bibinfo{booktitle}{\emph{PODS}}.
\newblock


\bibitem[Ngo et~al\mbox{.}(2013)]%
        {skew}
\bibfield{author}{\bibinfo{person}{Hung~Q. Ngo}, \bibinfo{person}{Christopher
  Ré}, {and} \bibinfo{person}{Atri Rudra}.} \bibinfo{year}{2013}\natexlab{}.
\newblock \showarticletitle{Skew strikes back: new developments in the theory
  of join algorithms}.
\newblock \bibinfo{journal}{\emph{SIGMOD Rec.}} \bibinfo{volume}{42},
  \bibinfo{number}{4} (\bibinfo{year}{2013}), \bibinfo{pages}{5--16}.
\newblock


\bibitem[Olteanu et~al\mbox{.}(2010)]%
        {DBLP:conf/icde/OlteanuHK10}
\bibfield{author}{\bibinfo{person}{Dan Olteanu}, \bibinfo{person}{Jiewen
  Huang}, {and} \bibinfo{person}{Christoph Koch}.}
  \bibinfo{year}{2010}\natexlab{}.
\newblock \showarticletitle{Approximate confidence computation in probabilistic
  databases}. In \bibinfo{booktitle}{\emph{ICDE}}. \bibinfo{pages}{145--156}.
\newblock


\bibitem[Olteanu and Schleich(2016)]%
        {factorized-db}
\bibfield{author}{\bibinfo{person}{Dan Olteanu} {and}
  \bibinfo{person}{Maximilian Schleich}.} \bibinfo{year}{2016}\natexlab{}.
\newblock \showarticletitle{Factorized Databases}.
\newblock \bibinfo{journal}{\emph{SIGMOD Rec.}} \bibinfo{volume}{45},
  \bibinfo{number}{2} (\bibinfo{year}{2016}), \bibinfo{pages}{5--16}.
\newblock


\bibitem[pdbench utility(2008)]%
        {pdbench}
pdbench utility \bibinfo{year}{2008}\natexlab{}.
\newblock \bibinfo{title}{pdbench}.
\newblock
\newblock
\urldef\tempurl%
\url{http://pdbench.sourceforge.net/}
\showURL{%
\tempurl}
\newblock
\shownote{Accessed: 2020-12-15}.


\bibitem[Rekatsinas et~al\mbox{.}(2017)]%
        {DBLP:journals/pvldb/RekatsinasCIR17}
\bibfield{author}{\bibinfo{person}{Theodoros Rekatsinas}, \bibinfo{person}{Xu
  Chu}, \bibinfo{person}{Ihab~F. Ilyas}, {and} \bibinfo{person}{Christopher
  R{\'{e}}}.} \bibinfo{year}{2017}\natexlab{}.
\newblock \showarticletitle{HoloClean: Holistic Data Repairs with Probabilistic
  Inference}.
\newblock \bibinfo{journal}{\emph{Proc. {VLDB} Endow.}} \bibinfo{volume}{10},
  \bibinfo{number}{11} (\bibinfo{year}{2017}), \bibinfo{pages}{1190--1201}.
\newblock


\bibitem[Roy et~al\mbox{.}(2011)]%
        {roy-11-f}
\bibfield{author}{\bibinfo{person}{Sudeepa Roy}, \bibinfo{person}{Vittorio
  Perduca}, {and} \bibinfo{person}{Val Tannen}.}
  \bibinfo{year}{2011}\natexlab{}.
\newblock \showarticletitle{Faster query answering in probabilistic databases
  using read-once functions}. In \bibinfo{booktitle}{\emph{ICDT}}.
\newblock


\bibitem[Ré and Suciu(2007)]%
        {RS07}
\bibfield{author}{\bibinfo{person}{C. Ré} {and} \bibinfo{person}{D. Suciu}.}
  \bibinfo{year}{2007}\natexlab{}.
\newblock \showarticletitle{Materialized views in probabilistic databases: for
  information exchange and query optimization}. In
  \bibinfo{booktitle}{\emph{VLDB}}. \bibinfo{pages}{51--62}.
\newblock


\bibitem[Sa et~al\mbox{.}(2017)]%
        {DBLP:journals/vldb/SaRR0W0Z17}
\bibfield{author}{\bibinfo{person}{Christopher~De Sa},
  \bibinfo{person}{Alexander Ratner}, \bibinfo{person}{Christopher R{\'{e}}},
  \bibinfo{person}{Jaeho Shin}, \bibinfo{person}{Feiran Wang},
  \bibinfo{person}{Sen Wu}, {and} \bibinfo{person}{Ce Zhang}.}
  \bibinfo{year}{2017}\natexlab{}.
\newblock \showarticletitle{Incremental knowledge base construction using
  DeepDive}.
\newblock \bibinfo{journal}{\emph{{VLDB} J.}} \bibinfo{volume}{26},
  \bibinfo{number}{1} (\bibinfo{year}{2017}), \bibinfo{pages}{81--105}.
\newblock


\bibitem[Sen et~al\mbox{.}(2010)]%
        {sen-10-ronfqevpd}
\bibfield{author}{\bibinfo{person}{Prithviraj Sen}, \bibinfo{person}{Amol
  Deshpande}, {and} \bibinfo{person}{Lise Getoor}.}
  \bibinfo{year}{2010}\natexlab{}.
\newblock \showarticletitle{Read-Once Functions and Query Evaluation in
  Probabilistic Databases}.
\newblock \bibinfo{journal}{\emph{PVLDB}} \bibinfo{volume}{3},
  \bibinfo{number}{1} (\bibinfo{year}{2010}), \bibinfo{pages}{1068--1079}.
\newblock


\bibitem[Senellart(2018)]%
        {S18a}
\bibfield{author}{\bibinfo{person}{Pierre Senellart}.}
  \bibinfo{year}{2018}\natexlab{}.
\newblock \showarticletitle{Provenance and Probabilities in Relational
  Databases}.
\newblock \bibinfo{journal}{\emph{SIGMOD Record}} \bibinfo{volume}{46},
  \bibinfo{number}{4} (\bibinfo{year}{2018}), \bibinfo{pages}{5--15}.
\newblock


\bibitem[Valiant(1979)]%
        {valiant-79-cenrp}
\bibfield{author}{\bibinfo{person}{Leslie~G. Valiant}.}
  \bibinfo{year}{1979}\natexlab{}.
\newblock \showarticletitle{The Complexity of Enumeration and Reliability
  Problems}.
\newblock \bibinfo{journal}{\emph{SIAM J. Comput.}} \bibinfo{volume}{8},
  \bibinfo{number}{3} (\bibinfo{year}{1979}), \bibinfo{pages}{410--421}.
\newblock


\bibitem[Van~den Broeck and Suciu(2017)]%
        {VS17}
\bibfield{author}{\bibinfo{person}{Guy Van~den Broeck} {and}
  \bibinfo{person}{Dan Suciu}.} \bibinfo{year}{2017}\natexlab{}.
\newblock \showarticletitle{Query Processing on Probabilistic Data: A Survey}.
\newblock  (\bibinfo{year}{2017}).
\newblock


\bibitem[Williams(2018)]%
        {virgi-survey}
\bibfield{author}{\bibinfo{person}{Virginia~Vassilevska Williams}.}
  \bibinfo{year}{2018}\natexlab{}.
\newblock \showarticletitle{Some Open Problems in Fine-Grained Complexity}.
\newblock \bibinfo{journal}{\emph{{SIGACT} News}} \bibinfo{volume}{49},
  \bibinfo{number}{4} (\bibinfo{year}{2018}), \bibinfo{pages}{29--35}.
\newblock
\urldef\tempurl%
\url{https://doi.org/10.1145/3300150.3300158}
\showDOI{\tempurl}


\bibitem[Yang et~al\mbox{.}(2015)]%
        {yang:2015:pvldb:lenses}
\bibfield{author}{\bibinfo{person}{Ying Yang}, \bibinfo{person}{Niccolò
  Meneghetti}, \bibinfo{person}{Ronny Fehling}, \bibinfo{person}{Zhen~Hua Liu},
  \bibinfo{person}{Dieter Gawlick}, {and} \bibinfo{person}{Oliver Kennedy}.}
  \bibinfo{year}{2015}\natexlab{}.
\newblock \showarticletitle{Lenses: An On-Demand Approach to ETL}.
\newblock \bibinfo{journal}{\emph{PVLDB}} \bibinfo{volume}{8},
  \bibinfo{number}{12} (\bibinfo{year}{2015}), \bibinfo{pages}{1578--1589}.
\newblock


\end{thebibliography}

 \clearpage
 \appendix
 \normalsize




\onecolumn
\section{Missing details from Section~\ref{sec:background}}\label{sec:proofs-background}
\subsection{Polynomials}
\begin{Definition}[Degree]\label{def:degree-of-poly}
The degree of polynomial $\genpoly(\vct{X})$ is the largest $\sum_{\tup\in S}d_\tup
$ for all $\vct{d}\in\inset{0,\ldots,\hideg}^S$
 such that $c_{(d_1,\dots,d_n)}\ne 0$. 
 We denote the degree of $\genpoly$ as $\deg\inparen{\genpoly}$.
\end{Definition}
As an example, the degree of the polynomial $X^2+2XY^2+Y^2$ is $3$.
Product terms in lineage arise only from join operations (\Cref{fig:nxDBSemantics}), so intuitively, the degree of a lineage polynomial is analogous to the largest number of joins needed to produce a result tuple.

\subsection{Background details for proof of~\Cref{prop:expection-of-polynom}}\label{app:subsec:background-nxdbs}
\subsubsection{$\semK$-relations and \abbrNXPDB\xplural}\label{subsec:supp-mat-background}\label{subsec:supp-mat-krelations}

We can use $\semK$-relations to model bags. A \emph{$\semK$-relation}~\cite{DBLP:conf/pods/GreenKT07} is a relation whose tuples are annotated with elements from a commutative semiring $\semK = \inset{\domK, \addK, \multK, \zeroK, \oneK}$.  A commutative semiring is a structure with a domain $\domK$ and associative and commutative binary operations $\addK$ and $\multK$ such that $\multK$ distributes over $\addK$, $\zeroK$ is the identity of $\addK$, $\oneK$ is the identity of $\multK$, and $\zeroK$ annihilates all elements of $\domK$ when combined by $\multK$.
Let $\udom$ be a countable domain of values.
Formally, an n-ary $\semK$-relation $\rel$ over $\udom$ is a function $\rel: \udom^n \to \domK$ with finite support $\support{\rel} = \{ \tup \mid \rel(\tup) \neq \zeroK \}$.  A $\semK$-database is defined similarly, where we view the $\semK$-database (relation) as a function mapping tuples to their respective annotations.
$\raPlus$ query semantics over $\semK$-relations are analogous to the lineage construction semantics of \Cref{fig:nxDBSemantics}, with the exception of replacing $+$ with $\addK$ and $\cdot$ with $\multK$.

Consider the semiring $\semN = \inset{\domN,+,\times,0,1}$ of natural numbers. $\semN$-databases model bag semantics by annotating each tuple with its multiplicity. A  probabilistic $\semN$-database ($\semN$-PDB) is a PDB where each possible world is an $\semN$-database. We study the problem of computing statistical moments for query results over such databases.  Given an $\semN$-\abbrPDB $\pdb = (\idb, \pd)$, ($\raPlus$) query $\query$, and possible result tuple $\tup$, we sum $\query(\db)(\tup)\cdot\pd\inparen{\db}$ for all $\db \in \idb$ to compute the expected multiplicity of $\tup$.  Intuitively, the expectation of $\query(\db)(t)$ is the number of duplicates of $t$ we expect to find in result of query $\query$.

Let $\semNX$ denote the set of polynomials over variables $\vct{X}=(X_1,\dots,X_n)$ with natural number coefficients and exponents.
Consider now the semiring (abusing notation) $\semNX = \inset{\semNX, +, \cdot, 0, 1}$ whose domain is $\semNX$, with the standard addition and multiplication of polynomials.
We define an \abbrNXPDB $\pxdb$ as the tuple $(\db_{\semNX}, \pd)$, where $\semNX$-database $\db_{\semNX}$ is paired with the probability distribution $\pd$ across the set of possible worlds \emph{represented} by $\db_{\semNX}$, i.e. the one induced from $\mathcal{P}_{\semNX}$, the probability distribution over $\vct{X}$.  Note that the notation is slightly abused since the first element of the pair is an encoded set of possible worlds, i.e. $\db_{\semNX}$ is the \dbbaseName.
We denote by $\nxpolyqdt$ the annotation of tuple $t$ in the result of $\query(\db_{\semNX})(t)$, and as before, interpret it as a function $\nxpolyqdt: \{0,1\}^{|\vct X|} \rightarrow \semN$ from vectors of variable assignments to the corresponding value of the annotating polynomial.
\abbrNXPDB\xplural and a function $\rmod$ (which transforms an \abbrNXPDB to an equivalent $\semN$-PDB) are both formalized next.


To justify the use of $\semNX$-databases, we need to show that we can encode any $\semN$-PDB in this way and that the query semantics over this representation coincides with query semantics over its respective $\semN$-PDB. For that it will be opportune to define representation systems for $\semN$-PDBs.

\begin{Definition}[Representation System]\label{def:representation-syste}
  A representation system for $\semN$-PDBs is a tuple $(\reprs, \rmod)$ where $\reprs$ is a set of representations and $\rmod$ associates with each $\repr \in \reprs$ an $\semN$-PDB $\pdb$. We say that a representation system is \emph{closed} under a class of queries $\qClass$ if for any query $\query \in \qClass$ and $\repr \in \reprs$ we have:
  \[ \rmod(\query(\repr)) = \query(\rmod(\repr)) \]

  A representation system is \emph{complete} if for every $\semN$-PDB $\pdb$ there exists $\repr \in \reprs$ such that:
  \[ \rmod(\repr) = \pdb \]

\end{Definition}

As mentioned above we will use $\semNX$-databases paired with a probability distribution as a representation system, referring to such databases as \abbrNXPDB\xplural.
Given \abbrNXPDB $\pxdb$, one can think of the of $\pd$ as the probability distribution across all worlds $\inset{0, 1}^\numvar$.  Denote a particular world to be $\vct{W}$.  For convenience let $\assign_\vct{W}: \pxdb\rightarrow\pndb$ be a function that computes the corresponding $\semN$-\abbrPDB upon assigning all values $W_i \in \vct{W}$ to $X_i \in \vct{X}$ of $\db_{\semNX}$.  Note the one-to-one correspondence between elements $\vct{W}\in\inset{0, 1}^\numvar$ to the worlds encoded by $\db_{\semNX}$ when $\vct{W}$ is assigned to $\vct{X}$ (assuming a domain of $\inset{0, 1}$ for each $X_i$).   
We can think of $\assign_\vct{W}(\pxdb)\inparen{\tup}$ as the semiring homomorphism $\semNX \to \semN$ that applies the assignment $\vct{W}$ to all variables $\vct{X}$ of a polynomial and evaluates the resulting expression in $\semN$.

\begin{Definition}[$\rmod\inparen{\pxdb}$]\label{def:semnx-pdbs}
  Given an \abbrNXPDB$\pxdb$, we compute its equivalent $\semN$-\abbrPDB $\pndb = \rmod\inparen{\pxdb} = \inparen{\idb, \pd'}$ as:
  \begin{align*}
    \idb      & = \{ \assign_{\vct{W}}(\pxdb) \mid \vct{W} \in  \{0,1\}^n \} \\
    \forall \db \in \idb: \probOf(\db) & = \sum_{\vct{W} \in \{0,1\}^n: \assign_{\vct{W}}(\pxdb) = \db} \probOf(\vct{W})
  \end{align*}
\end{Definition}

For instance, consider a $\pxdb$ consisting of a single tuple $\tup_1 = (1)$ annotated with $X_1 + X_2$ with probability distribution $\probOf([0,0]) = 0$, $\probOf([0,1]) = 0$, $\probOf([1,0]) = 0.3$ and $\probOf([1,1]) = 0.7$. This \abbrNXPDB encodes two possible worlds (with non-zero probability) that we denote using their world vectors.
\[
  D_{[0,1]}(\tup_1) = 1 \hspace{0.3cm} \mathbf{and} \hspace{0.3cm} D_{[1,1]}(\tup_1) = 2
\]
Importantly, as the following proposition shows, any finite $\semN$-PDB can be encoded as an \abbrNXPDB and \abbrNXPDB\xplural are closed under $\raPlus$\cite{DBLP:conf/pods/GreenKT07}.

\begin{Proposition}\label{prop:semnx-pdbs-are-a-}
\abbrNXPDB\xplural are a complete representation system for $\semN$-PDBs that is closed under $\raPlus$ queries.
\end{Proposition}

\begin{proof}
To prove that \abbrNXPDB\xplural are complete consider the following construction that for any $\semN$-PDB $\pdb = (\idb, \pd)$ produces an \abbrNXPDB $\pxdb = (\db_{\semNX}, \pd')$  such that $\rmod(\pxdb) = \pdb$. Let $\idb = \{D_1, \ldots, D_{\abs{\idb}}\}.$ 
 For each world $D_i$ we create a corresponding variable $X_i$.
In $\db_{\semNX}$ we assign each tuple $\tup$ the polynomial:
  \[
 \db_{\semNX}(\tup) = \sum_{i=1}^{\abs{\idb}} D_i(\tup)\cdot X_{i}
  \]
The probability distribution $\pd'$ assigns all world vectors zero probability except for $\abs{\idb}$ world vectors (representing the possible worlds) $\vct{W}_i$. All elements of $\vct{W}_i$ are zero except for the position corresponding to variables $X_{i}$ which is set to $1$. Unfolding definitions it is trivial to show that $\rmod(\pxdb) = \pdb$. Thus, \abbrNXPDB\xplural are a complete representation system.

Since $\semNX$ is the free object in the variety of semirings, Birkhoff's HSP theorem implies that any assignment $\vct{X} \to \semN$, which includes as a special case the assignments $\assign_{\vct{W}}$ used here, uniquely extends to the semiring homomorphism alluded to above, $\assign_\vct{W}\inparen{\pxdb}\inparen{\tup}: \semNX \to \semN$. For a polynomial $\assign_\vct{W}\inparen{\pxdb}\inparen{\tup}$ substitutes variables based on $\vct{W}$ and then evaluates the resulting expression in $\semN$. For instance, consider the polynomial $\pxdb\inparen{\tup} = \poly = X + Y$ and assignment $\vct{W} := X = 0, Y=1$. We get $\assign_\vct{W}\inparen{\pxdb}\inparen{\tup} = 0 + 1 = 1$. 
Closure under $\raPlus$ queries follows from this and from \cite{DBLP:conf/pods/GreenKT07}'s Proposition 3.5, which states that semiring homomorphisms commute with queries over $\semK$-relations.
\qed
\end{proof}

\subsubsection{\tis and \bis in the \abbrNXPDB model}\label{subsec:supp-mat-ti-bi-def}

Two important subclasses of \abbrNXPDB\xplural that are of interest to us are the bag versions of tuple-independent databases (\tis) and block-independent databases (\bis). Under set semantics, a \ti is a deterministic database $\db$ where each tuple $\tup$ is assigned a probability $\prob_\tup$. The set of possible worlds represented by a \ti $\db$ is all subsets of $\db$. The probability of each world is the product of the probabilities of all tuples that exist with one minus the probability of all tuples of $\db$ that are not part of this world, i.e., tuples are treated  as independent  random events. In a \bi, we also  assign each tuple a  probability,  but  additionally partition  $\db$ into blocks. The possible worlds of a \bi $\db$ are all subsets  of $\db$ that contain at most one tuple  from each block.  Note then that the tuples sharing the same block are disjoint, and the sum of the probabilitites of all the tuples in the same block $\block$ is at most $1$.
The probability of such a world is the product of the probabilities of all tuples present in the world and the product of the probabilities that no tuple is present in each block $\block$ for which no tuple exists in that world. 
For bag \tis and \bis, we define the probability of a tuple to  be the probability that the tuple exists with multiplicity at least $1$.

In this work, we define \tis and \bis as subclasses of \abbrNXPDB\xplural defined over variables $\vct{X}$ (\Cref{def:semnx-pdbs}) where $\vct{X}$ can be partitioned into blocks that satisfy the conditions of a \ti or \bi (stated formally in \Cref{subsec:tidbs-and-bidbs}).
In this work, we consider one further deviation from the standard: We use bag semantics for queries.
Even though tuples cannot occur more than once in the input \ti or \bi, they can occur with a multiplicity larger than one in the result of a query.
Since in \tis and \bis, there is a one-to-one correspondence between tuples in the database and variables, we can interpret a vector $\vct{W} \in \{0,1\}^n$ as denoting which tuples exist in the possible world $\assign_{\vct{W}}(\pxdb)$ (the ones where $W_i = 1$).
For BIDBs specifically, note that at most one of the bits corresponding to tuples in each of the $\numblock$ blocks will be set (i.e., for any pair of bits $W_j$, $W_{j'}$ that are part of the same block $\block_i \supseteq \{t_{j}, t_{j'}\}$, at most one of them will be set).
Denote the vector $\vct{p}$ to be a vector whose elements are the individual probabilities $\prob_i$ of each tuple $\tup_i$.  Given \abbrPDB $\pdb$ where $\pd$ is the distribution induced by $\vct{p}$, which we will denote $\pd^{\inparen{\vct{\prob}}}$.
\begin{align}\label{eq:tidb-expectation}
\expct_{\vct{W} \sim \pd^{(\vct{p})}}\pbox{\poly(\vct{W})}
  = \sum\limits_{\substack{\vct{W} \in \{0, 1\}^\numvar\\ \suchthat W_j,W_{j'} = 1 \rightarrow \not \exists \block_i \supseteq \{t_{j}, t_{j'}\}}} \poly(\vct{W})\prod_{\substack{j \in [\numvar]\\ \suchthat W_j = 1}}\prob_j \prod_{\substack{i\in\pbox{\numblock}\suchthat\\ \forall \tup_j \in \block_i, W_j = 0}}
  \left(1 - \sum_{\tup_{j} \in \block_i}\prob_j\right)
\end{align}
Recall that tuple blocks in a TIDB always have size 1, so the outer summation of \cref{eq:tidb-expectation} is over the full set of vectors.

\subsection{Proof of~\Cref{prop:ctidb-reduct}}
	\begin{proof}[Proof of~\Cref{prop:ctidb-reduct}]
		We first need to prove that any \abbrCTIDB $\pdb$ can be reduced to the \abbrOneBIDB created by~\Cref{prop:ctidb-reduct}.  By definition, any $\tup\in\tupset$ can be present $c'\in\pbox{\bound}$ times in the possible worlds it appears in, with a disjoint probability distribution across the multiplicities $\pbox{\bound}$.  Then the construction of~\Cref{prop:ctidb-reduct} of the block of tuples $\inset{\intuple{\tup, j}_{j\in\pbox{\bound}}}$ for each $\tup\in\tupset$ indeed encodes the disjoint behavior across multiplicities.  $\pdb$ further requires that all $\tup\in\tupset$ are independent, a property which is enforced by independence constraint across all blocks of tuples in~\Cref{def:one-bidb}.  Then the construction of $\pdb'$ in~\Cref{prop:ctidb-reduct} is an equivalent representation of $\pdb$.
		
		Next we need to show that the distributions over $\pdb$ and $\pdb'$ are equivalent.  The distribution $\bpd$ is a distribution disjoint across the set of multiplicities $\pbox{\bound}$ and independent across all $\tup\in\tupset$.  By definition,~\Cref{prop:ctidb-reduct} creates $\pdb'$ by producing a block of disjoint tuples $\tup_j = \intuple{\tup, j}$ for each $\tup\in\tupset$ and $j\in\pbox{\bound}$, where $\vct{\prob} = \inparen{\prob_{\tup, j}}_{\tup\in\tupset, j\in\pbox{\bound}}$.  Since the probability vector $\vct{\prob_{\pdb}} = \inparen{\prob_{\tup, j}}_{\tup\in\tupset, j\in\pbox{\bound}}$ \emph{and} each $\prob_{\tup, j}, \prob_{\tup, j'}$ for $j\neq j'\in\pbox{\bound}$ are disjoint, the distributions are hence the same.
	\end{proof}

\subsection{Equating Expectations of Tuple Multiplicities and Polynomials} 
\label{subsec:expectation-of-polynom-proof}

\begin{Proposition}[Expectation of polynomials]\label{prop:expection-of-polynom}
Given a \abbrBPDB $\pdb = (\Omega,\bpd)$, $\raPlus$ query $\query$, and lineage polynomial $\apolyqdt$ for arbitrary result tuple $\tup$, 
we have (denoting $\randDB$ as the random variable over $\Omega$):
  $ \expct_{\randDB \sim \bpd}[\query(\randDB)(t)] = \expct_{\vct{\randWorld}\sim \pdassign}\pbox{\apolyqdt\inparen{\vct{\randWorld}}}. $
\end{Proposition}

Although \Cref{prop:expection-of-polynom} follows, e.g., as an obvious consequence of~\cite{IL84a}'s Theorem 7.1, we are unaware of any prior formal proof for bag-probabilistic databases.

\begin{proof}
We need to prove for $\semN$-PDB $\pdb = (\idb,\pd)$ and \abbrNXPDB $\pxdb = (\db_{\semNX},\pd')$ where $\rmod(\pxdb) = \pdb$ that $\expct_{\randDB\sim \pd}[\query(\db)(t)] = \expct_{\vct{W} \sim \pd'}\pbox{\nxpolyqdt(\vct{W})}$
By expanding $\nxpolyqdt$ and the expectation we have:
\begin{align*}
\expct_{\vct{W} \sim \pd'}\pbox{\poly(\vct{W})}
& = \sum_{\vct{W} \in \{0,1\}^n}\probOf(\vct{W}) \cdot Q(\db_{\semNX})(t)(\vct{W})\\
\intertext{From $\rmod(\pxdb) = \pdb$, we have that the range of $\assign_{\vct{W}(\pxdb)}$ is $\idb$, so}
& = \sum_{\db \in \idb}\;\;\sum_{\vct{W} \in \{0,1\}^n : \assign_{\vct{W}}(\pxdb) = \db}\probOf(\vct{W}) \cdot Q(\db_{\semNX})(t)(\vct{W})\\
\intertext{The inner sum is only over $\vct{W}$ where $\assign_{\vct{W}}(\pxdb) = \db$ (i.e., $Q(\db_{\semNX})(t)(\vct{W}) = \query(\db)(t))$}
& = \sum_{\db \in \idb}\;\;\sum_{\vct{W} \in \{0,1\}^n : \assign_{\vct{W}}(\pxdb) = \db}\probOf(\vct{W}) \cdot \query(\db)(t)\\
\intertext{By distributivity of $+$ over $\times$}
& = \sum_{\db \in \idb}\query(\db)(t)\sum_{\vct{W} \in \{0,1\}^n : \assign_{\vct{W}}(\pxdb) = \db}\probOf(\vct{W})\\
\intertext{From the definition of $\pd$ in \cref{def:semnx-pdbs}, given $\rmod(\pxdb) = \pdb$, we get}
& = \sum_{\db \in \idb}\query(\db)(t) \cdot \probOf(D) \quad = \expct_{\randDB \sim \pd}[\query(\db)(t)]
\end{align*}
\qed
\end{proof}

\subsection{Construction of the lineage (polynomial) for an $\raPlus$ query $\query$ over $\gentupset'$}
\label{fig:lin-poly-bidb}

The following rules are analogous to \Cref{fig:nxDBSemantics}, differing only in the base case. They are presented here for completeness.

\begin{align*}
\poly'\pbox{\project_A\inparen{\query}, \gentupset', \tup_j} =
    &~\sum_{\substack{\tup_{j'},\\\project_{A}\inparen{\tup_{j'}} = \tup_j}}\poly'\pbox{\query, \gentupset',        \tup_{j'}}\\
\poly'\pbox{\query_1\union\query_2, \gentupset', \tup_j} = 
    &\qquad\poly'\pbox{\query_1, \gentupset', \tup_j}+\poly'\pbox{\query_2, \gentupset', \tup_j}\\
\poly'\pbox{\select_\theta\inparen{\query}, \gentupset', \tup_j} =
    &~\begin{cases}\theta = 1 &\poly'\pbox{\query, \gentupset', \tup_j}\\\theta = 0& 0\\\end{cases} \\
\poly'\pbox{\query_1\join\query_2, \gentupset', \tup_j} = 
    &\qquad \poly'\pbox{\query_1, \gentupset', \project_{attr\inparen{\query_1}}\inparen{\tup_j}}\\             &\qquad\cdot\poly'\pbox{\query_2, \gentupset', \project_{attr\inparen{\query_2}}              \inparen{\tup_j}}\\
\poly'\pbox{\rel,\gentupset', \tup_j} =& j\cdot X_{\tup, j}.
\end{align*}\\

\subsection{Proposition~\ref{proposition:q-qtilde}}\label{app:subsec-prop-q-qtilde}
\noindent Note the following fact:

\begin{Proposition}\label{proposition:q-qtilde} For any \bi-lineage polynomial $\poly(X_1, \ldots, X_\numvar)$ and all $\vct{W}$ such that $\probOf\pbox{\vct{W}} > 0$, 
it holds that
$
    \poly(\vct{W}) = \rpoly(\vct{W}).
$
\end{Proposition}

\begin{proof}
Note that any $\poly$ in factorized form is equivalent to its \abbrSMB expansion.  For each term in the expanded form, further note that for all $b \in \{0, 1\}$ and all $e \geq 1$, $b^e = b$.  By definition (see~\Cref{def:reduced-poly-one-bidb}), $\rpoly\inparen{\vct{X}}$ is the \abbrSMB expansion of $\poly\inparen{\vct{X}}$ followed by reducing every exponent $e > 1$ to $1$ and eliminating all cross terms for the \abbrBIDB case.  Note that it must be that no cross terms exist in $\poly\inparen{\vct{X}}$, since by the proposition statement, $\probOf\pbox{\vct{W}} > 0$.  Thus, since all monomials are indeed the same, it follows that $\poly\inparen{\vct{W}} = \rpoly\inparen{\vct{W}}$.
\qed
\end{proof}

\subsection{Proof for~\Cref{lem:bin-bidb-phi-eq-redphi}}\label{subsec:proof-exp-poly-rpoly}
\begin{proof}
Let $\poly$ be a polynomial of $\numvar$ variables with highest degree $= \hideg$, defined as follows: 
\[\poly(X_1,\ldots, X_\numvar) = \sum_{\vct{d} \in \{0,\ldots, \hideg\}^\numvar}c_{\vct{d}}\cdot \prod_{i = 1}^\numvar X_i^{d_i},\]

where $\tupset'$ has $\numvar$ tuples, we can equivalently write $\prod\limits_{\tup\in\tupset'}X_\tup^{d_\tup}$ for the product term.

Let the boolean function $\isInd{\cdot}$ take $\vct{d}$ as input and return true if there does not exist any dependent variables in the monomial encoded by $\vct{d}$, i.e., for any block $\block\in\tupset'$, $\not\exists \tup, \tup' \in\block~|~\vct{d}_\tup, \vct{d}_{\tup'}\geq 1$.


Then, given \abbrOneBIDB $\pdb$, query $\query$, and polynomial $\poly\inparen{\vct{W}} = \poly\pbox{\query, \tupset, \tup}$, in expectation we have

\begin{align}
\expct_{\vct{\randWorld}}\pbox{\poly(\vct{\randWorld})} &= \expct_{\vct{\randWorld}}\pbox{\sum_{\substack{\vct{d} \in \{0,\ldots,\hideg\}^\numvar\\\wedge~\isInd{\vct{d}}}}c_{\vct{d}}\cdot \prod_{\tup\in\tupset'} \randWorld_i^{d_i} + \sum_{\substack{\vct{d} \in \{0,\ldots, \hideg\}^\numvar\\\wedge ~\neg\isInd{\vct{d}}}} c_{\vct{d}}\cdot\prod_{\tup\in\tupset'}\randWorld_i^{d_i}}\label{p1-s1a}\\
&= \sum_{\substack{\vct{d} \in \{0,\ldots,\hideg\}^\numvar\\\wedge~\isInd{\vct{d}}}}c_{\vct{d}}\cdot \expct_{\vct{\randWorld}}\pbox{\prod_{\tup\in\tupset'} \randWorld_i^{d_i}} + \sum_{\substack{\vct{d} \in \{0,\ldots, \hideg\}^\numvar\\\wedge ~\neg\isInd{\vct{d}}}} c_{\vct{d}}\cdot\expct_{\vct{\randWorld}}\pbox{\prod_{\tup\in\tupset'}\randWorld_i^{d_i}}\label{p1-s1b}\\
&= \sum_{\substack{\vct{d} \in \{0,\ldots,\hideg\}^\numvar\\~\wedge\isInd{\vct{d}}}}c_{\vct{d}}\cdot \expct_{\vct{\randWorld}}\pbox{\prod_{\tup\in\tupset'} \randWorld_i^{d_i}}\label{p1-s1c}\\
&= \sum_{\substack{\vct{d} \in \{0,\ldots,\hideg\}^\numvar\\\wedge~\isInd{\vct{d}}}}c_{\vct{d}}\cdot \prod_{\tup\in\tupset'} \expct_{\vct{\randWorld}}\pbox{\randWorld_i^{d_i}}\label{p1-s2}\\
&= \sum_{\substack{\vct{d} \in \{0,\ldots,\hideg\}^\numvar\\\wedge~\isInd{\vct{d}}}}c_{\vct{d}}\cdot \prod_{\tup\in\tupset'} \expct_{\vct{\randWorld}}\pbox{\randWorld_i}\label{p1-s3}\\
&= \sum_{\substack{\vct{d} \in \{0,\ldots,\hideg\}^\numvar\\\wedge~\isInd{\vct{d}}}}c_{\vct{d}}\cdot \prod_{\tup\in\tupset'} \prob_i\label{p1-s4}\\
&= \rpoly(\prob_1,\ldots, \prob_\numvar).\label{p1-s5}
\end{align}
\Cref{p1-s1a} is the result of substituting in the definition of $\poly$ given above.  Then we arrive at \cref{p1-s1b} by linearity of expectation.  Next, \cref{p1-s1c} is the result of the independence constraint of \abbrBIDB\xplural, specifically that any monomial composed of dependent variables, i.e., variables from the same block $\block$, has a probability of $0$.  \Cref{p1-s2} is obtained by the fact that all variables in each monomial are independent, which allows for the expectation to be pushed through the product.  In \cref{p1-s3}, recall the lineage construction semantics of~\Cref{fig:lin-poly-bidb} for a \abbrOneBIDB, where the annotation of a tuple $\tup$ with multiplicity $j$ is written as $j^{d_\tup}\cdot X_{\tup}^{d_\tup}$ such that $\domain\inparen{\vct{W}_\tup} = \inset{0, 1}$.  Then $c_{\vct{d}}$ absorbs all such $j^{d_\tup}$ factors.  Since $\randWorld_\tup \in \{0, 1\}$ it is the case that for any exponent $e \geq 1$, $\vct{W}_\tup^e = \vct{W}_\tup$.  Next, in \cref{p1-s4} the expectation of a tuple is indeed its probability.

Finally, it can be verified that \Cref{p1-s5} follows since \cref{p1-s4} satisfies the construction of $\rpoly(\prob_1,\ldots, \prob_\numvar)$ in \Cref{def:reduced-poly}.
\qed
\end{proof}

\subsection{Proof For Corollary~\ref{cor:expct-sop}}
\begin{proof}
Note that~\Cref{lem:tidb-reduce-poly} shows that $\expct\pbox{\poly} =$ $\rpoly(\prob_1,\ldots, \prob_\numvar)$.  Therefore, if $\poly$ is already in \abbrSMB form, one only needs to compute $\poly(\prob_1,\ldots, \prob_\numvar)$ ignoring exponent terms (note that such a polynomial is $\rpoly(\prob_1,\ldots, \prob_\numvar)$), which indeed has $\bigO{\abs{\poly}}$ computations.
\qed
\end{proof}

\subsection{Definition of $\polyf(\cdot)$}
\label{def:poly-func}

We formally define the relationship of circuits with polynomials.  While the definition assumes one sink for notational convenience, it easily generalizes to the multiple sinks case.

$\polyf(\circuit)$ maps the sink of circuit $\circuit$ to its corresponding polynomial in \abbrSMB.  $\polyf(\cdot)$ is recursively defined on $\circuit$ as follows, with addition and multiplication following the standard interpretation for polynomials:
\begin{equation*}
  \polyf(\circuit) = \begin{cases}
          \polyf(\circuit_\lchild) + \polyf(\circuit_\rchild)     &\text{ if \circuit.\type } = \circplus\\
          \polyf(\circuit_\lchild) \cdot \polyf(\circuit_\rchild)   &\text{ if \circuit.\type } = \circmult\\
          \circuit.\val                 &\text{ if \circuit.\type } = \var \text{ OR } \tnum.
        \end{cases}
\end{equation*}

$\circuit$ need not encode $\poly\inparen{\vct{X}}$ in the same, default \abbrSMB representation.  For instance, $\circuit$ could encode the factorized representation $(X + 2Y)(2X - Y)$ of $\poly\inparen{\vct{X}} = 2X^2+3XY-2Y^2$, as shown in \Cref{fig:circuit}, while $\polyf(\circuit) = \poly\inparen{\vct{X}}$ is always the equivalent \abbrSMB representation.


\section{Missing details from Section~\ref{sec:hard}}
\label{app:single-mult-p}

\subsection{\Cref{lem:pdb-for-def-qk}}
\begin{Lemma}\label{lem:pdb-for-def-qk}
Assuming that each $v \in \vset$ has degree $\geq 1$,\footnote{This is WLOG, since any vertex with degree $0$ can be dropped without affecting the result of our hard query.} the \abbrPDB relations encoding the edges for $\poly_G^\kElem$ of \Cref{def:qk} can be computed in $\bigO{\numedge}$ time.
\end{Lemma}
\begin{proof}[Proof of \Cref{lem:pdb-for-def-qk}]
Only two relations need be constructed, one for the set $\vset$ and one for the set $\edgeSet$.  By a simple linear scan, each can be constructed in time $\bigO{\numedge + \numvar}$.  Given that the degree of each $v \in \vset$ is at least $1$, we have that $m\ge \Omega(n)$,
and this yields the claimed runtime.
\qed
\end{proof}

\subsection{Proof of \Cref{lem:tdet-om}}
\begin{proof}
By the recursive defintion of $\qruntimenoopt{\cdot, \cdot}$ (see \Cref{sec:gen}), we have the following equation for our hard query $\query$ when $k = 1$, (we denote this as $\query^1$).
\begin{equation*}
\qruntimenoopt{\query^1, \tupset} = \abs{\tupset.\vset} + \abs{\tupset.\edgeSet} + \abs{\tupset.\vset} + \jointime{\tupset.\vset , \tupset.\edgeSet , \tupset.\vset}.
\end{equation*}
We argue that $\jointime{\tupset.\vset , \tupset.\edgeSet , \tupset.\vset}$ is at most $O(\numedge)$ by noting that there exists an algorithm that computes $\tupset.\vset\join\tupset.\edgeSet\join\tupset.\vset$ in the same runtime\footnote{Indeed the trivial algorithm that computes the obvious pair-wise joins  has the claimed runtime. That is, we first compute $\tupset.\vset\join\tupset.\edgeSet$, which takes $O(m)$ (assuming $\tupset.\vset$ is stored in hash map) since tuples in $\tupset.\vset$ can only filter tuples in $\tupset.\edgeSet$. The resulting subset of tuples in $\tupset.\edgeSet$ are then again joined (on the right) with $\tupset.\vset$, which by the same argument as before also takes $O(m)$ time, as desried.}.  Then by the assumption of \Cref{lem:pdb-for-def-qk} (each $v \in \vset$ has degree $\geq 1$), the sum of the first three terms is $\bigO{\numedge}$.  We then obtain that $\qruntimenoopt{\query^1, \tupset} = \bigO{\numedge} + \bigO{\numedge} = \bigO{\numedge}$.  For $\query^k = \query_1^1 \times\cdots\times\query_k^1$, we have the recurrence $\qruntimenoopt{\query^k, \tupset} = \qruntimenoopt{\query_1^1, \tupset} + \cdots +\qruntimenoopt{\query_k^1, \tupset} + \jointime{\query_1^1,\cdots,\query_k^1}$.  Since $\query^1$ outputs a count, computing the join $\query_1^1\join\cdots\join\query_k^1$ is just multiplying $k$ numbers, which takes $O(k)$ time. Thus, we have
\[\qruntimenoopt{\query^k, \tupset} \le k\cdot O(m)+O(k)\le O(km),\]
as desired.
\qed
\end{proof}
\subsection{\Cref{lem:qEk-multi-p}}
\noindent The following lemma reduces the problem of counting $\kElem$-matchings in a graph to our problem (and proves \Cref{thm:mult-p-hard-result}):
\begin{Lemma}\label{lem:qEk-multi-p}
Let $\prob_0,\ldots, \prob_{2\kElem}$ be distinct values in $(0, 1]$.  Then given the values $\rpoly_{G}^\kElem(\prob_i,\ldots, \prob_i)$ for $0\leq i\leq 2\kElem$, the number of $\kElem$-matchings in $G$ can be computed in $\bigO{\kElem^3}$ time.
\end{Lemma}

\subsection{Proof of Lemma~\ref{lem:qEk-multi-p}}\label{subsec:c2k-proportional}

\begin{proof}
We first argue that $\rpoly_{G}^\kElem(\prob,\ldots, \prob) = \sum\limits_{i = 0}^{2\kElem} c_i \cdot \prob^i$.  First, since $\poly_G(\vct{X})$ has degree $2$, it follows that $\poly_G^\kElem(\vct{X})$ has degree $2\kElem$.  By definition, $\rpoly_{G}^{\kElem}(\vct{X})$ sets every exponent $e > 1$ to $e = 1$, which means that $\degree(\rpoly_{G}^\kElem)\le \degree(\poly_G^\kElem)= 2k$. Thus, if we think of $\prob$ as a variable, then $\rpoly_{G}^{\kElem}(\prob,\dots,\prob)$ is a univariate polynomial of degree at most $\degree(\rpoly_{G}^\kElem)\le 2k$. Thus, we can write
\begin{equation*}
\rpoly_{G}^{\kElem}(\prob,\ldots, \prob) = \sum_{i = 0}^{2\kElem} c_i \prob^i
\end{equation*}
We note that $c_i$ is {\em exactly} the number of monomials in the SMB expansion of $\poly_{G}^{\kElem}(\vct{X})$ composed of $i$ distinct variables.\footnote{Since $\rpoly_G^\kElem(\vct{X})$ does not have any monomial with degree $< 2$, it is the case that $c_0 = c_1 = 0$ but for the sake of simplcity we will ignore this observation.}

Given that we then have $2\kElem + 1$ distinct values of $\rpoly_{G}^\kElem(\prob,\ldots, \prob)$ for $0\leq i\leq2\kElem$, it follows that we have a linear system of the form $\vct{M} \cdot \vct{c} = \vct{b}$ where the $i$th row of $\vct{M}$ is $\inparen{\prob_i^0\ldots\prob_i^{2\kElem}}$, $\vct{c}$ is the coefficient vector $\inparen{c_0,\ldots, c_{2\kElem}}$, and $\vct{b}$ is the vector such that $\vct{b}[i] = \rpoly_{G}^\kElem(\prob_i,\ldots, \prob_i)$.  In other words, matrix $\vct{M}$ is the Vandermonde matrix, from which it follows that we have a matrix with full rank (the $p_i$'s are distinct), and we can solve the linear system in $O(k^3)$ time (e.g., using Gaussian Elimination) to determine $\vct{c}$ exactly. 
 Thus, after $O(k^3)$ work, we know $\vct{c}$ and in particular, $c_{2k}$ exactly. 
 
Next, we show why we can compute $\numocc{G}{\kmatch}$ from $c_{2k}$ in $O(1)$ additional time.
We claim that $c_{2\kElem}$ is $\kElem! \cdot \numocc{G}{\kmatch}$.  This can be seen intuitively by looking at the expansion of the original factorized representation 
\[\poly_{G}^\kElem(\vct{X}) = \sum_{\substack{(i_1, j_1),\cdots,(i_\kElem, j_\kElem) \in E}}X_{i_1}X_{j_1}\cdots X_{i_\kElem}X_{j_\kElem},\] 
where a unique $\kElem$-matching in the multi-set of product terms has $\kElem$ distinct $\inparen{i_\ell, j_\ell}$ index pairs.  Further, any monomial composed of such a distinct set of $2\kElem$ variables will be produced $\kElem!$ times in a $\kElem$-wise product of the sum of a set of elements with itself.  This is true because each (identical) product term contains each of the $\kElem$ distinct elements, giving us $\kElem\cdot\kElem-1\cdots 1 = \kElem!$ permutations of a distinct $\kElem$-matching.  
  
Since, as noted earlier, $c_{2\kElem}$ represents the number of monomials with $2\kElem$ distinct variables, then it must be that $c_{2\kElem}$ is the overall number of $\kElem$-matchings.  And since we have $\kElem!$ copies of each distinct $\kElem$-matching, it follows that
$c_{2\kElem}= \kElem! \cdot \numocc{G}{\kmatch}$.
Thus, simply dividing $c_{2\kElem}$ by $\kElem!$ gives us $\numocc{G}{\kmatch}$, as needed. \qed 
\end{proof}

\subsection{Proof of Theorem~\ref{thm:mult-p-hard-result}}
\begin{proof}
For the sake of contradiction, assume we can solve our problem in $\littleo{\kmatchtime}$ time. Given a graph $G$ by \Cref{lem:pdb-for-def-qk} we can compute the \abbrPDB encoding in $\bigO{\numedge}$ time. Then after we run our algorithm on $\rpoly_G^\kElem$, we get $\rpoly_{G}^\kElem(\prob_i,\ldots, \prob_i)$ for every $0\leq i\leq 2\kElem$ in additional $\bigO{k}\cdot \littleo{\kmatchtime}$ time. \Cref{lem:qEk-multi-p} then computes the number of $k$-matchings in $G$ in $O(\kElem^3)$ time.  Adding the runtime of all of these steps, we have an algorithm for computing the number of $k$-matchings that runs in time
\begin{align}
 &\bigO{\numedge} + \bigO{k}\cdot \littleo{\kmatchtime} + O(\kElem^3)\label{eq:proof-omega-kmatch2}\\
&\le \littleo{\kmatchtime}\label{eq:proof-omega-kmatch4}.
\end{align}
We obtain \Cref{eq:proof-omega-kmatch4} from the facts that $k$ is fixed (related to $m$) and the assumption that $\kmatchtime\ge\omega(m)$.
Thus we obtain the contradiction that we can achieve a runtime $\littleo{\kmatchtime}$ that is better than the optimal time $\kmatchtime$ required to compute $k$-matchings.
\qed
\end{proof}

\subsection{Subgraph Notation and $O(1)$ Closed Formulas}

We need all the possible edge patterns in an arbitrary $G$ with at most three distinct edges. We have already seen $\tri,\threepath$ and $\threedis$, so we define the remaining patterns:

\begin{itemize}
	\item Single Edge $\left(\ed\right)$
	\item 2-path ($\twopath$)
	\item 2-matching ($\twodis$)
	\item 3-star ($\oneint$)--this is the graph that results when all three edges share exactly one common endpoint.  The remaining endpoint for each edge is disconnected from any endpoint of the remaining two edges.
	\item Disjoint Two-Path ($\twopathdis$)--this subgraph consists of a two-path and a remaining disjoint edge.
\end{itemize}

For any graph $G$, the following formulas for $\numocc{G}{H}$ compute their respective patterns exactly  in $O(\numedge)$ time, with $d_i$ representing the degree of vertex $i$ (proofs are in \Cref{app:easy-counts}):
\begin{align}
	&\numocc{G}{\ed} = \numedge, \label{eq:1e}\\
	&\numocc{G}{\twopath} = \sum_{i \in V} \binom{d_i}{2} \label{eq:2p}\\						
	&\numocc{G}{\twodis} = \sum_{(i, j) \in E} \frac{\numedge - d_i - d_j + 1}{2}\label{eq:2m}\\
	&\numocc{G}{\oneint} = \sum_{i \in V} \binom{d_i}{3}\label{eq:3s}\\
	&\numocc{G}{\twopathdis} + 3\numocc{G}{\threedis} = \sum_{(i, j) \in E} \binom{\numedge - d_i - d_j + 1}{2}\label{eq:2pd-3d}\\
	&\numocc{G}{\threepath} + 3\numocc{G}{\tri} = \sum_{(i, j) \in E} (d_i - 1) \cdot (d_j - 1)\label{eq:3p-3tri}
\end{align}
\subsection{Proofs of \Cref{eq:1e}-\Cref{eq:3p-3tri}}
\label{app:easy-counts}

The proofs for \Cref{eq:1e}, \Cref{eq:2p} and \Cref{eq:3s} are immediate.

\begin{proof}[Proof of \Cref{eq:2m}]
For edge $(i, j)$ connecting arbitrary vertices $i$ and $j$, finding all other edges in $G$ disjoint to $(i, j)$ is equivalent to finding all edges that are not connected to either vertex $i$ or $j$.  The number of such edges is $m - d_i - d_j + 1$, where we add $1$ since edge $(i, j)$ is removed twice when subtracting both $d_i$ and $d_j$.  Since the summation is iterating over all edges such that a pair $\left((i, j), (k, \ell)\right)$ will also be counted as $\left((k, \ell), (i, j)\right)$, division by $2$ then eliminates this double counting.  Note that $m$ and $d_i$ for all $i \in V$ can be computed in one pass over the set of edges by simply maintaining counts for each quantity.  Finally, the summation is also one traversal through the set of edges where each operation is either a lookup ($O(1)$ time) or an addition operation (also $O(1)$) time.
\qed
\end{proof}

\begin{proof}[Proof of \Cref{eq:2pd-3d}]
\Cref{eq:2pd-3d} is true for similar reasons.  For edge $(i, j)$, it is necessary to find two additional edges, disjoint or connected.  As in our argument for \Cref{eq:2m}, once the number of edges disjoint to $(i, j)$ have been computed, then we only need to consider all possible combinations of two edges from the set of disjoint edges, since it doesn't matter if the two edges are connected or not.    Note, the factor $3$ of $\threedis$ is necessary to account for the triple counting of $3$-matchings, since it is indistinguishable to the closed form expression which of the remaining edges are either disjoint or connected to each of the edges in the {\emph{initial}} set of edges disjoint to the edge under consideration.  Observe that the disjoint case will be counted $3$ times since each edge of a $3$-path is visited once, and the same $3$-path counted in each visitation.  For the latter case however, it is true that since the two path in $\twopathdis$ is connected, there will be no multiple counting by the fact that the summation automatically disconnects the current edge, meaning that a two matching at the current vertex under consideration will not be counted.  Thus, $\twopathdis$ will only be counted once, precisely when the single disjoint edge is visited in the summation.  The sum over all such edge combinations is precisely then $\numocc{G}{\twopathdis} + 3\numocc{G}{\threedis}$.  Note that all factorials can be computed in $O(m)$ time, and then the remaining operations in computing each combination $\binom{n}{2}$ are constant, thus yielding the claimed $O(m)$ run time.
\qed
\end{proof}
\begin{proof}[Proof of \Cref{eq:3p-3tri}]
To  compute $\numocc{G}{\threepath}$, note that for an arbitrary edge $(i, j)$, a 3-path exists for edge pair $(i, \ell)$ and $(j, k)$ where $i, j, k, \ell$ are distinct.  Further, the quantity $(d_i - 1) \cdot (d_j - 1)$ represents the number of 3-edge subgraphs with middle edge $(i, j)$ and outer edges $(i, \ell), (j, k)$ such that $\ell \neq j$ and $k \neq i$.  When $k = \ell$, the resulting subgraph is a triangle, and when $k \neq \ell$, the subgraph is a 3-path.  Summing over all edges (i, j) gives \Cref{eq:3p-3tri} by observing that each triangle is counted thrice, while each 3-path is counted just once.  For reasons similar to \Cref{eq:2m}, all $d_i$ can be computed in $O(m)$ time and each summand can then be computed in $O(1)$ time, yielding an overall $O(m)$ run time.
\qed
\end{proof}

\subsection{Tools to prove \Cref{th:single-p-hard}}

Note that $\rpoly_{G}^3(\prob,\ldots, \prob)$ as a polynomial in $\prob$ has degree at most six.  \Cref{lem:qE3-exp} shows the exact coefficients of $\rpoly^3\inparen{\prob,\ldots,\prob}$, which we prove next.
\subsubsection{Proof for \Cref{lem:qE3-exp}}
\begin{proof}
By definition we have that
		\[\poly_{G}^3(\vct{X}) = \sum_{\substack{(i_1, j_1), (i_2, j_2), (i_3, j_3) \in E}}~\; \prod_{\ell = 1}^{3}X_{i_\ell}X_{j_\ell}.\]
Hence $\poly_{G}^3(\vct{X})$ has degree six. Note that the monomial $\prod_{\ell = 1}^{3}X_{i_\ell}X_{j_\ell}$ will contribute to the coefficient of $\prob^\nu$ in $\rpoly_{G}^3(\vct{X})$, where $\nu$ is the number of distinct variables in the monomial.
Let $e_1 = (i_1, j_1), e_2 = (i_2, j_2),$ and $e_3 = (i_3, j_3)$.  
We compute $\rpoly_{G}^3(\vct{X})$ by considering each of the three forms that the triple $(e_1, e_2, e_3)$ can take.

\textsc{case 1:} $e_1 = e_2 = e_3$ (all edges are the same).  When we have that $e_1 = e_2 = e_3$, then the monomial corresponds to $\numocc{G}{\ed}$. There are exactly $\numedge$ such triples, each with a $\prob^2$ factor in $\rpoly_{G}^3\left(\prob,\ldots, \prob\right)$.

\textsc{case 2:}  This case occurs when there are two distinct edges of the three, call them $e$ and $e'$.  When there are two distinct edges, there is then the occurence when $2$ variables in the triple $(e_1, e_2, e_3)$ are bound to $e$.  There are three combinations for this occurrence in $\poly_{G}^3(\vct{X})$.  Analogusly, there are three such occurrences in $\poly_{G}^3(\vct{X})$ when there is only one occurrence of $e$, i.e. $2$ of the variables in $(e_1, e_2, e_3)$ are $e'$.   
This implies that all $3 + 3 = 6$ combinations of two distinct edges $e$ and $e'$ contribute to the same monomial in $\rpoly_{G}^3$. 
Since $e\ne e'$, this case produces the following edge patterns: $\twopath, \twodis$, which contribute $6\prob^3$ and $6\prob^4$ respectively to  $\rpoly_{G}^3\left(\prob,\ldots, \prob\right)$.

\textsc{case 3:} All $e_1,e_2$ and $e_3$ are distinct.  For this case, we have $3! = 6$ permutations of $(e_1, e_2, e_3)$, each of which contribute to the same monomial in the \textsc{SMB} representation of $\poly_{G}^3(\vct{X})$.  This case consists of the following edge patterns: $\tri, \oneint, \threepath, \twopathdis, \threedis$, which contribute $6\prob^3, 6\prob^4, 6\prob^4, 6\prob^5$ and $6\prob^6$ respectively to  $\rpoly_{G}^3\left(\prob,\ldots, \prob\right)$.
\qed
\end{proof}


Since $\prob$ is fixed, \Cref{lem:qE3-exp} gives us one linear equation in $\numocc{G}{\tri}$ and $\numocc{G}{\threedis}$ (we can handle the other counts due to equations (\ref{eq:1e})-(\ref{eq:3p-3tri})). However, we need to generate one more independent linear equation in these two variables. Towards this end we generate another graph related to $G$:
\begin{Definition}\label{def:Gk}
For $\ell \geq 1$, let graph $\graph{\ell}$ be a graph generated from an arbitrary graph $G$, by replacing every edge $e$ of $G$ with an $\ell$-path, such that all inner vertices of an $\ell$-path replacement edge have degree $2$.\footnote{Note that $G\equiv \graph{1}$.}
\end{Definition}
We will prove \Cref{th:single-p-hard} by the following reduction:
\begin{Theorem}\label{th:single-p}
Fix $\prob\in (0,1)$. Let $G$ be a graph on $\numedge$ edges.
If we can compute $\rpoly_{G}^3(\prob,\dots,\prob)$ exactly in $T(\numedge)$ time, then we can exactly compute $\numocc{G}{\tri}$ 
in $O\inparen{T(\numedge) + \numedge}$ time.
\end{Theorem}
For clarity, we repeat the notion of $\numocc{G}{H}$ to mean the count of subgraphs in $G$ isomorphic to $H$.
The following lemmas relate these counts in $\graph{2}$ to $\graph{1}$ and are useful in proving \Cref{lem:lin-sys}.

\begin{Lemma}\label{lem:3m-G2}
The $3$-matchings in graph $\graph{2}$ satisfy the identity:
\begin{align*}
\numocc{\graph{2}}{\threedis} &= 8 \cdot \numocc{\graph{1}}{\threedis} + 6 \cdot \numocc{\graph{1}}{\twopathdis}\\
&+ 4 \cdot \numocc{\graph{1}}{\oneint} + 4 \cdot \numocc{\graph{1}}{\threepath} + 2 \cdot \numocc{\graph{1}}{\tri}.
\end{align*}
\end{Lemma}

\begin{Lemma}\label{lem:tri}
For $\ell > 1$ and any graph $\graph{\ell}$, $\numocc{\graph{\ell}}{\tri} = 0$.
\end{Lemma}


\subsection{Proofs for \Cref{lem:3m-G2}, \Cref{lem:tri}, and \Cref{lem:lin-sys}}\label{subsec:proofs-struc-lemmas}
Before proceeding, let us introduce a few more helpful definitions.

\begin{Definition}[$\esetType{\ell}$]\label{def:ed-nota}
For $\ell > 1$, we use $\esetType{\ell}$ to denote the set of edges in $\graph{\ell}$.  For any graph $\graph{\ell}$, its edges are denoted by the a pair $(e, b)$, such that $b \in \{0,\ldots, \ell-1\}$ where $(e,0),\dots,(e,\ell-1)$ is the $\ell$-path that replaces the edge $e$ for $e\in \esetType{1}$.
\end{Definition}

\begin{Definition}[$\eset{\ell}$]
Given an arbitrary subgraph $\sg{1}$ of $\graph{1}$, let $\eset{1}$ denote the set of edges in $\sg{1}$.  Define then $\eset{\ell}$ for $\ell > 1$ as the set of edges in the generated subgraph $\sg{\ell}$ (i.e. when we apply \Cref{def:Gk} to $S$ to generate $\sg{\ell}$).
\end{Definition}

For example, consider $\sg{1}$ with edges $\eset{1} = \{e_1\}$.  Then the edge set of $\sg{2}$ is defined as $\eset{2} = \{(e_1, 0), (e_1, 1)\}$.
\begin{Definition}[$\binom{\edgeSet}{t}$ and $\binom{\edgeSet}{\leq t}$]\label{def:ed-sub}
Let $\binom{E}{t}$ denote the set of subsets in $E$ with exactly $t$ edges.  In a similar manner, $\binom{E}{\leq t}$ is used to mean the subsets of $E$ with $t$ or fewer edges.
\end{Definition}

The following function $f_\ell$  is a mapping from every $3$-edge shape in $\graph{\ell}$ to its `projection' in $\graph{1}$.
\begin{Definition}\label{def:fk}
Let $f_\ell: \binom{\esetType{\ell}}{3} \rightarrow \binom{\esetType{1}}{\leq3}$ be defined as follows.  For any element $s \in \binom{\esetType{\ell}}{3}$ such that $s = \pbrace{(e_1, b_1), (e_2, b_2), (e_3, b_3)}$, define:
\[ f_\ell\left(\pbrace{(e_1, b_1), (e_2, b_2), (e_3, b_3)}\right) = \pbrace{e_1, e_2, e_3}.\]
\end{Definition}

\begin{Definition}[$f_\ell^{-1}$]\label{def:fk-inv}
For an arbitrary subgraph $\sg{1}$ of $\graph{1}$ with at most $m \leq 3$ edges, the inverse function $f_\ell^{-1}: \binom{\esetType{1}}{\leq 3}\rightarrow 2^{\binom{\esetType{\ell}}{3}}$ takes $\eset{1}$ and outputs the set of all elements $s \in \binom{\eset{\ell}}{3}$ such that 
$f_\ell(s) = \eset{1}$.  
\end{Definition}

Note, importantly, that when we discuss $f_\ell^{-1}$, that each \textit{edge} present in $\eset{1}$ must have an edge in each $s\in f_\ell^{-1}(\eset{1})$ that projects down to it.  In particular, if $|\eset{1}| = 3$, then it must be the case that each $s\in f_\ell^{-1}(\eset{1})$ consists of the following set of edges: $\{ (e_i, b_1), (e_j, b_2), (e_m, b_3) \}$, where $i,j$ and $m$ are distinct.  

We are now ready to prove the structural lemmas. 
To prove the structural lemmas, we will 
count the number of occurrences of $\tri$ and $\threedis$ in $\graph{\ell}$.  This is accomplished by computing how many  $\threedis$ and $\tri$ subgraphs appear in $f_\ell^{-1}(\eset{1})$ for each $S\in\binom{E_1}{\le 3}$.

\subsubsection{Proof of Lemma \ref{lem:3m-G2}}
\begin{proof}
For each subset  $\eset{1}\in \binom{E_1}{\le 3}$, we count the number of {\emph{$3$-matchings }}in the $3$-edge subgraphs of $\graph{2}$ in $f_2^{-1}(\eset{1})$.  Denote the set of $3$-matchings in $f_2^{-1}(\eset{1})$ as $f_{2,\threedis}^{-1}\inparen{\eset{1}}$ and its size as $\abs{f_{2, \threedis}^{-1}\inparen{\eset{1}}}$.  We first consider the case of $\eset{1} \in \binom{E_1}{3}$, where $\eset{1}$ is composed of the edges $e_1, e_2, e_3$ and $f_2^{-1}(\eset{1})$ is the set $s$ of all $3$-edge subsets of edges $e_i \in \{(e_1, 0), (e_1, 1), (e_2, 0), (e_2, 1),$ $(e_3, 0), (e_3, 1)\}$ such that $f_\ell(s) = \{e_1, e_2, e_3\}$.  For the case that $f_2^{-1}\inparen{\eset{1}} = f_{2, \threedis}^{-1}\inparen{\eset{1}}$ we have that $\abs{f_{2, \threedis}^{-1}\inparen{\eset{1}}} = 8$.

We do a case analysis based on the subgraph $\sg{1}$ induced by $\eset{1}$. 
\begin{itemize}
	\item $3$-matching ($\threedis$)
\end{itemize}
When $\sg{1}$ is isomorphic to $\threedis$, it is the case that edges in $\eset{2}$ are {\em not} disjoint only for the pairs $(e_i, 0), (e_i, 1)$ for $i\in \{1,2,3\}$.  By definition, each set of edges in $f_2^{-1}\inparen{\eset{1}}$ is a three matching and $\abs{f_{2, \threedis}^{-1}\inparen{\eset{1}}} = 8$.

\begin{itemize}
	\item Disjoint Two-Path ($\twopathdis$)
\end{itemize}
For $\sg{1}$ isomorphic to $\twopathdis$ edges $e_2, e_3$ form a $2$-path with $e_1$ being disjoint.  This means that in $\sg{2}$ edges $(e_2, 0), (e_2, 1), (e_3, 0), (e_3, 1)$ form a $4$-path while $(e_1, 0), (e_1, 1)$ is its own disjoint $2$-path.  We can pick either $(e_1, 0)$ or $(e_1, 1)$ for the first edge in the $3$-matching, while it is necessary to have a $2$-matching from path $(e_2, 0),\ldots(e_3, 1)$.  Note that the $4$-path allows for three possible $2$-matchings, specifically,
\begin{equation*}
\pbrace{(e_2, 0), (e_3, 0)}, \pbrace{(e_2, 0), (e_3, 1)}, \pbrace{(e_2, 1), (e_3, 1)}.
\end{equation*}

Since these two selections can be made independently, $\abs{f_{2, \threedis}^{-1}\inparen{\eset{1}}} = 2 \cdot 3 = 6$.

\begin{itemize}
	\item $3$-star ($\oneint$)
\end{itemize}
When $\sg{1}$ is isomorphic to $\oneint$, the inner edges $(e_i, 1)$ of $\sg{2}$ are all connected, and the outer edges $(e_i, 0)$ are all disjoint.  Note that for a valid $3$-matching it must be the case that at most one inner edge can be part of the set of disjoint edges.  For the case of when exactly one inner edge is chosen, there exist $3$ possiblities, based on which inner edge is chosen.  Note that if $(e_i, 1)$ is chosen, the matching has to choose $(e_j, 0)$ for $j \neq i$ and $(e_{j'}, 0)$ for $j' \neq i, j' \neq j$.  The remaining possible 3-matching occurs when all 3 outer edges are chosen, and $\abs{f_{2, \threedis}^{-1}\inparen{\eset{1}}} = 4$.  

\begin{itemize}
	\item $3$-path ($\threepath$)
\end{itemize}
When $\sg{1}$ is isomorphic to $\threepath$ it is the case that all edges beginning with $e_1$ and ending with $e_3$ are successively connected.  This means that the edges of $\eset{2}$ form a $6$-path.  For a $3$-matching to exist in $f_2^{-1}(\eset{1})$, we cannot pick both $(e_i,0)$ and $(e_i,1)$ or both $(e_i, 1)$ and $(e_j, 0)$ where $j = i + 1$. 
 There are four such possibilities: $\pbrace{(e_1, 0), (e_2, 0), (e_3, 0)}$, $\pbrace{(e_1, 0), (e_2, 0), (e_3, 1)}$, $\pbrace{(e_1, 0), (e_2, 1), (e_3, 1)},$ $\pbrace{(e_1, 1), (e_2, 1),  (e_3, 1)}$ and $\abs{f^{-1}_{2, \threedis}\inparen{\eset{1}}} = 4.$
\begin{itemize}
	\item Triangle ($\tri$)
\end{itemize}
For $\sg{1}$ isomorphic to $\tri$, note that it is the case that the edges in $\eset{2}$ are connected in a successive manner, but this time in a cycle, such that $(e_1, 0)$ and $(e_3, 1)$ are also connected.  While this is similar to the discussion of the three path above, the first and last edges are not disjoint.
This rules out both subsets of $(e_1, 0), (e_2, 0), (e_3, 1)$ and $(e_1, 0), (e_2, 1), (e_3, 1)$, so that $\abs{f_{2, \threedis}^{-1}\inparen{\eset{1}}} = 2$.

\noindent Let us now consider when $\eset{1} \in \binom{E_1}{\leq 2}$, i.e. fixed subgraphs among
\begin{itemize}
	\item $2$-matching ($\twodis$), $2$-path ($\twopath$), $1$ edge ($\ed$)
\end{itemize}
When $|\eset{1}| = 2$, we can only pick one from each of two pairs, $\pbrace{(e_1, 0), (e_1, 1)}$ and $\pbrace{(e_2, 0), (e_2, 1)}$.  The third edge choice in $\eset{2}$ will break the disjoint property of a $3$-matching. Thus, a $3$-matching cannot exist in $f_2^{-1}(\eset{1})$.  A similar argument holds for $|\eset{1}| = 1$, where the output of $f_2^{-1}$ is $\{\emptyset\}$ since there are not enough edges in the input to produce any other output.  

Observe that all of the arguments above focused solely on the property of subgraph $\sg{1}$ being isomorphmic.  In other words, all $\eset{1}$ of a given ``shape'' yield the same number of $3$-matchings in $f_2^{-1}(\eset{1})$, and this is why we get the required identity using the above case analysis.
\qed
\end{proof}

\subsubsection{Proof of \Cref{lem:tri}}
\begin{proof}
The number of triangles in $\graph{\ell}$ for $\ell \geq 2$ will always be $0$ for the simple fact that all cycles in $\graph{\ell}$ will have at least six edges.
\qed
\end{proof}

\subsubsection{Proof of \Cref{lem:lin-sys}}

\begin{proof}
The proof consists of two parts.  First we need to show that a vector $\vct{b}$ satisfying the linear system exists and further can be computed in $O(m)$ time.  Second we need to show that $\numocc{G}{\tri}, \numocc{G}{\threedis}$ can indeed be computed in time $O(1)$.

The lemma claims that for $\vct{M} = 
\begin{pmatrix}
1 - 3p                                     &       -(3\prob^2 - \prob^3)\\
10(3\prob^2 - \prob^3)		&       10(3\prob^2 - \prob^3)
\end{pmatrix}$, $\vct{x} = 
\begin{pmatrix}
\numocc{G}{\tri}]\\
\numocc{G}{\threedis}
\end{pmatrix}$
satisfies the linear system $\vct{M} \cdot \vct{x} = \vct{b}$.

To prove the first step, we use \Cref{lem:qE3-exp} to derive the following equality (dropping the superscript and referring to $G^{(1)}$ as $G$):
\begin{align}
\numocc{G}{\ed}\prob^2 &+ 6\numocc{G}{\twopath}\prob^3 + 6\numocc{G}{\twodis}\prob^4 + 6\numocc{G}{\tri}\prob^3 + 6\numocc{G}{\oneint}\prob^4 \nonumber\\
&+ 6\numocc{G}{\threepath}\prob^4 + 6\numocc{G}{\twopathdis}\prob^5 + 6\numocc{G}{\threedis}\prob^6 = \rpoly_{G}^3(\prob,\ldots, \prob)\label{eq:lem-qE3-exp}\\
\numocc{G}{\tri}&+\numocc{G}{\threepath}\prob+\numocc{G}{\twopathdis}\prob^2+\numocc{G}{\threedis}\prob^3\nonumber\\
&= \frac{\rpoly_{G}^3(\prob,\ldots, \prob)}{6\prob^3} - \frac{\numocc{G}{\ed}}{6\prob} - \numocc{G}{\twopath}-\numocc{G}{\twodis}\prob-\numocc{G}{\oneint}\prob\label{eq:b1-alg-1}\\
\numocc{G}{\tri}(1-3p) &- \numocc{G}{\threedis}(3\prob^2 -\prob^3) = \nonumber\\
\frac{\rpoly_{G}^3(\prob,\ldots, \prob)}{6\prob^3} &- \frac{\numocc{G}{\ed}}{6\prob} - \numocc{G}{\twopath}-\numocc{G}{\twodis}\prob-\numocc{G}{\oneint}\prob\nonumber\\
&-\left[\numocc{G}{\threepath}\prob+3\numocc{G}{\tri}\prob\right]-\left[\numocc{G}{\twopathdis}\prob^2+3\numocc{G}{\threedis}\prob^2\right]\label{eq:b1-alg-2}
\end{align}
\Cref{eq:lem-qE3-exp} is the result of \Cref{lem:qE3-exp}.  We obtain the remaining equations through standard algebraic manipulations.  

Note that the LHS of \Cref{eq:b1-alg-2} is obtained using \cref{eq:2pd-3d} and \cref{eq:3p-3tri} and is indeed the product $\vct{M}[1] \cdot \vct{x}[1]$.  Further note that this product is equal to the RHS of \Cref{eq:b1-alg-2}, where every term is computable in $O(m)$ time (by equations (\ref{eq:1e})-(\ref{eq:3p-3tri})).  We set $\vct{b}[1]$ to the RHS of \Cref{eq:b1-alg-2}.

We follow the same process in deriving an equality for $G^{(2)}$.  Replacing occurrences of $G$ with $G^{(2)}$, we obtain an equation (below) of the form of \cref{eq:b1-alg-2} for $G^{(2)}$.  Substituting identities from \cref{lem:3m-G2} and \Cref{lem:tri} we obtain
\begin{align}
0-\left(8\numocc{G}{\threedis}\right.&\left.+6\numocc{G}{\twopathdis}+4\numocc{G}{\oneint}+4\numocc{G}{\threepath}+2\numocc{G}{\tri}\right)(3\prob^2 -\prob^3)=\nonumber\\
&\frac{\rpoly_{\graph{2}}^3(\prob,\ldots, \prob)}{6\prob^3} - \frac{\numocc{\graph{2}}{\ed}}{6\prob} - \numocc{\graph{2}}{\twopath}-\numocc{\graph{2}}{\twodis}\prob-\numocc{\graph{2}}{\oneint}\prob\nonumber\\
&-\left[\numocc{\graph{2}}{\twopathdis}\prob^2+3\numocc{\graph{2}}{\threedis}\prob^2\right]-\left[\numocc{\graph{2}}{\threepath}\prob + 3\numocc{\graph{2}}{\tri}\prob\right]\label{eq:b2-sub-lem}\\
(10\numocc{G}{\tri} &+ 10{G}{\threedis})(3\prob^2 -\prob^3) = \nonumber\\
&\frac{\rpoly_{\graph{2}}^3(\prob,\ldots, \prob)}{6\prob^3} - \frac{\numocc{\graph{2}}{\ed}}{6\prob} - \numocc{\graph{2}}{\twopath}-\numocc{\graph{2}}{\twodis}\prob-\numocc{\graph{2}}{\oneint}\prob\nonumber\\
&-\left[\numocc{\graph{2}}{\threepath}\prob+3\numocc{\graph{2}}{\tri}\prob\right]-\left[\numocc{\graph{2}}{\twopathdis}\prob^2-3\numocc{\graph{2}}{\threedis}\prob^2\right]\nonumber\\
&+\left(4\numocc{G}{\oneint}+\left[6\numocc{G}{\twopathdis}+18\numocc{G}{\threedis}\right]+\left[4\numocc{G}{\threepath}+12\numocc{G}{\tri}\right]\right)(3\prob^2 - \prob^3)\label{eq:b2-final}   
\end{align}
The steps to obtaining \cref{eq:b2-final} are analogous to the derivation of~\Cref{eq:b1-alg-2}.  As in the previous derivation, note that the LHS of \Cref{eq:b2-final} is the same as $\vct{M}[2]\cdot \vct{x}[2]$.  The RHS of \Cref{eq:b2-final} has terms all computable (by equations (\ref{eq:1e})-(\ref{eq:3p-3tri})) in $O(m)$ time.  Setting $\vct{b}[2]$ to the RHS then completes the proof of step 1.

Note that if $\vct{M}$ has full rank then one can compute $\numocc{G}{\tri}$ and $\numocc{G}{\threedis}$ in $O(1)$ using Gaussian elimination.

To show that $\vct{M}$ indeed has full rank, we show in what follows that $\dtrm{\vct{M}}\ne 0$ for every $\prob\in (0,1)$.
$\dtrm{\vct{M}} = $
\begin{align}
&\begin{vmatrix}
1-3\prob				&-(3\prob^2 - \prob^3)\\
10(3\prob^2 - \prob^3) 	&10(3\prob^2 - \prob^3)
\end{vmatrix}
= (1-3\prob)\cdot 10(3\prob^2-\prob^3) +10(3\prob^2-\prob^3)\cdot(3\prob^2 - \prob^3)\nonumber\\
&=10(3\prob^2-\prob^3)\cdot(1-3\prob+3\prob^2-\prob^3) = 10(3\prob^2-\prob^3)\cdot(-\prob^3+3\prob^2-3\prob + 1)\nonumber\\
&=10\prob^2(3 - \prob)\cdot(1-\prob)^3\label{eq:det-final}
\end{align}

From  \Cref{eq:det-final} it can easily be seen that the roots of $\dtrm{\vct{M}}$ are $0, 1,$ and $3$.  Hence there are no roots in $(0, 1)$ and  \Cref{lem:lin-sys} follows.
\qed
\end{proof}

\subsection{Proof of \Cref{th:single-p}}
\begin{proof}
We can compute $\graph{2}$ from $\graph{1}$ in $O(m)$ time. Additionally, if in time $O(T(m))$, we have $\rpoly_{\graph{\ell}}^3(\prob,\dots,\prob)$ for $\ell\in [2]$, then the theorem follows by \Cref{lem:lin-sys}.
\qed
\end{proof}

\subsection{Proof of \Cref{th:single-p-hard}}
\begin{proof}
For the sake of contradiction, assume that for any $G$, we can compute $\rpoly_{G}^3(\prob,\dots,\prob)$ in $o\inparen{m^{1+\eps_0}}$ time.
Let $G$ be the input graph. 
Then by \Cref{th:single-p} we can compute $\numocc{G}{\tri}$ in further time $o\inparen{m^{1+\eps_0}}+O(m)$. Thus, the overall, reduction takes $o\inparen{m^{1+\eps_0}}+O(m)= o\inparen{m^{1+\eps_0}}$ time, which violates \Cref{conj:graph}.
\qed
\end{proof}

In other words, if \Cref{th:single-p} holds, then so must \Cref{th:single-p-hard}.

\section{Missing Details from Section~\ref{sec:algo}}\label{sec:proofs-approx-alg}
In the following definitions and examples, we use the following polynomial as an example:
\begin{equation}
\label{eq:poly-eg}
\poly(X, Y) = 2X^2 + 3XY - 2Y^2.
\end{equation}

\begin{Definition}[Pure Expansion]
The pure expansion of a polynomial $\poly$ is formed by computing all product of sums occurring in $\poly$, without combining like monomials.  The pure expansion of $\poly$ generalizes \Cref{def:smb} by allowing monomials $m_i = m_j$ for $i \neq j$.
\end{Definition}
Note that similar in spirit to \Cref{def:reduced-poly-one-bidb}, $\expansion{\circuit}$ \Cref{def:expand-circuit} reduces all variable exponents $e > 1$ to $e = 1$.  Further, it is true that $\expansion{\circuit}$ encodes the pure expansion of $\circuit$.

\begin{Example}[Example of Pure Expansion]\label{example:expr-tree-T}
Consider the factorized representation $(X+ 2Y)(2X - Y)$ of the polynomial in \Cref{eq:poly-eg}.
Its circuit $\circuit$ is illustrated in \Cref{fig:circuit}.
The pure expansion of the product is $2X^2 - XY + 4XY - 2Y^2$.  As an additional example of \Cref{def:expand-circuit}, $\expansion{\circuit}=[(X, 2), (XY, -1), (XY, 4), (Y, -2)]$.
\end{Example}
$\expansion{\circuit}$ effectively\footnote{The minor difference here is that $\expansion{\circuit}$ encodes the \emph{reduced} form of the pure expansion of the compressed representation, as opposed to the \abbrSMB representation} encodes the \emph{reduced} form of $\polyf\inparen{\circuit}$, decoupling each monomial into a set of variables $\monom$ and a real coefficient $\coef$.
However, unlike the constraint on the input $\poly$ to compute $\rpoly$, the input circuit $\circuit$ does not need to be in \abbrSMB/SOP form.

\begin{Example}[Example for \Cref{def:positive-circuit}]\label{ex:def-pos-circ}
Using the same factorization from \Cref{example:expr-tree-T}, $\polyf(\abs{\circuit}) = (X + 2Y)(2X + Y) = 2X^2 +XY +4XY + 2Y^2 = 2X^2 + 5XY + 2Y^2$.  Note that this \textit{is not} the same as the polynomial from \Cref{eq:poly-eg}.  As an example of the slight abuse of notation we alluded to in~\Cref{sec:algo}, $\abs{\circuit}\inparen{1,\ldots, 1} =2\inparen{1}^2 + 5\inparen{1}\inparen{1} + 2\inparen{1}^2 = 9$. 
\end{Example}

\begin{Definition}[Subcircuit]
A subcircuit of a circuit $\circuit$ is a circuit \subcircuit such that \subcircuit is a DAG \textit{subgraph} of the DAG representing \circuit.  The sink of \subcircuit has exactly one gate \gate.
\end{Definition}



The following results assume input circuit \circuit computed from an arbitrary $\raPlus$ query $\query$ and arbitrary \abbrBIDB $\pdb$.  We refer to \circuit as a \abbrBIDB circuit.
\begin{Theorem}\label{lem:approx-alg}
Let \circuit be an arbitrary \abbrBIDB circuit 
and define $\poly(\vct{X})=\polyf(\circuit)$ and let $k=\degree(\circuit)$.
Then an estimate $\mathcal{E}$ of $\rpoly(\prob_1,\ldots, \prob_\numvar)$ can be computed in time
{\small
\[O\left(\left(\size(\circuit) + \frac{\log{\frac{1}{\conf}}\cdot \abs{\circuit}^2(1,\ldots, 1)\cdot  k\cdot \log{k} \cdot \depth(\circuit))}{\inparen{\error}^2\cdot\rpoly^2(\prob_1,\ldots, \prob_\numvar)}\right)\cdot\multc{\log\left(\abs{\circuit}(1,\ldots, 1)\right)}{\log\left(\size(\circuit)\right)}\right)\]
}
such that
\begin{equation}
\label{eq:approx-algo-bound}
\probOf\left(\left|\mathcal{E} - \rpoly(\prob_1,\dots,\prob_\numvar)\right|> \error \cdot \rpoly(\prob_1,\dots,\prob_\numvar)\right) \leq \conf.
\end{equation}
\end{Theorem}
The slight abuse of notation seen in $\abs{\circuit}\inparen{1,\ldots,1}$ is explained after \Cref{def:positive-circuit} and an example is given in \Cref{ex:def-pos-circ}.  The only difference in the use of this notation in \Cref{lem:approx-alg} is that we include an additional exponent to square the quantity.

\subsection{Proof of Theorem \ref{lem:approx-alg}}\label{sec:proof-lem-approx-alg}
\begin{algorithm}[t]
	\caption{$\approxq(\circuit, \vct{p}, \conf, \error)$}
	\label{alg:mon-sam}
	\begin{algorithmic}[1]
		\Require \circuit: Circuit
		\Require $\vct{p} = (\prob_1,\ldots, \prob_\numvar)$ $\in [0, 1]^N$
		\Require $\conf$ $\in [0, 1]$
		\Require $\error$ $\in [0, 1]$
		\Ensure \vari{acc} $\in \mathbb{R}$

		\State $\accum \gets 0$\label{alg:mon-sam-global1}
		\State $\numsamp \gets \ceil{\frac{2 \log{\frac{2}{\conf}}}{\error^2}}$\label{alg:mon-sam-global2}
		\State $(\circuit_\vari{mod}, \vari{size}) \gets $ \onepass($\circuit$)\label{alg:mon-sam-onepass}\Comment{$\onepass$ is \Cref{alg:one-pass-iter}}
	
		\For{$\vari{i} \in 1 \text{ to }\numsamp$}\label{alg:sampling-loop}\Comment{Perform the required number of samples}
			\State $(\vari{M}, \vari{sgn}_\vari{i}) \gets  $ \sampmon($\circuit_\vari{mod}$)\label{alg:mon-sam-sample}\Comment{\sampmon is \Cref{alg:sample}. Note that $\vari{sgn}_\vari{i}$ is the \emph{sign} of the monomial's coefficient and \emph{not} the coefficient itself}
			\If{$\vari{M}$ has at most one variable from each block}\label{alg:check-duplicate-block}
				\State $\vari{Y}_\vari{i} \gets \prod_{X_j\in\vari{M}}p_j$\label{alg:mon-sam-assign1}\Comment{\vari{M} is the sampled monomial's set of variables (cref. \cref{subsec:sampmon-remarks})}
				\State $\vari{Y}_\vari{i} \gets \vari{Y}_\vari{i} \times\; \vari{sgn}_\vari{i}$\label{alg:mon-sam-product}
			\State $\accum \gets \accum + \vari{Y}_\vari{i}$\Comment{Store the sum over all samples}\label{alg:mon-sam-add}
			\EndIf
		\EndFor

		\State  $\vari{acc} \gets \vari{acc} \times \frac{\vari{size}}{\numsamp}$\label{alg:mon-sam-global3}
		\State \Return \vari{acc}
	\end{algorithmic}
\end{algorithm}
We prove \Cref{lem:approx-alg} constructively by presenting an algorithm \approxq (\Cref{alg:mon-sam}) which has the desired runtime and computes an approximation with the desired approximation guarantee. Algorithm \approxq uses auxiliary algorithm \onepass to compute weights on the edges of a circuit. These weights are then used to sample a set of monomials of $\poly(\circuit)$ from the circuit $\circuit$ by traversing the circuit using the weights to ensure that monomials are sampled with an appropriate probability. The correctness of \approxq relies on the correctness (and runtime behavior) of auxiliary algorithms \onepass and \sampmon that we state in the following lemmas (and prove later in this part of the appendix).

\begin{Lemma}\label{lem:one-pass}
The $\onepass$ function completes in time:
$$O\left(\size(\circuit) \cdot \multc{\log\left(\abs{\circuit(1\ldots, 1)}\right)}{\log{\size(\circuit)}}\right)$$
  $\onepass$ guarantees two post-conditions:  First, for each subcircuit $\vari{S}$ of $\circuit$, we have that $\vari{S}.\vari{partial}$ is set to $\abs{\vari{S}}(1,\ldots, 1)$.  Second, when $\vari{S}.\type  = \circplus$, \subcircuit.\lwght $= \frac{\abs{\subcircuit_\linput}(1,\ldots, 1)}{\abs{\subcircuit}(1,\ldots, 1)}$ and likewise for \subcircuit.\rwght.
\end{Lemma}
To prove correctness of \Cref{alg:mon-sam}, we use the following fact that follows from the above lemma: for the modified circuit ($\circuit_{\vari{mod}}$) output by \onepass, $\circuit_{\vari{mod}}.\vari{partial}=\abs{\circuit}(1,\dots,1)$.

\AH{I don't think the word \emph{only} is needed.}
\begin{Lemma}\label{lem:sample}
The function $\sampmon$ completes in time
$$O(\log{k} \cdot k \cdot \depth(\circuit)\cdot\multc{\log\left(\abs{\circuit}(1,\ldots, 1)\right)}{\log{\size(\circuit)}})$$
 where $k = \degree(\circuit)$.  The function returns every $\left(\monom, sign(\coef)\right)$ for $(\monom, \coef)\in \expansion{\circuit}$ with probability $\frac{|\coef|}{\abs{\circuit}(1,\ldots, 1)}$.
\end{Lemma}

With the above two lemmas, we are ready to argue the following result: 
\begin{Theorem}\label{lem:mon-samp}
For any $\circuit$ with
$\degree(\polyf(|\circuit|)) = k$, algorithm \ref{alg:mon-sam} outputs an estimate $\vari{acc}$ of $\rpoly(\prob_1,\ldots, \prob_\numvar)$ such that
\[\probOf\left(\left|\vari{acc} - \rpoly(\prob_1,\ldots, \prob_\numvar)\right|\geq \error \cdot \abs{\circuit}(1,\ldots, 1)\right) \leq \conf,\]
 in $O\left(\left(\size(\circuit)+\frac{\log{\frac{1}{\conf}}}{\error^2} \cdot k \cdot\log{k} \cdot \depth(\circuit)\right)\cdot \multc{\log\left(\abs{\circuit}(1,\ldots, 1)\right)}{\log{\size(\circuit)}}\right)$ time.
\end{Theorem}

Before proving \Cref{lem:mon-samp}, we use it to argue the claimed runtime of our main result, \Cref{lem:approx-alg}.

\begin{proof}[Proof of \Cref{lem:approx-alg}]
Set $\mathcal{E}=\approxq({\circuit}, (\prob_1,\dots,\prob_\numvar),$ $\conf, \error')$, where
\[\error' = \error \cdot \frac{\rpoly(\prob_1,\ldots, \prob_\numvar)}{\abs{{\circuit}}(1,\ldots, 1)},\]
 which achieves the claimed error bound on $\mathcal{E}$ (\vari{acc}) trivially due to the assignment to $\error'$ and \cref{lem:mon-samp}, since $\error' \cdot \abs{\circuit}(1,\ldots, 1) = \error\cdot\frac{\rpoly(\prob_1,\ldots, \prob_\numvar)}{\abs{\circuit}(1,\ldots, 1)} \cdot \abs{\circuit}(1,\ldots, 1) = \error\cdot\rpoly(\prob_1,\ldots, \prob_\numvar)$.

The claim on the runtime follows from \Cref{lem:mon-samp} since

\begin{align*}
\frac 1{\inparen{\error'}^2}\cdot \log\inparen{\frac 1\conf}=&\frac{\log{\frac{1}{\conf}}}{\error^2 \left(\frac{\rpoly(\prob_1,\ldots, \prob_N)}{\abs{{\circuit}}(1,\ldots, 1)}\right)^2}\\
= &\frac{\log{\frac{1}{\conf}}\cdot \abs{{\circuit}}^2(1,\ldots, 1)}{\error^2 \cdot \rpoly^2(\prob_1,\ldots, \prob_\numvar)}.
\end{align*}
\qed
\end{proof}

Let us now prove \Cref{lem:mon-samp}:
\subsection{Proof of Theorem \ref{lem:mon-samp}}\label{app:subsec-th-mon-samp}
\begin{proof}
Consider now the random variables $\randvar_1,\dots,\randvar_\numsamp$, where each $\randvar_\vari{i}$ is the value of $\vari{Y}_{\vari{i}}$ in \cref{alg:mon-sam} after \cref{alg:mon-sam-product} is executed. Overloading $\isInd{\cdot}$ to receive monomial input (recall $\encMon$ is the monomial composed of the variables in the set $\monom$), we have
\[\randvar_\vari{i}= \indicator{\inparen{\isInd{\encMon}}}\cdot \prod_{X_i\in \var\inparen{v}} p_i,\]
where the indicator variable handles the check in \Cref{alg:check-duplicate-block}
Then for random variable $\randvar_i$, it is the case that
\begin{align*}
\expct\pbox{\randvar_\vari{i}} &= \sum\limits_{(\monom, \coef) \in \expansion{{\circuit}} }\frac{\indicator{\inparen{\isInd{\encMon}}}\cdot c\cdot\prod_{X_i\in \var\inparen{v}} p_i }{\abs{{\circuit}}(1,\dots,1)} \\
&= \frac{\rpoly(\prob_1,\ldots, \prob_\numvar)}{\abs{{\circuit}}(1,\ldots, 1)},
\end{align*}
where in the first equality we use the fact that $\vari{sgn}_{\vari{i}}\cdot \abs{\coef}=\coef$ and the second equality follows from \Cref{eq:tilde-Q-bi} with $X_i$ substituted by $\prob_i$.

Let $\empmean = \frac{1}{\samplesize}\sum_{i = 1}^{\samplesize}\randvar_\vari{i}$.  It is also true that

\[\expct\pbox{\empmean}
= \frac{1}{\samplesize}\sum_{i = 1}^{\samplesize}\expct\pbox{\randvar_\vari{i}}
= \frac{\rpoly(\prob_1,\ldots, \prob_\numvar)}{\abs{{\circuit}}(1,\ldots, 1)}.\]

Hoeffding's inequality states that if we know that each $\randvar_i$ (which are all independent) always lie in the intervals $[a_i, b_i]$, then it is true that
\begin{equation*}
\probOf\left(\left|\empmean - \expct\pbox{\empmean}\right| \geq \error\right) \leq 2\exp{\left(-\frac{2\samplesize^2\error^2}{\sum_{i = 1}^{\samplesize}(b_i -a_i)^2}\right)}.
\end{equation*}

Line~\ref{alg:mon-sam-sample} shows that $\vari{sgn}_\vari{i}$ has a value in $\{-1, 1\}$ that is multiplied with $O(k)$ $\prob_i\in [0, 1]$, which implies the range for each $\randvar_i$ is $[-1, 1]$.
Using Hoeffding's inequality, we then get:
\begin{equation*}
\probOf\left(~\left| \empmean - \expct\pbox{\empmean} ~\right| \geq \error\right) \leq 2\exp{\left(-\frac{2\samplesize^2\error^2}{2^2 \samplesize}\right)} = 2\exp{\left(-\frac{\samplesize\error^2}{2 }\right)}\leq \conf,
\end{equation*}
where the last inequality dictates our choice of $\samplesize$ in \Cref{alg:mon-sam-global2}.
\AH{Why does the $\geq$ sign change to $>$?}
For the claimed probability bound of $\probOf\left(\left|\vari{acc} - \rpoly(\prob_1,\ldots, \prob_\numvar)\right|\geq \error \cdot \abs{\circuit}(1,\ldots, 1)\right) \leq \conf$, note that in the algorithm, \vari{acc} is exactly $\empmean \cdot \abs{\circuit}(1,\ldots, 1)$.  Multiplying the rest of the terms by the additional factor $\abs{\circuit}(1,\ldots, 1)$ yields the said bound.

This concludes the proof for the first claim of theorem~\ref{lem:mon-samp}.  Next, we prove the claim on the runtime.

\paragraph*{Run-time Analysis}
The runtime of the algorithm is dominated first by \Cref{alg:mon-sam-onepass} which has $O\left({\size(\circuit)}\cdot \multc{\log\left(\abs{\circuit}(1,\ldots, 1)\right)}{\log\left(\size(\circuit)\right)}\right)$ runtime by \Cref{lem:one-pass}.  There are then $\samplesize$ iterations of the loop in \Cref{alg:sampling-loop}. Each iteration's run time is dominated by the call to \sampmon in \Cref{alg:mon-sam-sample} (which by \Cref{lem:sample} takes $O\left(\log{k} \cdot k \cdot {\depth(\circuit)}\cdot \multc{\log\left(\abs{\circuit}(1,\ldots, 1)\right)}{\log\left(\size(\circuit)\right)}\right)$
) and the check \Cref{alg:check-duplicate-block}, which by the subsequent argument takes $O(k\log{k})$ time. We sort the $O(k)$ variables by their block IDs and then check if there is a duplicate block ID or not. Combining all the times discussed here gives us the desired overall runtime.
\qed
\end{proof}

\subsection{Proof of \Cref{cor:approx-algo-const-p}}
\begin{proof}
The result follows by first noting that by definition of $\gamma$, we have
\[\rpoly(1,\dots,1)= (1-\gamma)\cdot \abs{{\circuit}}(1,\dots,1).\]
Further, since each $\prob_i\ge \prob_0$ and $\poly(\vct{X})$ (and hence $\rpoly(\vct{X})$) has degree at most $k$, we have that
\[ \rpoly(1,\dots,1) \ge \prob_0^k\cdot \rpoly(1,\dots,1).\]
The above two inequalities implies $\rpoly(1,\dots,1) \ge \prob_0^k\cdot (1-\gamma)\cdot \abs{{\circuit}}(1,\dots,1)$.
Applying this bound in the runtime bound in \Cref{lem:approx-alg} gives the first claimed runtime. The final runtime of $O_k\left(\frac 1{\eps^2}\cdot\size(\circuit)\cdot \log{\frac{1}{\conf}}\cdot \multc{\log\left(\abs{\circuit}(1,\ldots, 1)\right)}{\log\left(\size(\circuit)\right)}\right)$ follows by noting that $\depth({\circuit})\le \size({\circuit})$ and absorbing all factors that just depend on $k$.
\qed
\end{proof}



\subsection{$\onepass$ Remarks}

Please note that it is \textit{assumed} that the original call to \onepass consists of a call on an input circuit \circuit such that the values of members \prt, \lwght and \rwght have been initialized to Null across all gates.

The evaluation of $\abs{\circuit}(1,\ldots, 1)$ can be defined recursively, as follows (where $\circuit_\linput$ and $\circuit_\rinput$ are the `left' and `right' inputs of $\circuit$ if they exist):

{\small
\begin{align}
\label{eq:T-all-ones}
\abs{\circuit}(1,\ldots, 1) = \begin{cases}
						\abs{\circuit_\linput}(1,\ldots, 1) \cdot \abs{\circuit_\rinput}(1,\ldots, 1)	&\textbf{if }\circuit.\type = \circmult\\
						\abs{\circuit_\linput}(1,\ldots, 1) + \abs{\circuit_\rinput}(1,\ldots, 1)		&\textbf{if }\circuit.\type = \circplus \\
						 |\circuit.\val|											&\textbf{if }\circuit.\type = \tnum\\
						1													&\textbf{if }\circuit.\type = \var.
					\end{cases}
\end{align}
}

It turns out that for proof of \Cref{lem:sample}, we need to argue that when $\circuit.\type = +$, we indeed have
\begin{align}
\label{eq:T-weights}
\circuit.\lwght &\gets \frac{\abs{\circuit_\linput}(1,\ldots, 1)}{\abs{\circuit_\linput}(1,\ldots, 1) + \abs{\circuit_\rinput}(1,\ldots, 1)};\\
\circuit.\rwght &\gets \frac{\abs{\circuit_\rinput}(1,\ldots, 1)}{\abs{\circuit_\linput}(1,\ldots, 1)+ \abs{\circuit_\rinput}(1,\ldots, 1)}
\end{align}

\subsection{$\onepass$ Example}
\begin{figure}[h!]
\centering
	\begin{tikzpicture}[thick]
		\node[tree_node] (a1) at (1, 0) {$\boldsymbol{Y}$};
		\node[tree_node] (b1) at (3, 0) {$\boldsymbol{-1}$};
		\node[tree_node] (a2) at (-0.75, 0) {$\boldsymbol{X}$};
		\node[tree_node] (b2) at (1.6,1.25) {$\boldsymbol{\circmult}$};
		\node[tree_node] (c2) at (2.9, 1.25) {$\boldsymbol{\circmult}$};
		\node[tree_node] (a3) at (0.7, 2.5) {$\boldsymbol{\circplus}$};
		\node[tree_node] (b3) at (1.6, 2.5) {$\boldsymbol{\circplus}$};
		\node[tree_node] (a4) at (1.5, 3.75) {$\boldsymbol{\circmult}$};
		\node[tree_node] (b4) at (2.8, 4) {$\boldsymbol{\circplus}$};
		\node[above right=0.15cm of b4, inner sep=0pt, font=\bfseries](labelC) {$\circuit$};

		\draw[->] (a1) edge[right] node{$\frac{1}{2}$} (a3);
		\draw[->] (b1) -- (b2);
		\draw[->] (a1) -- (b2);
		\draw[->] (a1) edge[bend left=15] (c2);
		\draw[->] (a1) edge[bend right=15] (c2);
		\draw[->] (a2) edge[left] node{$\frac{1}{2}$} (a3);
		\draw[->] (a2) edge[below] node{$\frac{1}{2}$} (b3);
		\draw[->] (b2) edge[right] node{$\frac{1}{2}$} (b3);
		\draw[->] (c2) edge[right] node{$\frac{1}{5}$} (b4);
		\draw[->] (a3) -- (a4);
		\draw[->] (b3) -- (a4);
		\draw[->] (a4) edge[above] node{$\frac{4}{5}$} (b4);
		\draw[black] (b4) -- (labelC);
	\end{tikzpicture}
		\caption{Weights computed by $\onepass$ in  \Cref{example:one-pass}.}

		\label{fig:expr-tree-T-wght}
\end{figure}
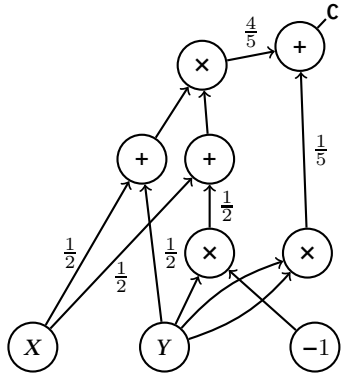

\begin{Example}\label{example:one-pass}
 Let $\etree$ encode the expression $(X + Y)(X - Y) + Y^2$.  After one pass, \Cref{alg:one-pass-iter} would have computed the following weight distribution.  For the two inputs of the sink gate $\circuit$, $\circuit.\lwght = \frac{4}{5}$ and $\circuit.\rwght = \frac{1}{5}$.  Similarly, for $\stree$ denoting the left input $\circuit_{\lchild}$ of \circuit, $\stree.\lwght = \stree.\rwght = \frac{1}{2}$.  This is depicted in \Cref{fig:expr-tree-T-wght}. 
\end{Example}

\begin{algorithm}[h!]
	\caption{\onepass$(\circuit)$}
	\label{alg:one-pass-iter}
	\begin{algorithmic}[1]
		\Require \circuit: Circuit
		\Ensure \circuit: Annotated Circuit
		\Ensure \vari{sum} $\in \domN$
		\For{\gate in \topord(\circuit)}\label{alg:one-pass-loop}\Comment{\topord($\cdot$) is the topological order of \circuit}
			\If{\gate.\type $=$ \var}
				\State \gate.\prt $\gets 1$\label{alg:one-pass-var}
			\ElsIf{\gate.\type $=$ \tnum}
				\State \gate.\prt $\gets \abs{\gate.\val}$\label{alg:one-pass-num}
			\ElsIf{\gate.\type $= \circmult$}		
				\State \gate.\prt $\gets \gate_\linput.\prt \times \gate_\rinput.\prt$\label{alg:one-pass-mult}
			\Else 
				\State \gate.\prt $\gets \gate_\linput.\prt + \gate_\rinput.\prt$\label{alg:one-pass-plus}
				\State \gate.\lwght $\gets \frac{\gate_\linput.\prt}{\gate.\prt}$\label{alg:one-pass-lwght}
				\State \gate.\rwght $\gets \frac{\gate_\rinput.\prt}{\gate.\prt}$\label{alg:one-pass-rwght}
			\EndIf
			\State \vari{sum} $\gets \gate.\prt$
		\EndFor
		\State \Return (\vari{sum}, $\circuit$) 
	\end{algorithmic}
\end{algorithm}

\subsection{Proof of \onepass (\Cref{lem:one-pass})}\label{sec:proof-one-pass}
\begin{proof}
We prove the correct computation of \prt, \lwght, \rwght values on \circuit by induction over the number of iterations in  the topological order \topord (line~\ref{alg:one-pass-loop}) of the input circuit \circuit.  \topord follows the standard definition of a topological ordering over the DAG structure of \circuit.

For the base case, we have only one gate, which by definition is a source gate and must be either \var or \tnum.  In this case, as per \cref{eq:T-all-ones}, lines~\ref{alg:one-pass-var} and~\ref{alg:one-pass-num} correctly compute \circuit.\prt as $1$.

For the inductive hypothesis, assume that \onepass correctly computes \subcircuit.\prt, \subcircuit.\lwght, and \subcircuit.\rwght for all gates \gate in \circuit with $k \geq 0$ iterations over \topord.  
We now prove for $k + 1$ iterations that \onepass correctly computes the \prt, \lwght, and \rwght values for each gate $\gate_\vari{i}$ in \circuit for $i \in [k + 1]$.
The $\gate_\vari{k + 1}$ must be in the last ordering of all gates $\gate_\vari{i}$.  When \size(\circuit) > 1, if $\gate_{k+1}$ is a leaf node, we are back to the base case.  Otherwise $\gate_{k + 1}$ is an internal node 
which requires binary input.

When $\gate_{k+1}.\type = \circplus$, then by line~\ref{alg:one-pass-plus} $\gate_{k+1}$.\prt $= \gate_{{k+1}_\lchild}$.\prt $+ \gate_{{k+1}_\rchild}$.\prt, a correct computation, as per \cref{eq:T-all-ones}.  Further, lines~\ref{alg:one-pass-lwght} and~\ref{alg:one-pass-rwght} compute $\gate_{{k+1}}.\lwght = \frac{\gate_{{k+1}_\lchild}.\prt}{\gate_{{k+1}}.\prt}$ and analogously for $\gate_{{k+1}}.\rwght$.  All values needed for each computation have been correctly computed by the inductive hypothesis.

When $\gate_{k+1}.\type = \circmult$, then line~\ref{alg:one-pass-mult} computes $\gate_{k+1}.\prt = \gate_{{k+1}_\lchild.\prt} \circmult \gate_{{k+1}_\rchild}.\prt$, which indeed by \cref{eq:T-all-ones} is correct.  This concludes the proof of correctness.

\paragraph*{Runtime Analysis}
It is known that $\topord(G)$ is computable in linear time.  There are $\size(\circuit)$ iterations.  Each iteration has runtime $O\left( \multc{\log\left(\abs{\circuit(1\ldots, 1)}\right)}{\log\inparen{\size(\circuit)}}\right)$ time.  This can be seen since each of all the numbers which the algorithm computes is at most $\abs{\circuit}(1,\dots,1)$. Hence, by definition each such operation takes $\multc{\log\left(\abs{\circuit(1\ldots, 1)}\right)}{\log{\size(\circuit)}}$ time, which proves the claimed runtime.
\qed
\end{proof}

\subsection{\sampmon Remarks}\label{subsec:sampmon-remarks}
\begin{algorithm}[t]
	\caption{\sampmon(\circuit)}
	\label{alg:sample}
	\begin{algorithmic}[1]
		\Require \circuit: Circuit
		\Ensure \vari{vars}: TreeSet
		\Ensure \vari{sgn} $\in \{-1, 1\}$
		\Comment{\Cref{alg:one-pass-iter} should have been run before this one} 
		\State $\vari{vars} \gets \emptyset$ \label{alg:sample-global1}
		\If{$\circuit.\type = +$}\Comment{Sample at every $+$ node}
			\State $\circuit_{\vari{samp}} \gets$ Sample from left input ($\circuit_{\linput}$) and right input ($\circuit_{\rinput}$) w.p. $\circuit.\vari{Lweight}$ and $\circuit.\vari{Rweight}$. \label{alg:sample-plus-bsamp} \Comment{Each call to \sampmon uses fresh randomness}
			\State $(\vari{v}, \vari{s}) \gets \sampmon(\circuit_{\vari{samp}})$\label{alg:sample-plus-traversal}
			\State $\Return ~(\vari{v}, \vari{s})$
		\ElsIf{$\circuit.\type = \times$}\Comment{Multiply the sampled values of all inputs}
			\State $\vari{sgn} \gets 1$\label{alg:sample-global2}
			\For {$input$ in $\circuit.\vari{input}$}\label{alg:sample-times-for-loop}
				\State $(\vari{v}, \vari{s}) \gets \sampmon(input)$
				\State $\vari{vars} \gets \vari{vars} \cup \{\vari{v}\}$\label{alg:sample-times-union}
				\State $\vari{sgn} \gets \vari{sgn} \times \vari{s}$\label{alg:sample-times-product}
			\EndFor
			\State $\Return ~(\vari{vars}, \vari{sgn})$
		\ElsIf{$\circuit.\type = \tnum$}\Comment{The leaf is a coefficient}
			\State $\Return ~\left(\{\}, \func{sgn}(\circuit.\val)\right)$\label{alg:sample-num-return}\Comment{$\func{sgn}(\cdot)$ outputs $-1$ for \circuit.\val $\geq 1$ and $-1$ for \circuit.\val $\leq -1$}\label{alg:sample-num-leaf}
		\ElsIf{$\circuit.\type = var$}
			\State $\Return~\left(\{\circuit.\val\}, 1\right)	$\label{alg:sample-var-return}
		\EndIf
	\end{algorithmic}
\end{algorithm}
We briefly describe the top-down traversal of \sampmon.  When \circuit.\type $= +$, the input to be visited is sampled from the weighted distribution precomputed by \onepass.
When a \circuit.\type$= \times$ node is visited, both inputs are visited.
The algorithm computes two properties: the set of all variable leaf nodes visited, and the product of the signs of visited coefficient leaf nodes.
We will assume the TreeSet data structure to maintain sets with logarithmic time insertion and linear time traversal of its elements.
While we would like to take advantage of the space efficiency gained in using a circuit \circuit instead an expression tree \etree, we do not know that such a method exists when computing a sample of the input polynomial representation.  

The efficiency gains of circuits over trees is found in the capability of circuits to only require space for each \emph{distinct} term in the compressed representation.  This saves space in such polynomials containing non-distinct terms multiplied or added to each other, e.g., $x^4$.  However, to avoid biased sampling, it is imperative to sample from both inputs of a multiplication gate, independently, which is indeed the approach of \sampmon.  

\subsection{Proof of \sampmon (\Cref{lem:sample})}\label{sec:proof-sample-monom}
\begin{proof}

We first need to show that $\sampmon$ samples a valid monomial $\encMon$ by sampling and returning a set of variables $\monom$, such that $(\monom, \coef)$ is in $\expansion{\circuit}$ and $\encMon$ is indeed a monomial of the $\rpoly\inparen{\vct{X}}$ encoded in \circuit.  We show this via induction over the depth of \circuit.
For the base case, let the depth $d$ of $\circuit$ be $1$.  We have that the single gate is either a constant $\coef$ for which by line~\ref{alg:sample-num-return} we return $\{~\}$, or we have that $\circuit.\type = \var$ and $\circuit.\val = x$, and  by line~\ref{alg:sample-var-return} we return $\{x\}$.  By \cref{def:expand-circuit}, both cases return a valid $\monom$ for some $(\monom, \coef)$ from $\expansion{\circuit}$, and the base case is proven.

\AH{I think it is slightly confusing to say that depth $= 0$ in view of the definition of depth in S.4.  To say $k = 0$ is also strange, since, for a single join, we have that $k = 2$.}

For the inductive hypothesis, assume that for $d \leq k$ for some $k \geq 1$, that it is indeed the case that $\sampmon$ returns a valid monomial.

For the inductive step, let us take a circuit $\circuit$ with $d = k + 1$.  Note that each input has depth $d - 1 \leq k$, and by inductive hypothesis both of them sample a valid monomial.  Then the sink can be either a $\circplus$ or $\circmult$ gate.  For the case when $\circuit.\type = \circplus$, line~\ref{alg:sample-plus-bsamp} of $\sampmon$ will choose one of the inputs of the source.  By inductive hypothesis it is the case that some valid monomial is being randomly sampled from each of the inputs.  Then it follows when $\circuit.\type = \circplus$ that a valid monomial is sampled by $\sampmon$.  When the $\circuit.\type = \circmult$, line~\ref{alg:sample-times-union} computes the set union of the monomials returned by the two inputs of the sink, and it is trivial to see by \cref{def:expand-circuit} that $\encMon$ is a valid monomial encoded by some $(\monom, \coef)$ of $\expansion{\circuit}$.

We will next prove by induction on the depth $d$ of $\circuit$ that for $(\monom,\coef) \in \expansion{\circuit}$, $\monom$ is sampled with a probability $\frac{|\coef|}{\abs{\circuit}\polyinput{1}{1}}$.

For the base case $d = 1$, by definition~\ref{def:circuit} we know that the $\size\inparen{\circuit} = 1$ and \circuit.\type$=$ \tnum or \var.  For either case, the probability of the value returned is $1$ since there is only one value to sample from.  When \circuit.\val $= x$, the algorithm always return the variable set $\{x\}$.  When $\circuit.\type = \tnum$, \sampmon will always return the variable set $\emptyset$.

\AH{I don't think this is technically right, since \sampmon returns a tuple of two values.}

For the inductive hypothesis, assume that for $d \leq k$ and $k \geq 1$ $\sampmon$ indeed returns $\monom$ in $(\monom, \coef)$ of $\expansion{\circuit}$ with probability $\frac{|\coef|}{\abs{\circuit}\polyinput{1}{1}}$.

We prove now for $d = k + 1$ the inductive step holds.  It is the case that the sink of $\circuit$ has two inputs $\circuit_\linput$ and $\circuit_\rinput$.  Since $\circuit_\linput$ and $\circuit_\rinput$ are both depth $d - 1 \leq k$, by inductive hypothesis, $\sampmon$ will return $\monom_\linput$ in $(\monom_\lchild, \coef_\lchild)$ of $\expansion{\circuit_\linput}$ and $\monom_\rinput$ in $(\monom_\rchild, \coef_\rchild)$ of $\expansion{\circuit_\rinput}$, from $\circuit_\linput$ and $\circuit_\rinput$ with probability $\frac{|\coef_\lchild|}{\abs{\circuit_\linput}\polyinput{1}{1}}$ and $\frac{|\coef_\rchild|}{\abs{\circuit_\rinput}\polyinput{1}{1}}$.

Consider the case when $\circuit.\type = \circmult$.  For the term $(\monom, \coef)$ from $\expansion{\circuit}$ that is being sampled it is the case that $\monom = \monom_\lchild \cup \monom_\rchild$, where $\monom_\lchild$ is coming from $\circuit_\linput$ and $\monom_\rchild$ from $\circuit_\rinput$.  The probability that \sampmon$(\circuit_{\lchild})$ returns $\monom_\lchild$ is $\frac{|\coef_{\monom_\lchild}|}{|\circuit_\linput|(1,\ldots, 1)}$ and $\frac{|\coef_{\monom_\rchild}|}{\abs{\circuit_\rinput}\polyinput{1}{1}}$ for $\monom_\rchild$.  Since both $\monom_\lchild$ and $\monom_\rchild$ are sampled with independent randomness, the final probability for sample $\monom$ is then $\frac{|\coef_{\monom_\lchild}| \cdot |\coef_{\monom_\rchild}|}{|\circuit_\linput|(1,\ldots, 1) \cdot |\circuit_\rinput|(1,\ldots, 1)}$.  For $(\monom, \coef)$ in $\expansion{\circuit}$, by \cref{def:expand-circuit} it is indeed the case that $|\coef| = |\coef_{\monom_\lchild}| \cdot |\coef_{\monom_\rchild}|$ and that (as shown in \cref{eq:T-all-ones}) $\abs{\circuit}(1,\ldots, 1) = |\circuit_\linput|(1,\ldots, 1) \cdot |\circuit_\rinput|(1,\ldots, 1)$, and therefore $\monom$ is sampled with correct probability $\frac{|\coef|}{\abs{\circuit}(1,\ldots, 1)}$.

For the case when $\circuit.\type = \circplus$, \sampmon ~will sample $\monom$ from one of its inputs.  By inductive hypothesis we know that any $\monom_\lchild$ in $\expansion{\circuit_\linput}$ and any $\monom_\rchild$ in $\expansion{\circuit_\rinput}$ will both be sampled with correct probability $\frac{|\coef_{\monom_\lchild}|}{\abs{\circuit_{\lchild}}(1,\ldots, 1)}$ and $\frac{|\coef_{\monom_\rchild}|}{|\circuit_\rinput|(1,\ldots, 1)}$, where either $\monom_\lchild$ or $\monom_\rchild$ will equal $\monom$, depending on whether $\circuit_\linput$ or $\circuit_\rinput$ is sampled.  Assume that $\monom$ is sampled from $\circuit_\linput$, and note that a symmetric argument holds for the case when $\monom$ is sampled from $\circuit_\rinput$.  Notice also that the probability of choosing $\circuit_\linput$ from $\circuit$ is $\frac{\abs{\circuit_\linput}\polyinput{1}{1}}{\abs{\circuit_\linput}\polyinput{1}{1} + \abs{\circuit_\rinput}\polyinput{1}{1}}$ as computed by $\onepass$.  Then, since $\sampmon$ goes top-down, and each sampling choice is independent (which follows from the randomness in the root of $\circuit$ being independent from the randomness used in its subtrees), the probability for $\monom$ to be sampled from $\circuit$ is equal to the product of the probability that $\circuit_\linput$ is sampled from $\circuit$ and $\monom$ is sampled in $\circuit_\linput$, and
\begin{align*}
&\probOf(\sampmon(\circuit) = \monom) = \\
&\probOf(\sampmon(\circuit_\linput) = \monom) \cdot \probOf(SampledChild(\circuit) = \circuit_\linput)\\
&= \frac{|\coef_\monom|}{|\circuit_\linput|(1,\ldots, 1)} \cdot \frac{\abs{\circuit_\linput}(1,\ldots, 1)}{|\circuit_\linput|(1,\ldots, 1) + |\circuit_\rinput|(1,\ldots, 1)}\\
&= \frac{|\coef_\monom|}{\abs{\circuit}(1,\ldots, 1)},
\end{align*}
and we obtain the desired result.

Lastly, we show by simple induction of the depth $d$ of \circuit that \sampmon indeed returns the correct sign value of $\coef$ in $(\monom, \coef)$.

In the base case, $\circuit.\type = \tnum$ or $\var$.  For the former by~\Cref{alg:sample-num-leaf}, \sampmon correctly returns the sign value of the gate.  For the latter by~\Cref{alg:sample-var-return}, \sampmon returns the correct sign of $1$, since a variable is a neutral element, and $1$ is the multiplicative identity, whose product with another sign element will not change that sign element.

For the inductive hypothesis, we assume for a circuit of depth $d \leq k$ and $k \geq 1$ that the algorithm correctly returns the sign value of $\coef$.

Similar to before, for a depth \AH{Why do we use $d = k + 1$ for the inductive cases above?}
$d \leq k + 1$, it is true that $\circuit_\linput$ and $\circuit_\rinput$ both return the correct sign of $\coef$.  For the case that $\circuit.\type = \circmult$, the sign value of both inputs are multiplied, which is the correct behavior by \cref{def:expand-circuit}.  When $\circuit.\type = \circplus$, only one input of $\circuit$ is sampled, and the algorithm  returns the correct sign value of $\coef$ by inductive hyptothesis.

\paragraph*{Run-time Analysis}
It is easy to check that except for lines~\ref{alg:sample-plus-bsamp} and~\ref{alg:sample-times-union}, all lines take $O(1)$ time.  Consider an execution of \cref{alg:sample-times-union}. We note that we will be adding a given set of variables to some set at most once: since the sum of the sizes of the sets at a given level is at most $\degree(\circuit)$, each gate visited takes $O(\log{\degree(\circuit)})$.  For \Cref{alg:sample-plus-bsamp}, note that we pick $\circuit_\linput$ with probability $\frac a{a+b}$ where $a=\circuit.\vari{Lweight}$ and $b=\circuit.\vari{Rweight}$. We can implement this step by picking a random number $r\in[a+b]$ and then checking if $r\le a$. It is easy to check that $a+b\le \abs{\circuit}(1,\dots,1)$. This means we need to add and compare $\log{\abs{\circuit}(1,\ldots, 1)}$-bit numbers, which can certainly be done in time $\multc{\log\left(\abs{\circuit(1\ldots, 1)}\right)}{\log{\size(\circuit)}}$ (note that this is an over-estimate). 
Denote \cost(\circuit) (\Cref{eq:cost-sampmon}) to be an upper bound of the number of gates visited by \sampmon.  Then the runtime is $O\left(\cost(\circuit)\cdot \log{\degree(\circuit)}\cdot \multc{\log\left(\abs{\circuit(1\ldots, 1)}\right)}{\log{\size(\circuit)}}\right)$.

\AH{We don't really justify why we can bound the number of recursive calls as we claim in what follows.}
Since there can be at most $k = \degree\inparen{\circuit}$ nodes visited at every level of the circuit, and each of the first $d - 1$ levels (going from the sink to the source nodes) will contain at least one recursive call, we can upperbound the number of recursive calls in $\sampmon$ by $O\left((\degree(\circuit) + 1)\right.$$\left.\cdot\right.$ $\left.\depth(\circuit)\right)$, which by the above will prove the claimed runtime of~\Cref{lem:sample}.  

Let \cost$(\cdot)$ be a function that models an upper bound on the number of gates that can be visited in the run of \sampmon.  We define \cost$(\cdot)$ recursively as follows.

\begin{equation}
	\cost(\circuit) =
		\begin{cases}
			1 + \cost(\circuit_\linput) + \cost(\circuit_\rinput) & \textbf{if } \text{\circuit.\type = }\circmult\\
			1 + \max\left(\cost(\circuit_\linput), \cost(\circuit_\rinput)\right) & \textbf{if } \text{\circuit.\type = \circplus}\\
			1 & \textbf{otherwise}
		\end{cases}\label{eq:cost-sampmon}
\end{equation}

First note that the number of gates visited in \sampmon is $\leq\cost(\circuit)$.  To show that \cref{eq:cost-sampmon} upper bounds the number of nodes visited by \sampmon, note that when \sampmon visits a gate such that \circuit.\type $ =\circmult$, line~\ref{alg:sample-times-for-loop} visits each input of \circuit, as defined in (\ref{eq:cost-sampmon}).  For the case when \circuit.\type $= \circplus$, line~\ref{alg:sample-plus-bsamp} visits exactly one of the input gates, which may or may not be the subcircuit with the maximum number of gates traversed, which makes \cost$(\cdot)$ an upperbound.  Finally, it is trivial to see that when \circuit.\type $\in \{\var, \tnum\}$, i.e., a source gate, that only one gate is visited.

We prove the following inequality holds.
\begin{equation}
2\left(\degree(\circuit) + 1\right) \cdot \depth(\circuit) + 1 \geq \cost(\circuit)\label{eq:strict-upper-bound}
\end{equation} 

Note that \cref{eq:strict-upper-bound} implies the claimed runtime.  
\AH{If the claimed runtime is from the first paragraph, then I don't follow.}

We prove \cref{eq:strict-upper-bound} for the number of gates traversed in \sampmon using induction over $\depth(\circuit)$.  Recall how degree is defined in \cref{def:degree}.
\AH{In the following, by~\Cref{def:size-depth}, we would have that $\depth\inparen{\circuit} = 1$ \emph{technically}.}
For the base case $\degree(\circuit) \in \inset{0, 1}, \depth(\circuit) = 1$, $\cost(\circuit) = 1$, and it is trivial to see that the inequality $2\degree(\circuit) \cdot \depth(\circuit) + 1 \geq \cost(\circuit)$ holds.

For the inductive hypothesis, we assume the bound holds for any circuit where $\ell \geq \depth(\circuit) \geq 0$.
Now consider the case when \sampmon has an arbitrary circuit \circuit input with $\depth(\circuit) = \ell + 1$.  By definition \circuit.\type $\in \{\circplus, \circmult\}$. Note that since $\depth(\circuit) \geq 2$, \circuit must have input(s).  Further we know that by the inductive hypothesis the inputs $\circuit_i$ for $i \in \{\linput, \rinput\}$ of the sink gate \circuit uphold the bound
\begin{equation}
2\left(\degree(\circuit_i) + 1\right)\cdot \depth(\circuit_i) + 1 \geq \cost(\circuit_i).\label{eq:ih-bound-cost}
\end{equation}
In particular, since for any $i$, \cref{eq:ih-bound-cost} holds, then it immediately follows that an inequality whose operands consist of a sum of the aforementioned inequalities must also hold.  This is readily seen in the inequality of \cref{eq:times-middle} and \cref{eq:times-rhs}, where $2\inparen{\degree(\circuit_\linput) + 1}\cdot \depth(\circuit_\linput) \geq \cost(\circuit_\linput)$, likewise for $\circuit_\rinput$, and $1\geq 1$.
It is also true that $\depth(\circuit_\linput) \leq \depth(\circuit) - 1$ and $\depth(\circuit_\rinput) \leq \depth(\circuit) - 1$.  

If \circuit.\type $= \circplus$, then $\degree(\circuit) = \max\left(\degree(\circuit_\linput), \degree(\circuit_\rinput)\right)$.  Otherwise \circuit.\type = $\circmult$ and $\degree(\circuit) = \degree(\circuit_\linput) + \degree(\circuit_\rinput) + 1$.  In either case it is true that $\depth(\circuit) = \max\inparen{\depth(\circuit_\linput), \depth(\circuit_\rinput)} + 1$.

If \circuit.\type $= \circmult$, then, by \cref{eq:cost-sampmon}, substituting values, the following should hold,  
\begin{align}
&2\left(\degree(\circuit_\linput) + \degree(\circuit_\rinput) + 2\right) \cdot \left(\max(\depth(\circuit_\linput), \depth(\circuit_\rinput)) + 1\right) + 1 \label{eq:times-lhs}\\
&\qquad\geq 2\left(\degree(\circuit_\linput) + 1\right) \cdot \depth(\circuit_\linput) + 2\left(\degree(\circuit_\rinput) + 1\right)\cdot \depth(\circuit_\rinput) + 3\label{eq:times-middle} \\
&\qquad\geq 1 + \cost(\circuit_\linput) + \cost(\circuit_\rinput) = \cost(\circuit) \label{eq:times-rhs}.
\end{align}

To prove (\ref{eq:times-middle}), first, \cref{eq:times-lhs} expands to,   
\begin{equation}
2\degree(\circuit_\linput)\cdot\depth_{\max} + 2\degree(\circuit_\rinput)\cdot\depth_{\max} + 4\depth_{\max} + 2\degree(\circuit_\linput) +  2\degree(\circuit_\rinput) + 4 + 1\label{eq:times-lhs-expanded}
\end{equation}
where $\depth_{\max}$ is used to denote the maximum depth of the two input subcircuits.  \Cref{eq:times-middle} expands to
\begin{equation}
2\degree(\circuit_\linput)\cdot\depth(\circuit_\linput) + 2\depth(\circuit_\linput) + 2\degree(\circuit_\rinput)\cdot\depth(\circuit_\rinput) + 2\depth(\circuit_\rinput) + 3\label{eq:times-middle-expanded}
\end{equation}

Putting \Cref{eq:times-lhs-expanded} and \Cref{eq:times-middle-expanded} together we get
\begin{align}
&2\degree(\circuit_\linput)\cdot\depth_{\max} + 2\degree(\circuit_\rinput)\cdot\depth_{\max} + 4\depth_{\max} + 2\degree(\circuit_\linput) +  2\degree(\circuit_\rinput) + 5\nonumber\\
&\qquad\geq 2\degree(\circuit_\linput)\cdot\depth(\circuit_\linput) + 2\degree(\circuit_\rinput)\cdot\depth(\circuit_\rinput) + 2\depth(\circuit_\linput) + 2\depth(\circuit_\rinput) + 3.\label{eq:times-lhs-middle}
\end{align}


Now to justify (\ref{eq:times-rhs}) which holds for the following reasons.  First, \cref{eq:times-rhs} 
is the result of \Cref{eq:cost-sampmon} when $\circuit.\type = \circmult$.  \Cref{eq:times-middle} 
is then produced by substituting the upperbound of (\ref{eq:ih-bound-cost}) for each $\cost(\circuit_i)$, trivially establishing the upper bound of (\ref{eq:times-rhs}).  This proves \cref{eq:strict-upper-bound} for the $\circmult$ case.

For the case when \circuit.\type $= \circplus$, substituting values yields
\begin{align}
&2\left(\max(\degree(\circuit_\linput), \degree(\circuit_\rinput)) + 1\right) \cdot \left(\max(\depth(\circuit_\linput), \depth(\circuit_\rinput)) + 1\right) +1\label{eq:plus-lhs-inequality}\\
&\qquad \geq \max\left(2\left(\degree(\circuit_\linput) + 1\right) \cdot \depth(\circuit_\linput) + 1, 2\left(\degree(\circuit_\rinput) + 1\right) \cdot \depth(\circuit_\rinput) +1\right) + 1\label{eq:plus-middle}\\
&\qquad \geq 1 + \max(\cost(\circuit_\linput), \cost(\circuit_\rinput)) = \cost(\circuit)\label{eq:plus-rhs}
\end{align}

To prove  (\ref{eq:plus-middle}), \cref{eq:plus-lhs-inequality} expands to
\begin{equation}
2\degree_{\max}\depth_{\max} + 2\degree_{\max} + 2\depth_{\max} + 2 + 1.\label{eq:plus-lhs-expanded}
\end{equation}

Since $\degree_{\max} \cdot \depth_{\max} \geq \degree(\circuit_i)\cdot \depth(\circuit_i),$ the following upperbounds the expansion of \cref{eq:plus-middle}:
\begin{equation}
2\degree_{\max}\depth_{\max} + 2\depth_{\max} + 2 
\label{eq:plus-middle-expanded}
\end{equation}
Putting it together we obtain the following for (\ref{eq:plus-middle}):
\begin{align}
&2\degree_{\max}\depth_{\max} + 2\degree_{\max} + 2\depth_{\max} + 3\nonumber\\
&\qquad \geq 2\degree_{\max}\depth_{\max} + 2\depth_{\max} + 2, \label{eq:plus-upper-bound-final}
\end{align}
where it can be readily seen that the inequality stands and (\ref{eq:plus-upper-bound-final}) follows.  This proves (\ref{eq:plus-middle}).

Similar to the case of $\circuit.\type = \circmult$, (\ref{eq:plus-rhs}) follows by equations $(\ref{eq:cost-sampmon})$ and $(\ref{eq:ih-bound-cost})$.

This proves (\ref{eq:strict-upper-bound}) as desired. 
\qed
\end{proof}


\subsection{\Cref{lem:ctidb-gamma},~\Cref{lem:val-ub},~\Cref{cor:approx-algo-punchline}, and Proof of~\Cref{cor:approx-algo-punchline-ctidb}} 

\begin{Lemma}
\label{lem:ctidb-gamma}
Given $\raPlus$ query $\query$ and \abbrCTIDB $\pdb$, let \circuit be the circuit computed by $\query\inparen{\tupset}$.  Then, for the reduced \abbrOneBIDB $\pdb'$ there exists an equivalent circuit \circuit' obtained from $\query\inparen{\tupset'}$, such that $\gamma\inparen{\circuit'}\leq 1 - \bound^{-\inparen{k-1}}$ with $\size\inparen{\circuit'} \leq \size\inparen{\circuit} + \bigO{\numvar\bound}$ 
 and $\depth\inparen{\circuit'} = \depth\inparen{\circuit} + \bigO{\log{\bound}}$.
\end{Lemma}

We briefly connect the runtime in \Cref{eq:approx-algo-runtime} to the algorithm outline earlier (where we ignore the dependence on $\multc{\cdot}{\cdot}$, which is needed to handle the cost of arithmetic operations over integers). The $\size(\circuit)$ comes from the time taken to run \onepass once (\onepass essentially computes $\abs{\circuit}(1,\ldots, 1)$ using the natural circuit evaluation algorithm on $\circuit$). We make $\frac{\log{\frac{1}{\conf}}}{\inparen{\error'}^2\cdot(1-\gamma)^2\cdot \prob_0^{2k}}$ many calls to \sampmon (each of which essentially traces $O(k)$ random sink to source paths in $\circuit$ all of which by definition have length at most $\depth(\circuit)$).

Finally, we address the $\multc{\log\left(\abs{\circuit}(1,\ldots, 1)\right)}{\log\left(\size(\circuit)\right)}$ term in the runtime. 
\begin{Lemma}
\label{lem:val-ub}
For any \emph{\abbrOneBIDB} circuit $\circuit$ with $\degree(\circuit)=k$, we have
$\abs{\circuit}(1,\ldots, 1)\le 2^{2^k\cdot \depth(\circuit)}.$
Further, if $\circuit$ is a tree, then we have $\abs{\circuit}(1,\ldots, 1)\le  \size(\circuit)^{O(k)}.$
\end{Lemma}

Note that the above implies that with the assumption $\prob_0>0$ and $\gamma<1$ are absolute constants from \Cref{cor:approx-algo-const-p}, then the runtime there simplifies to $O_k\left(\frac 1{\inparen{\error'}^2}\cdot\size(\circuit)^2\cdot \log{\frac{1}{\conf}}\right)$ for general circuits $\circuit$. If $\circuit$ is a tree, then the runtime simplifies to $O_k\left(\frac 1{\inparen{\error'}^2}\cdot\size(\circuit)\cdot \log{\frac{1}{\conf}}\right)$, which then answers \Cref{prob:intro-stmt} with yes for such circuits.

Finally, note that by \Cref{prop:circuit-depth} and \Cref{lem:circ-model-runtime} for any $\raPlus$ query $\query$, there exists a circuit $\circuit^*$ for $\apolyqdt$ such that $\depth(\circuit^*)\le O_{|Q|}(\log{n})$ and $\size(\circuit)\le O_k\inparen{\qruntime{\query, \tupset, \bound}}$. Using this along with \Cref{lem:val-ub}, \Cref{cor:approx-algo-const-p} and the fact that $n\le \qruntime{\query, \tupset, \bound}$, we have the following corollary:
\begin{Corollary}
\label{cor:approx-algo-punchline}
Let $\query$ be an $\raPlus$ query and $\pdb$ be a \emph{\abbrOneBIDB} with $p_0>0$ and $\gamma<1$, where $p_0,\gamma$ as in \Cref{cor:approx-algo-const-p}, are absolute constants. Let $\poly(\vct{X})=\apolyqdt$ for any result tuple $\tup$ with $\deg(\poly)=k$. Then one can compute an approximation satisfying \Cref{eq:approx-algo-bound-main} in time $O_{k,|Q|,\error',\conf}\inparen{\qruntime{\optquery{\query}, \tupset, \bound}}$ (given $\query,\tupset$ and $p_i$ for each $i\in [n]$ that defines $\pd$).
\end{Corollary}

\subsection{Proof of~\Cref{lem:ctidb-gamma}}

\begin{proof}
The circuit \circuit' is built from \circuit in the following manner.  For each input gate $\gate_i$ with $\gate_i.\val = X_\tup$, replace $\gate_i$ with the circuit \subcircuit encoding the sum $\sum_{j = 1}^\bound j\cdot X_{\tup, j}$.  We argue that \circuit' is a valid circuit by the following facts.  Let $\pdb = \inparen{\worlds, \bpd}$ be the original \abbrCTIDB \circuit was generated from.  Then, by~\Cref{prop:ctidb-reduct} there exists a \abbrOneBIDB $\pdb' = \inparen{\onebidbworlds{\tupset'}, \bpd'}$, with $\tupset' = \inset{\intuple{\tup, j}~|~\tup\in\tupset, j\in\pbox{\bound}}$, from which the conversion from \circuit to \circuit' follows.  Both $\polyf\inparen{\circuit}$ and $\polyf\inparen{\circuit'}$ have the same expected multiplicity since (by~\Cref{prop:ctidb-reduct}) the distributions $\bpd$ and $\bpd'$ are equivalent and $\sum_{j=1}^\bound j\cdot\worldvec'_{\tup, j} = \worldvec_\tup$ for $\worldvec'\in\inset{0, 1}^{\bound\numvar}$ and $\worldvec\in\worlds$ such that $\worldvec_\tup\equiv\worldvec'_\tup$.  Finally, note that because there exists a (sub) circuit encoding $\sum_{j = 1}^\bound j\cdot X_{\tup, j}$ that is a \emph{balanced} binary tree, the above conversion implies the claimed size and depth bounds of the lemma.

Next we argue the claim on $\gamma\inparen{\circuit'}$.  Consider the list of expanded monomials $\expansion{\circuit}$ for \abbrCTIDB circuit \circuit.  Let 
 $\encMon = X_{\tup_1}^{d_1}\cdots X_{\tup_\ell}^{d_\ell}$ be an arbitrary monomial with $\ell$ variables and let (abusing notation) $\encMon' = \inparen{\sum_{j = 1}^{\bound}j\cdot X_{\tup_1, j}}^{d_1}\cdots\inparen{\sum_{j = 1}^{\bound}j\cdot X_{\tup_\ell, j}}^{d_\ell}$.  Then, for $f_\ell = \sum_{i = 1}^\ell d_i$, $\encMon$ induces the set of monomials $\inset{\prod_{i = 1}^{f_\ell} j_i\cdot X_{\tup_i, j_i}^{d_i}}_{j_i\in\pbox{\bound}}$ in the pure expansion of $\encMon'$.  
    Recall that a cancellation occurs in $\encMon'$ when there exists $\tup_{i, j}\neq\tup_{i, j'}$ in the same block $\block$ where variables $X_{\tup_i, j}, X_{\tup_i, j'}$ are in the set of variables $\monom_i'$ of $\monom_{\vari{m}_\vari{i}}\in\encMon'$.  Observe that cancellations can only occur for each $X_{\tup}^{d_\tup}\in \encMon$, where the expansion $\inparen{\sum_{j = 1}^\bound j\cdot X_{\tup, j}}^{d_\tup}$ represents the monomial $X_\tup^{d_\tup}$ in $\tupset'$.  Consider the number of cancellations for $\inparen{\sum_{j = 1}^\bound j\cdot X_{\tup, j}}^{d_t}$.  Then $\gamma \leq 1 - \bound^{-\inparen{d_\tup - 1}}$, since 
 for each element in the set of cross products $\inset{\bigtimes_{i\in\pbox{d_\tup}, j_i\in\pbox{\bound}}X_{\tup, j_i}}$ there are \emph{exactly} $\bound$ surviving elements with $j_1=\cdots=j_{d_\tup}=j$, i.e. $X_{t,j}^{d_\tup}$ for each $j\in\pbox{\bound}$.  The rest of the $\bound^{d_\tup}-c$ cross terms cancel.  Regarding all of $\encMon'$, it is the case that the proportion of non-cancellations for each $\inparen{\sum_{j = 1}^{\bound}j\cdot X_{\tup_i, j }}^{d_i}\in\encMon'$ multiply because non-cancelling terms for $\inparen{\sum_{j = 1}^{\bound}j\cdot X_{\tup_i, j}}^{d_i}$ can only be joined with non-cancelling terms of $\inparen{\sum_{j=1}^{\bound}X_{\tup_{i'}, j}}^{d_{i'}}\in\encMon'$ for $\tup\neq\tup'$.  This then yields the fraction of cancelled monomials $\gamma\le 1 - \prod_{i = 1}^{\ell}\bound^{-\inparen{d_i - 1}} \leq 1 - \bound^{-\inparen{k - 1}}$ where the inequalities take into account the fact that $f_\ell \leq k$.

Since this is true for arbitrary \monom, the bound follows for $\polyf\inparen{\circuit'}$.
\end{proof}
\qed

\subsection{Proof of \Cref{lem:val-ub}}\label{susec:proof-val-up}
\label{app:proof-lem-val-ub}

We will prove \Cref{lem:val-ub} by considering the two cases separately. We start by considering the case when $\circuit$ is a tree:
\begin{Lemma}
\label{lem:C-ub-tree}
Let $\circuit$ be a tree (i.e. the sub-circuits corresponding to two children of a node in $\circuit$ are completely disjoint). Then we have
\[\abs{\circuit}(1,\dots,1)\le \left(\size(\circuit)\right)^{\degree(\circuit)+1}.\]
\end{Lemma}
\begin{proof}[Proof of \Cref{lem:C-ub-tree}]
For notational simplicity define $N=\size(\circuit)$ and $k=\degree(\circuit)$.
We use induction on $\depth(\circuit)$ to show that $\abs{\circuit}(1,\ldots, 1) \leq N^{k+1 }$.
For the base case, we have that \depth(\circuit) $= 0$, and there can only be one node which must contain a coefficient or constant.  In this case, $\abs{\circuit}(1,\ldots, 1) = 1$, and \size(\circuit) $= 1$, and by \Cref{def:degree} it is the case that $0 \leq k = \degree\inparen{\circuit} \leq 1$, and it is true that $\abs{\circuit}(1,\ldots, 1) = 1 \leq N^{k+1} = 1^{k + 1} = 1$ for $k \in \inset{0, 1}$.

Assume for $\ell > 0$ an arbitrary circuit \circuit of $\depth(\circuit) \leq \ell$ that it is true that $\abs{\circuit}(1,\ldots, 1) \leq N^{k+1 }$.

For the inductive step we consider a circuit \circuit such that $\depth(\circuit) = \ell + 1$.  The sink can only be either a $\circmult$ or $\circplus$ gate.  Let $k_\linput, k_\rinput$ denote \degree($\circuit_\linput$) and \degree($\circuit_\rinput$) respectively.  Consider when sink node is $\circmult$.
 Then note that
\begin{align}
\abs{\circuit}(1,\ldots, 1) &= \abs{\circuit_\linput}(1,\ldots, 1)\cdot \abs{\circuit_\rinput}(1,\ldots, 1) \nonumber\\
&\leq (N-1)^{k_\linput+1} \cdot (N - 1)^{k_\rinput+1}\nonumber\\
 &= (N-1)^{k+1}\label{eq:sumcoeff-times-upper}\\
 &\leq N^{k + 1}.\nonumber
\end{align}
In the above the first inequality follows from the inductive hypothesis (and the fact that the size of either subtree is at most $N-1$) and \Cref{eq:sumcoeff-times-upper} follows by \cref{def:degree} which states that for $k = \degree(\circuit)$ we have $k=k_\linput+k_\rinput+1$.

For the case when the sink gate is a $\circplus$ gate, then for $N_\linput = \size(\circuit_\linput)$ and $N_\rinput = \size(\circuit_\rinput)$ we have
\begin{align}
\abs{\circuit}(1,\ldots, 1) &= \abs{\circuit_\linput}(1,\ldots, 1) \circplus \abs{\circuit_\rinput}(1,\ldots, 1) \nonumber\\
&\leq
N_\linput^{k+1} + N_\rinput^{k+1}\nonumber\\
&\leq (N-1)^{k+1 } \label{eq:sumcoeff-plus-upper}\\
&\leq N^{k+1}.\nonumber
\end{align}
In the above, the first inequality follows from the inductive hypothes and \cref{def:degree} (which implies the fact that $k_\linput,k_\rinput\le k$).  Note that the RHS of this inequality is maximized when the base and exponent of one of the terms is maximized.  The second inequality follows from this fact as well as the fact that since $\circuit$ is a tree we have $N_\linput+N_\rinput=N-1$ and, lastly, the fact that $k\ge 0$. This completes the proof.
\end{proof}

The upper bound in \Cref{lem:val-ub} for the general case is a simple variant of the above proof (but we present a proof sketch of the bound below for completeness):
\begin{Lemma}
\label{lem:C-ub-gen}
Let $\circuit$ be a (general) circuit. 
Then we have
\[\abs{\circuit}(1,\dots,1)\le 2^{2^{\degree(\circuit)}\cdot \depth(\circuit)}.\]
\end{Lemma}
\begin{proof}[Proof Sketch of \Cref{lem:C-ub-gen}]
We use the same notation as in the proof of \Cref{lem:C-ub-tree} and further define $d=\depth(\circuit)$. We will prove by induction on $\depth(\circuit)$ that $\abs{\circuit}(1,\ldots, 1) \leq 2^{2^k\cdot d }$. The base case argument is similar to that in the proof of \Cref{lem:C-ub-tree}. In the inductive case we have that $d_\linput,d_\rinput\le d-1$.

For the case when the sink node is $\times$, we get that
\begin{align*}
\abs{\circuit}(1,\ldots, 1) &= \abs{\circuit_\linput}(1,\ldots, 1)\circmult \abs{\circuit_\rinput}(1,\ldots, 1) \\
&\leq {2^{2^{k_\linput}\cdot d_\linput}} \circmult {2^{2^{k_\rinput}\cdot d_\rinput}}\\
 &\leq 2^{2\cdot 2^{k-1}\cdot (d-1)}\\
 &\leq 2^{2^k d}.
\end{align*}
In the above the first inequality follows from inductive hypothesis while the second inequality follows from the fact that $k_\linput,k_\rinput\le k-1$ and $d_\linput, d_\rinput\le d-1$, where we substitute the upperbound into every respective term.

Now consider the case when the sink node is $+$, we get that
\begin{align*}
\abs{\circuit}(1,\ldots, 1) &= \abs{\circuit_\linput}(1,\ldots, 1) \circplus \abs{\circuit_\rinput}(1,\ldots, 1) \\
&\leq 2^{2^{k_\linput}\cdot d_\linput} + 2^{2^{k_\rinput}\cdot d_\rinput}\\
&\leq 2\cdot {2^{2^k(d-1)} } \\
&\leq 2^{2^kd}.
\end{align*}
In the above the first inequality follows from the inductive hypothesis while the second inequality follows from the facts that $k_\linput,k_\rinput\le k$ and $d_\linput,d_\rinput\le d-1$. The final inequality follows from the fact that $k\ge 0$.
\qed
\end{proof}

\begin{proof}[Proof of~\Cref{cor:approx-algo-punchline-ctidb}]
By~\Cref{lem:ctidb-gamma} and~\Cref{cor:approx-algo-punchline}, the proof follows.
\end{proof}
\qed

\subsection{Experimental Results}\label{app:subsec:experiment}

Recall that by definition of $\abbrBIDB$, a query result cannot be derived by a self-join between non-identical tuples belonging to the same block.  Note, that by \Cref{cor:approx-algo-const-p}, $\gamma$ must be a constant in order for \Cref{alg:mon-sam} to acheive linear time.  We would like to determine experimentally whether queries over $\abbrBIDB$ instances in practice generate a constant number of cancellations or not.  Such an experiment would ideally use a database instance with queries both considered to be typical representations of what is seen in practice. 

We ran our experiments using Windows 10 WSL Operating System with an Intel Core i7 2.40GHz processor and 16GB RAM.  All experiments used the PostgreSQL 13.0 database system.

For the data we used the MayBMS data generator~\cite{pdbench} tool to randomly generate uncertain versions of TPCH tables.  The queries computed over the database instance are $\query_1$, $\query_2$, and $\query_3$ from~\cite{4497507}, all of which are modified versions of TPC-H queries $\query_3$, $\query_6$, and $\query_7$ where all aggregations have been dropped.

As written, the queries disallow $\abbrBIDB$ cross terms.  We first ran all queries, noting the result size for each.  Next the queries were rewritten so as not to filter out the cross terms.  The comparison of the sizes of both result sets should then suggest in one way or another whether or not there exist many cross terms in practice.  As seen, the experimental query results contain little to no cancelling terms.  \Cref{fig:experiment-bidb-cancel} shows the result sizes of the queries, where column CF is the result size when all cross terms are filtered out, column CI shows the number of output tuples when the cancelled tuples are included in the result,  and the last column is the value of $\gamma$.  The experiments show $\gamma$ to be in a range between $[0, 0.1]\%$, indicating that only a negligible or constant (compare the result sizes of $\query_1 < \query_2$ and their respective $\gamma$ values) amount of tuples are cancelled in practice when running queries over a typical \abbrBIDB instance.  Interestingly, only one of the three queries had tuples that violated the \abbrBIDB constraint.

To conclude, the results in \Cref{fig:experiment-bidb-cancel} show experimentally that $\gamma$ is negligible in practice for BIDB queries.  We also observe that (i) tuple presence is independent across blocks, so the corresponding probabilities (and hence $\prob_0$) are independent of the number of blocks, and (ii) \bis model uncertain attributes, so block size (and hence $\gamma$) is a function of the ``messiness'' of a dataset, rather than its size.
Thus, we expect \Cref{cor:approx-algo-const-p} to hold in general.

\begin{figure}[ht]
		\begin{tabular}{ c | c c c}\label{tbl:cancel}
			Query & CF & CI & $\gamma$\\
			\hline
			 $\query_1$ & $46,714$ & $46,768$ & $0.1\%$\\
			 $\query_2$ & $179.917$ & $179,917$ & $0\%$\\
			 $\query_3$ & $11,535$ & $11,535$ & $0\%$\\
		\end{tabular}
	\caption{Number of Cancellations for Queries Over $\abbrBIDB$.}
	\label{fig:experiment-bidb-cancel}
\end{figure}

\section{Circuits}\label{app:sec-cicuits}
\subsection{Representing Polynomials with Circuits}\label{app:subsec-rep-poly-lin-circ}
\subsubsection{Circuits for query plans}
\label{sec:circuits-formal}
We now formalize circuits and the construction of circuits for $\raPlus$ queries.
As mentioned earlier, we represent lineage polynomials as arithmetic circuits over $\mathbb N$-valued variables with $+$, $\times$.
A circuit for query $Q$ and \abbrNXPDB $\pxdb$ \footnote{For background on \abbrNXPDB\xplural, see~\Cref{app:subsec:background-nxdbs}} is a directed acyclic graph $\tuple{V_{Q,\pxdb}, E_{Q,\pxdb}, \phi_{Q,\pxdb}, \ell_{Q,\pxdb}}$ with vertices $V_{Q,\pxdb}$ and directed edges $E_{Q,\pxdb} \subset {V_{Q,\pxdb}}^2$.
The sink function $\phi_{Q,\pxdb} : \udom^n \rightarrow V_{Q,\pxdb}$ is a partial function that maps the tuples of the $n$-ary relation $Q(\pxdb)$ to vertices.
We require that $\phi_{Q,\pxdb}$'s range be limited to sink vertices (i.e., vertices with out-degree 0).
A function $\ell_{Q,\pxdb} : V_{Q,\pxdb} \rightarrow \{\;+,\times\;\}\cup \mathbb N \cup \vct X$ assigns a label to each node: Source nodes (i.e., vertices with in-degree 0) are labeled with constants or variables (i.e., $\mathbb N \cup \vct X$), while the remaining nodes are labeled with the symbol $+$ or $\times$.
We require that vertices have an in-degree of at most two.
Note that we can construct circuits for \bis in time linear in the time required for deterministic query processing over a possible world of the \bi under the aforementioned assumption that $\abs{\pxdb} \leq c \cdot \abs{\db}$.


\subsection{Modeling Circuit Construction}

\newcommand{\bagdbof}{\textsc{bag}(\pxdb)}

We now connect the size of a circuit (where the size of a circuit is the number of vertices in the corresponding DAG) 
for a given $\raPlus$ query $Q$ and \abbrNXPDB $\pxdb$ to
the runtime $\qruntime{\query,\tupset, \bound}$ of the PDB's \dbbaseName $\tupset$.
\AH{@atri: do we use $\tupset$ or $\gentupset$ here?}
We do this formally by showing that the size of the circuit is asymptotically no worse than the corresponding runtime of a large class of deterministic query processing algorithms.

\newcommand{\getpoly}[1]{\textbf{lin}\inparen{#1}}
Each vertex $v \in V_{Q,\pxdb}$ in the arithmetic circuit for

\[\tuple{V_{Q,\pxdb}, E_{Q,\pxdb}, \phi_{Q,\pxdb}, \ell_{Q,\pxdb}}\]

encodes a polynomial, realized as

\[\getpoly{v} = \begin{cases}
\sum_{v' : (v',v) \in E_{Q,\pxdb}} \getpoly{v'} & \textbf{if } \ell(v) = +\\
\prod_{v' : (v',v) \in E_{Q,\pxdb}} \getpoly{v'} & \textbf{if } \ell(v) = \times\\
\ell(v) & \textbf{otherwise}
\end{cases}\]

We define the circuit for a $\raPlus$ query $\query$ recursively by cases as follows.  In each case, let $\tuple{V_{Q_i,\pxdb}, E_{Q_i,\pxdb}, \phi_{Q_{i},\pxdb}, \ell_{Q_i,\pxdb}}$ denote the circuit for subquery $Q_i$.  We implicitly include in all circuits a global zero node $v_0$ s.t., $\ell_{Q, \pxdb}(v_0) = 0$ for any $Q,\pxdb$.

\begin{algorithm}
\caption{\lincirc$(\query, \tupset, E, V, \ell)$}
\label{alg:lc}
  \begin{algorithmic}[1]
  \Require $\query$: query
  \Require $\tupset$: a \dbbaseName
  \Require $E, V, \ell$: accumulators for the edge list, vertex list, and vertex label list.
  \Ensure $\circuit = \tuple{V, E, \phi, \ell}$: a circuit encoding the lineage of each tuple in $\query(\tupset)$
  \If{$\query$ is $\rel$} \Comment{\textbf{Case 1}: $\query$ is a relation atom}
    \For{$t \in \tupset.\rel$}
      \State $V \leftarrow V \cup \{v_t\}$; $\ell \leftarrow \ell \cup \{\inparen{v_t, \rel\inparen{\tup}}\}$  \Comment{Allocate a fresh node $v_t$; note that when $\rel\inparen{\tup} = \bound\cdot X_\tup$ for $\bound > 1$, we assume the algorithm generates a $3$ node circuit encoding the multiplcation of $\bound\cdot X_\tup$, adding the new vertices, edges, and vertice/label pairs to their respective sets.}
      \State $\phi(t) \gets v_t$
    \EndFor
    \State\Return $\tuple{V, E, \phi, \ell}$
  \ElsIf{$\query$ is $\sigma_\theta(\query_1)$} \Comment{\textbf{Case 2}: $\query$ is a Selection}
    \State $\tuple{V, E, \phi', \ell} \gets \lincirc(\query_1, \tupset, V, E, \ell)$
    \For{$t \in \domain(\phi')$}
      \State \textbf{if }$\theta(t)$
              \textbf{ then } $\phi(t) \gets \phi'(t)$
              \textbf{ else } $\phi(t) \gets v_0$
    \EndFor
    \State\Return $\tuple{V, E, \phi, \ell}$
  \ElsIf{$\query$ is $\pi_{A}(\query_1)$} \Comment{\textbf{Case 3}: $\query$ is a Projection}
    \State $\tuple{V, E, \phi', \ell} \gets \lincirc(\query_1, \tupset, V, E, \ell)$
    \For{$t \in \pi_{A}(\query_1(\tupset))$}
      \State $V \leftarrow V \cup \{v_t\}$; $\ell \leftarrow \ell \cup \{(v_t, +)\}$\Comment{Allocate a fresh node $v_t$}
      \State $\phi(t) \leftarrow v_t$
    \EndFor
    \For{$t \in \query_1(\tupset)$}
      \State $E \leftarrow E \cup \{(\phi'(t), \phi(\pi_{A}t))\}$
    \EndFor
    \State Correct nodes with in-degrees $>2$ by appending an equivalent fan-in two tree instead
    \State\Return $\tuple{V, E, \phi, \ell}$
  \ElsIf{$\query$ is $\query_1 \cup \query_2$} \Comment{\textbf{Case 4}: $\query$ is a Bag Union}
    \State $\tuple{V', E', \phi_1, \ell'} \gets \lincirc(\query_1, \tupset, V, E, \ell)$
    \State $\tuple{V, E, \phi_2, \ell} \gets \lincirc(\query_2, \tupset, V', E', \ell')$
    \State $\phi \gets \phi_1 \cup \phi_2$\label{alg:lincirc-union-phi}
    \For{$t \in \domain(\phi_1) \cap \domain(\phi_2)$}\label{alg:lincirc-union-intersection}
      \State $V \leftarrow V \cup \{v_t\}$; $\ell \leftarrow \ell \cup \{(v_t, +)\}$\label{alg:lincirc-union-intersection-one} \Comment{Allocate a fresh node $v_t$}
      \State $\phi(t) \gets v_t$\label{alg:lincirc-union-intersection-two}
      \State $E \leftarrow E \cup \{(\phi_1(t), v_t), (\phi_2(t), v_t)\}$\label{alg:lincirc-union-intersection-three}
    \EndFor
    \State\Return $\tuple{V, E, \phi, \ell}$
  \ElsIf{$\query$ is $\query_1 \bowtie \ldots \bowtie \query_m$} \Comment{\textbf{Case 5}: $\query$ is a $m$-ary Join}
    \For{$i \in [m]$}
      \State $\tuple{V, E, \phi_i, \ell} \gets \lincirc(\query_i, \tupset, V, E, \ell)$
    \EndFor
    \For{$t \in \domain(\phi_1) \bowtie \ldots \bowtie \domain(\phi_m)$}
      \State $V \leftarrow V \cup \{v_t\}$; $\ell \leftarrow \ell \cup \{(v_t, \times)\}$ \Comment{Allocate a fresh node $v_t$}
      \State $\phi(t) \gets v_t$
      \State $E \leftarrow E \cup \comprehension{(\phi_i(\pi_{sch(\query_i(\tupset))}(t)), v_t)}{i \in [m]}$
    \EndFor
    \State Correct nodes with in-degrees $>2$ by appending an equivalent fan-in two tree instead
    \State\Return $\tuple{V, E, \phi, \ell}$

  \EndIf

  \end{algorithmic}
\end{algorithm}

\Cref{alg:lc} defines how the circuit for a query result is constructed.  Denote the set of active output tuples as $\domain\inparen{\phi}$.  We quickly review the number of vertices emitted in each case.

\caseheading{Base Relation}
This circuit has $\abs{\tupset.\rel}$ vertices.

\caseheading{Selection}
If we assume dead sinks are iteratively garbage collected,
this circuit has at most $|V_{Q_1,\pxdb}|$ vertices.

\caseheading{Projection}
This formulation will produce vertices with an in-degree greater than two, a problem that we correct by replacing every vertex with an in-degree over two by an equivalent fan-in two tree.  The resulting structure has at most $|{Q_1}|-1$ new vertices.
The corrected circuit thus has at most $|V_{Q_1,\pxdb}|+|{Q_1}|$ vertices.

\caseheading{Union}
This circuit has $|V_{Q_1,\pxdb}|+|V_{Q_2,\pxdb}|+|{Q_1} \cap {Q_2}|$ vertices.

\caseheading{$k$-ary Join}
As in projection, newly created vertices will have an in-degree of $k$, and a fan-in two tree is required.
There are $|{Q_1} \bowtie \ldots \bowtie {Q_k}|$ such vertices, so the corrected circuit has $|V_{Q_1,\pxdb}|+\ldots+|V_{Q_k,\pxdb}|+(k-1)|{Q_1} \bowtie \ldots \bowtie {Q_k}|$ vertices.

\subsubsection{Bounding circuit depth}
\label{sec:circuit-depth}

We first show that the depth of the circuit (\depth; \Cref{def:size-depth}) is bounded by the size of the query.  Denote by $|\query|$ the number of relational operators in query $\query$, which recall we assume is a constant.

\begin{Proposition}[Circuit depth is bounded]
\label{prop:circuit-depth}
Let $\query$ be a relational query and $\tupset$ be a \dbbaseName with $n$ tuples.  There exists a (lineage) circuit $\circuit^*$ encoding the lineage of all tuples $\tup \in \query(\tupset)$ for which
$\depth(\circuit^*) \leq O(k|\query|\log(n))$.
\end{Proposition}

\begin{proof}
We show that the bound of \Cref{prop:circuit-depth} holds for the circuit constructed by \Cref{alg:lc}.
First, observe that \Cref{alg:lc} is (recursively) invoked exactly once for every relational operator or base relation in $\query$; It thus suffices to show that a call to \Cref{alg:lc} adds at most $O_k(\log(n))$ to the depth of a circuit produced by any recursive invocation.
Second, observe that modulo the logarithmic fan-in of the projection and join cases, the depth of the output is at most one greater than the depth of any input (or at most 1 in the base case of relation atoms).
For the join case, the number of in-edges can be no greater than the join width, which itself is bounded by $k$.  The depth thus increases by at most a constant factor of $\lceil \log(k) \rceil = O_k(1)$.
For the projection case, observe that the fan-in is bounded by $|\query'(\dbbase)|$, which is in turn bounded by $n^k$.  The depth increase for any projection node is thus at most $\lceil \log(n^k)\rceil = O(k\log(n))$, as desired. 
\qed
\end{proof}

\subsubsection{Circuit size vs. runtime}
\label{sec:circuit-runtime}

\begin{Lemma}\label{lem:circ-model-runtime}
\label{lem:circuits-model-runtime}
Given a \abbrNXPDB $\pxdb$ with \dbbaseName $\tupset$, and an $\raPlus$ query  $Q$, the runtime of $Q$ over $\tupset$ has the same or greater complexity as the size of the lineage of $Q(\pxdb)$.  That is, we have $\abs{V_{Q,\pxdb}} \leq k\qruntime{\query, \tupset, \bound}+1$, where $k\ge 1$ is the maximal degree of  any  polynomial in $Q(\pxdb)$.
\end{Lemma}

\begin{proof}
We prove by induction that $\abs{V_{Q,\pxdb} \setminus \{v_0\}} \leq k\qruntime{\query, \tupset, \bound}$.  For clarity, we implicitly exclude $v_0$ in the proof below.

The base case is a base relation: $Q = R$ and is trivially true since $|V_{R,\pxdb}| = |\tupset.R|=\qruntime{\rel, \tupset, \bound}$ (note that here the degree $k=1$).
For the inductive step, we assume that we have circuits for subqueries $Q_1, \ldots, Q_m$ such that $|V_{Q_i,\pxdb}| \leq k_i\qruntime{\query_i,\tupset, \bound}$ where $k_i$ is the degree of $Q_i$.

\caseheading{Selection}
Assume that $Q = \sigma_\theta(Q_1)$.
In the circuit for $Q$, $|V_{Q,\pxdb}| = |V_{Q_1,\tupset}|$ vertices, so from the inductive assumption and $\qruntime{\query,\tupset, \bound} = \qruntime{\query_1,\tupset, \bound}$ by definition, we have $|V_{Q,\pxdb}| \leq k \qruntime{\query,\tupset, \bound} $.

\caseheading{Projection}
Assume that $Q = \pi_{\vct A}(Q_1)$.
The circuit for $Q$ has at most $|V_{Q_1,\pxdb}|+|{Q_1}|$ vertices.
\begin{align*}
|V_{Q,\pxdb}| & \leq |V_{Q_1,\pxdb}| + |Q_1|\\
\intertext{(From the inductive assumption)}
& \leq k\qruntime{\query_1,\tupset, \bound} + \abs{Q_1}\\
\intertext{(By definition  of $\qruntime{\query,\tupset, \bound}$)}
& \le k\qruntime{\query,\tupset, \bound}.
\end{align*}
\caseheading{Union}
Assume that $Q = Q_1 \cup Q_2$.
The circuit for $Q$ has $|V_{Q_1,\pxdb}|+|V_{Q_2,\pxdb}|+|{Q_1} \cap {Q_2}|$ vertices.
\begin{align*}
|V_{Q,\pxdb}| & \leq |V_{Q_1,\pxdb}|+|V_{Q_2,\pxdb}|+|{Q_1}|+|{Q_2}|\\
\intertext{(From the inductive assumption)}
& \leq k(\qruntime{\query_1,\tupset, \bound} + \qruntime{\query_2,\tupset, \bound}) + (|Q_1| + |Q_2|)
\intertext{(By definition of $\qruntime{\query, \tupset, \bound}$)}
& \leq k(\qruntime{\query,\tupset, \bound}).
\end{align*}

\caseheading{$m$-ary Join}
Assume that $Q = Q_1 \bowtie \ldots \bowtie Q_m$. Note that $k=\sum_{i=1}^m k_i\ge m$.
The circuit for $Q$ has $|V_{Q_1,\pxdb}|+\ldots+|V_{Q_k,\pxdb}|+(m-1)|{Q_1} \bowtie \ldots \bowtie {Q_k}|$ vertices.
\begin{align*}
|V_{Q,\pxdb}| & = |V_{Q_1,\pxdb}|+\ldots+|V_{Q_k,\pxdb}|+(m-1)|{Q_1} \bowtie \ldots \bowtie {Q_k}|\\
\intertext{From the inductive assumption and noting $\forall i: k_i \leq k$ and $m\le k$}
& \leq k\qruntime{\query_1,\tupset, \bound}+\ldots+k\qruntime{\query_k,\tupset, \bound}+\\
&\;\;\; (m-1)|{Q_1} \bowtie \ldots \bowtie {Q_m}|\\
& \leq k\left(\qruntime{\query_1, 	\tupset, \bound}+\ldots+\qruntime{\query_1, 	\tupset, \bound}+\right.\\
&\;\;\;\left.|{Q_1} \bowtie \ldots \bowtie {Q_m}|\right)\\
\intertext{(By definition of $\qruntime{\query,\tupset, \bound}$ and assumption on $\jointime{\cdot}$)}
& \le k\qruntime{\query,\tupset, \bound}.
\end{align*}

The property holds for all recursive queries, and the proof holds.
\qed
\end{proof}

\subsubsection{Runtime of \lincirc}
\label{sec:lc-runtime}

We next need to show that we can construct the circuit in time linear in the deterministic runtime.
\begin{Lemma}\label{lem:tlc-is-the-same-as-det}
Given a query $\query$ over a \dbbaseName $\tupset$ and the $\circuit^*$ output by \Cref{alg:lc}, the runtime $\timeOf{\lincirc}(\query,\tupset,\circuit^*) \le O(\qruntime{\query, \tupset, \bound})$.
\end{Lemma}
\begin{proof}
By analysis of \Cref{alg:lc}, invoked as $\circuit^*\gets\lincirc(\query, \tupset, \emptyset, \{v_0\}, \{(v_0, 0)\})$.

We assume that the vertex list $V$, edge list $E$, and vertex label list $\ell$ are mutable accumulators with $O(1)$ ammortized append.
We assume that the tuple to sink mapping $\phi$ is a linked hashmap, with $O(1)$ insertions and retrievals, and $O(n)$ iteration over the domain of keys.
We assume that the n-ary join $\domain(\phi_1) \bowtie \ldots \bowtie\domain(\phi_n)$ can be computed in time $\jointime{\domain(\phi_1), \ldots, \domain(\phi_n)}$ (\Cref{def:join-cost}) and that an intersection $\domain(\phi_1) \cap \domain(\phi_2)$ can be computed in time $O(|\domain(\phi_1)| + |\domain(\phi_2)|)$ (e.g., with a hash table).

Before proving our runtime bound, we first observe that $\qruntime{\query, \tupset, \bound} \geq \Omega(|\query(\db)|)$.
This is true by construction for the relation, projection, and union cases, by \Cref{def:join-cost} for joins, and by the observation that $|\sigma(R)| \leq |R|$.

We show that $\qruntime{\query, \tupset, \bound}$ is an upper-bound for the runtime of \Cref{alg:lc} by recursion.
The base case of a relation atom requires only an $O(|\tupset.R|)$ iteration over the source tuples.
For the remaining cases, we make the recursive assumption that for every subquery $\query'$, it holds that $O(\qruntime{\query', \tupset, \bound})$ bounds the runtime of \Cref{alg:lc}.

\caseheading{Selection}
Selection requires a recursive call to \Cref{alg:lc}, which by the recursive assumption is bounded by $O(\qruntime{\query', \tupset, \bound})$.
\Cref{alg:lc} requires a loop over every element of $\query'(\tupset)$.
By the observation above that $\qruntime{\query, \db, \bound} \geq \Omega(|\query(\db)|)$, this iteration is also bounded by $O(\qruntime{\query', \tupset, \bound})$.

\caseheading{Projection}
Projection requires a recursive call to \Cref{alg:lc}, which by the recursive assumption is bounded by $O(\qruntime{\query', \tupset})$, which in turn is a term in $\qruntime{\pi_{A}\query', \tupset, \bound}$.
What remains is an iteration over $\pi_{A}(\query(\tupset))$ (lines 13--16), an iteration over $\query'(\tupset)$ (lines 17--19), and the construction of a fan-in tree (line 20).
The first iteration is $O(|\query(\tupset)|) \leq O(\qruntime{\query, \tupset, \bound})$.
The second iteration and the construction of the bounded fan-in tree are both $O(|\query'(\tupset)|) \leq O(\qruntime{\query', \tupset}) \leq O(\qruntime{\query, \tupset, \bound}) $, by the the observation above that $\qruntime{\query, \db, \bound} \geq \Omega(|\query(\db)|)$.

\caseheading{Bag Union}
As above, the recursive calls explicitly correspond to terms in the expansion of $\qruntime{\query_1 \cup \query_2, \tupset, \bound}$.
Initializing $\phi$ (\Cref{alg:lincirc-union-phi}) can be accomplished in $O(\domain(\phi_1) + \domain(\phi_2)) = O(|\query_1(\tupset)| + |\query_2(\tupset)|) \leq O(\qruntime{\query_1, \tupset} + \qruntime{\query_2, \tupset, \bound})$.
The remainder requires computing $\query_1 \cap \query_2$ (\Cref{alg:lincirc-union-intersection}) and iterating over it (\Crefrange{alg:lincirc-union-intersection-one}{alg:lincirc-union-intersection-three}), which is $O(|\query_1| + |\query_2|)$ as noted above --- this directly corresponds to terms in $\qruntime{\query_1 \cup \query_2, \tupset, \bound}$.

\caseheading{$m$-ary Join}
As in the prior cases, recursive calls explicitly correspond to terms in our target runtime.
The remaining logic involves (i) computing $\domain(\phi_1) \bowtie \ldots \bowtie \domain(\phi_m)$, (ii) iterating over the results, and (iii) creating a fan-in tree.
Respectively, these are: \\
~(i)~$\jointime{\domain(\phi_1), \ldots, \domain(\phi_m)}$\\
~(ii)~$O(|\query_1(\tupset) \bowtie \ldots \bowtie \query_m(\tupset)|) \leq O(\jointime{\domain(\phi_1), \ldots, \domain(\phi_m)})$ (\Cref{def:join-cost})\\
~(iii)~$O(m|\query_1(\tupset) \bowtie \ldots \bowtie \query_m(\tupset)|)$ (as (ii), noting that $m \leq k = O(1)$)
\qed
\end{proof}

\section{Higher Moments}
\label{sec:momemts}
We make a simple observation to conclude the presentation of our results.
So far we have only focused on the expectation of $\poly$.
In addition, we could e.g. prove bounds of the probability of a tuple's multiplicity being at least $1$.
Progress can be made on this as follows:
For any positive integer $m$ we can compute the $m$-th moment of the multiplicities, allowing us to e.g. use the Chebyschev inequality or other high moment based probability bounds on the events we might be interested in.
We leave further investigations for future work.

\section{The Karp-Luby Estimator}
\label{sec:karp-luby}
Computing the marginal probability of a tuple in the output of a set-probabilistic database query has been studied extensively.
To the best of our knowledge, the current state of the art approximation algorithm for this problem is the Karp-Luby estimator~\cite{DBLP:journals/jal/KarpLM89}, which first appeared in MayBMS/Sprout~\cite{DBLP:conf/icde/OlteanuHK10}, and more recently as part of an online ``anytime'' approximation algorithm~\cite{FH13,heuvel-19-anappdsd}.  

The estimator works by observing that for any $\ell$ random binary (but not necessarily independent) events $\vct{W}_1, \ldots, \vct{W}_\ell$, the probability of at least one event occurring (i.e., $\probOf\inparen{\vct{W}_1 \vee \ldots \vee\vct{W}_\ell}$) is bounded from above by the sum of the independent event probabilities (i.e., $\probOf\inparen{\vct{W}_1 \vee \ldots \vee \vct{W}_\ell} \leq \probOf\inparen{\vct{W}_1} + \ldots + \probOf\inparen{\vct{W}_\ell}$).  
Starting from this (`easily' computable and large) value, the estimator proceeds to correct the estimate by estimating how much of an over-estimate it is.
Specifically, if $\mathcal P$ is the joint distribution over $\vct{W}$, the estimator computes an approximation of:
$$\mathcal O = \underset{\vct{W} \sim \mathcal P}{\expct}\Big[
\left|\comprehension{i}{\vct{W}_i = 1, i \in [\ell]}\right| 
\Big].$$
\AH{Why are we computing the cardinality of variables that equal 1?}
The accuracy of this estimate is improved by conditioning $\mathcal P$ on a $W_i$ chosen uniformly at random (which ensures that the sampled count will be at least 1) and correcting the resulting estimate by $\probOf\inparen{W_i}$.  With an estimate of $\mathcal O$, it can easily be verified that the probability of the disjunction can be computed as:
\AH{A bit confused on the above sentence:
\par
i) what is meant by conditioning $\mathcal{P}$ on a $W_i$,

ii) why is each $W_i$ a monomial?
}
$$\probOf\inparen{\vct{W}_1 \vee \ldots \vee\vct{W}_\ell} = \probOf\inparen{\vct{W}_1} + \ldots + \probOf\inparen{\vct{W}_\ell} - \mathcal O$$

The Karp-Luby estimator is employed on the \abbrSMB representation\footnote{Note that since we are in the set semantics, in the lineage polynomial/formula, addition is logical OR and multiplication is logical AND.} of $\circuit$ (to solve the set-PDB version of \Cref{prob:intro-stmt}), where each $W_i$ represents the event that one monomial is true.
By simple inspection, if there are $\ell$ monomials, this estimator has runtime $\Omega(\ell)$.  Further, a minimum of $\left\lceil\frac{3\cdot \ell\cdot \log(\frac{2}{\delta})}{\epsilon^2}\right\rceil$ invocations of the estimator are required to achieve  $1\pm\epsilon$ approximation with probability at least $1-\delta$~\cite{DBLP:conf/icde/OlteanuHK10}, entailing a runtime at least quadratic in $\ell$.
As an arbitrary lineage circuit $\circuit$ may encode $\Omega\inparen{|\circuit|^k}$ monomials, the worst case runtime is at least $\Omega\inparen{|\circuit|^{2k}}$ (where $k$ is the `degree'  of lineage polynomial encoded by $\circuit$). By contrast note that by the discussion after \Cref{lem:val-ub} we can solve \Cref{prob:intro-stmt} in time $O\inparen{|\circuit|^2}$ for all \abbrBIDB circuits {\em independent} of the degree $k$.



\section{Parameterized Complexity}\label{sec:param-compl}

In \Cref{sec:hard}, we utilized common conjectures from fine-grained complexity theory. The notion of $\sharpwonehard$ is a standard notion in {\em parameterized complexity}, which by now is a standard complexity tool in providing data complexity bounds on query processing results~\cite{param-comp}. E.g. the fact that $k$-matching is $\sharpwonehard$ implies that we cannot have an $n^{\Omega(1)}$ runtime. However, these results do not carefully track the exponent in the hardness result. E.g. $\sharpwonehard$ for the general $k$-matching problem does not imply anything specific for the $3$-matching problem. Similar questions have led to intense research into the new sub-field of {\em fine-grained complexity} (see~\cite{virgi-survey}), where we care about the exponent in our hardness assumptions as well-- e.g. \Cref{conj:graph} is based on the popular {\em Triangle detection hypothesis} in this area (cf.~\cite{triang-hard}).

\end{document}